\documentclass[12pt,a4paper]{article}
\usepackage{amsmath,color,subfigure}
\usepackage[dvips]{graphicx}
\usepackage{amsthm,amssymb}
\usepackage{bm}
\usepackage{psfrag}
\setkeys{Gin}{draft=false}
\usepackage{subfig}
\usepackage{rotating}

\begin{document}

%
%

\title{Energetics of PCMDI/CMIP3 Climate Models: Energy Budget and Meridional Enthalpy Transport}

\author{V. Lucarini, Dept. of Meteorology \& Dept. of Mathematics\\ University of Reading, Reading, RG6 6BB, UK \\ Email: \texttt{v.lucarini@reading.ac.uk} \\ F. Ragone, Meteorologisches Institut, KlimaCampus\\ University of Hamburg, Hamburg, Germany}

\maketitle

%
%


\begin{abstract}
We analyze the PCMDI/CMIP3 simulations performed by climate models (CMs) using pre-industrial and SRESA1B scenarios. Relatively large biases are present for most CMs when global energy budgets and when the atmospheric, oceanic, and land budgets are considered. Apparently, the biases do not result from transient effects, but depend on the imperfect closure of the energy cycle in the fluid components and on inconsistencies over land. Therefore, the planetary emission temperature is underestimated. This may explain the CMs' cold bias. In the pre-industrial scenario, CMs agree on the location in the mid-latitudes of the peaks of the meridional atmospheric enthalpy transport, while large discrepancies exist on the intensity. Disagreements on the location and intensity of the oceanic transport peaks are serious. With increased $CO_2$ concentration, a small poleward shift of the peak and an increase in the intensity of the atmospheric transport of up to 10\% are detected in both hemispheres. Instead, most CMs feature a decrease in the oceanic transport intensity in the northern hemisphere and an equatorward shift of the peak in both hemispheres. The Bjerkens compensation mechanism is active both on climatological and interannual time scales. The peak of the total meridional transport is typically around $35^\circ$ in both hemispheres and scenarios, whereas disagreements on the intensity are relevant. With increased $CO_2$ concentration, the total transport increases by up to 10\%, thus contributing to polar amplification. Advances in the representation of physical processes are definitely needed for providing a self-consistent representation of climate as a non-equilibrium thermodynamical system.
\end{abstract}

%
%

%

\section{Introduction}

The investigation of the global structural properties plays a central role for the provision of a unifying picture of the climate system (CS). Such an endeavor is of fundamental importance for understanding climate variability and climate change on a large variety of scales, which encompass major paleoclimatic shifts, almost regularly repeated events such as ice ages, as well as the ongoing and future anthropogenic climate change, as envisioned by the scientific programme proposed in the landmark book by Saltzmann \cite{Saltzmann}.

This effort has significant relevance also in the context of the ever-increasing attention paid by the scientific community to the quest for reliable metrics to be used for the validation of climate models (CMs) of various degrees of complexity, as explicitly requested by the 4th Assessment Report of the Intergovernmental Panel on Climate Change (IPCC4AR) \cite{IPCC} and for the definition of strategies aimed at the radical improvement of their performances, beyond incremental advances often obtained at the price of large increases in requested computing power. See the discussions in, e.g., \cite{DefCliSci,Held,IntercomparisonNorthWinterModels,DanubioGCM,Validation}. EU projects such as PRUDENCE, ENSEMBLES, DEMETER, US initiatives such as the Phase 3 of the Third Coupled Model Intercomparison Project developed within the Program for Climate Model Diagnosis and Intercomparison (PCMDI/CMIP3) and innovative volunteering activities as those organized by the UK climateprediction.net project have fueled great efforts aimed at auditing CMs' performances. The pursuit of a "quantum leap" in climate modeling - which definitely requires new scientific ideas rather than just faster supercomputers - is becoming a key issue in the climate community \cite{Shukla}.

In a modern perspective, the climate can be seen as a complex, non-equilibrium system, which generates entropy by irreversible processes, transforms moist static energy into mechanical energy \cite{Lorenz55,Lorenz} as if it were a heat engine \cite{Johnson2000}, and, when the external and internal parameters have fixed values, achieves a steady state by balancing the input and output of energy and entropy with the surrounding environment \cite{Stone78b,Peixoto,Ozawa}. The tools of phenomenological non-equilibrium thermodynamics \cite{Prigogine,Groot} seem very well suited in defining a new point of view for the analysis of the CS \cite{Kleidon,L09}, for understanding its variability and its large-scale processes, including the atmosphere-ocean coupling and the hydrological cycle. Moreover, such approach seems valuable for understanding the mechanisms involved in climate phase transitions observed at the so-called tipping points, i.e., conditions under which catastrophes may occur for small variations in the boundary conditions or in the internal parameters of the system \cite{tipping}. Recent results suggest that great progresses in this direction are definitely within reach \cite{Snowball,Climatechange}.

The energy driving the CS comes from the Sun in form of shortwave  radiation. Part of the radiation is scattered elastically back to space, with an efficiency which depends critically on the surface irradiated the incoming light. Clouds, ice- and snow-covered and barren surface have in general the ability to scatter back to space relevant portions of the incoming radiation, while the process is much less efficient for, e.g.,  forested lands and especially the ocean surface. Albedo is commonly referred to as the fraction of the scattered with respect to the incoming radiation, and is in general a two-dimensional time-dependent variable. The rest of the incoming radiation is absorbed by the fluid and solid components of the CS. The absorption process is not homogeneus in space and time, because of variability in the amount of incoming energy and because the interaction of radiation with matter depends critically on the nature and density of the irradiated medium. Most of the net incoming solar radiation is absorbed at surface and is redistributed throughtout the atmospheric column by a variety of means such as convection, latent and sensible heat fluxes, and longwave radiative exchanges within the atmosphere and between the atmosphere and the underlying surface. Organized and turbulent atmospheric motions and the hydrological cycle (including the formation and dissipation of clouds) are the direct result of this process of energy redistribution. Oceanic motions are also generated, basically as an outcome of the variable exchange of heat, water and momentum between the water masses and the overlying atmosphere. Energy is irradiated back to space in the form of longwave radiation approximately thermalized at temperatures typical of the CS, so that on the global average the net incoming solar radiation is approximately compensated by the outgoing longwave radiation. Since clouds have a great impact in both determining the amount of scattered solar radiation - because of their typically high albedo - and of outgoing longwave radiation - because of their typically high emissivity - they have a crucial role in modulating the local and global energy budget. On a planetary scale, large scale atmospheric and oceanic motions allow for the reduction of the temperature difference between high and low latitudes in both hemispheres with respect to what would be enforced by purely local radiative and convective processes. The low and high latitude regions differ largely in terms of net incoming radiation, because of astronomical factors and local albedo properties, and feature opposite sign of the net budget of the radiative flux at the top of the atmosphere (TOA). In general, we understand that the so-called differential heating, both in the small and in the large scales, is the basic ingredient of the non-equilibrium nature of the CS \cite{Lorenz,Peixoto}. The described processes have relevant impacts also in terms of the second law of thermodynamics, as in the CS entropy production is inextricably related to the irreversible heat transfer from high to low temperature regions \cite{Goody2000,Ozawa,Snowball,Climatechange}.

Therefore, it seems of the greatest importance to devote attention to the analysis of the consistency of the state-of-the-art CMs in the representation of the energy and entropy budget and transport in the CS as a whole as well as in its main subdomains. While the analysis of entropy budget will be the subject of a separate publication, we hereby wish to focus on the analysis of energetics of the CS in selected simulations of the CMs included in the IPCC4AR, by taking advantage of the immense data repository provided by the PCMDI/CMIP3 project.

In this work we focus on one side on the behavior of the CMs under steady state conditions, and thus take advantage of the data produced with the pre-industrial scenario, where CMs are integrated for several hundreds of years with fixed atmospheric composition with 280 ppm $CO_2$ concentration. On the other side we aim at understanding how different is the CMs' response to the $CO_2$ radiative forcing impacts. For this latter purpose, we have considered the CMs' outputs produced under the Special Report on Emission Scenario A1B (SRESA1B) scenario, which foresees a stable $CO_2$ concentration of 720 ppm after 2100 following a prolonged increase after 2000. The initial conditions of the SRESA1B runs are provided by the final state of the XX century runs, which are in turn initialized in the second half of the XIX century using the pre-industrial runs state as initial conditions. Since we wish to avoid the period of direct forcing and focus on the time frame when CMs start to relax towards the steady state, we analyze data referring to the XXII and XXIII century.


\subsection{Global Budgets in a Steady State}
Following and expanding previous analyses \cite{WildSolar,WildSurface,EEB09,TF09,TF10b}, we first perform a thorough audit of the CMs in their representation of the energy budget of the whole CS, and, separately, of the atmosphere, of the ocean, and of the land, in order to test how consistently steady state conditions are realized in the case of pre-industrial simulations. Moreover, by comparing the obtained results with the energy budgets obtained from the runs performed under SRESA1B conditions, we understand how the CMs describe the CS response to time-dependent $CO_2$ radiative forcing.

As discussed above, basic physical principles suggest that under generic boundary conditions the ultimately realised steady state obeys general zero-sum properties for energy fluxes \cite{Prigogine,Peixoto}. Therefore, when statistical stationarity is eventually obtained after transients have died out, a physically consistent - even if not necessarily realistic in the representation of each and every climatic feature - CM should feature a vanishing (when long time averages are considered) energy budget as a whole and for any of its subdomains. If this is not the case, the energy bias of the CM has to be attributed to unphysical energy sources or sinks. We remark that this property has nothing to do with the so-called operation of \textit{model tuning}: a CM with a bad parametric tuning probably features unaccurate or even unrealistic climate properties, but, if the CM is energetically consistent and steady state is attained, the energy budget for any subdomain should be vanishing on the average.

\begin{figure}[htbp] 
   \begin{center}

   \includegraphics[width=0.5\textwidth,angle=270]{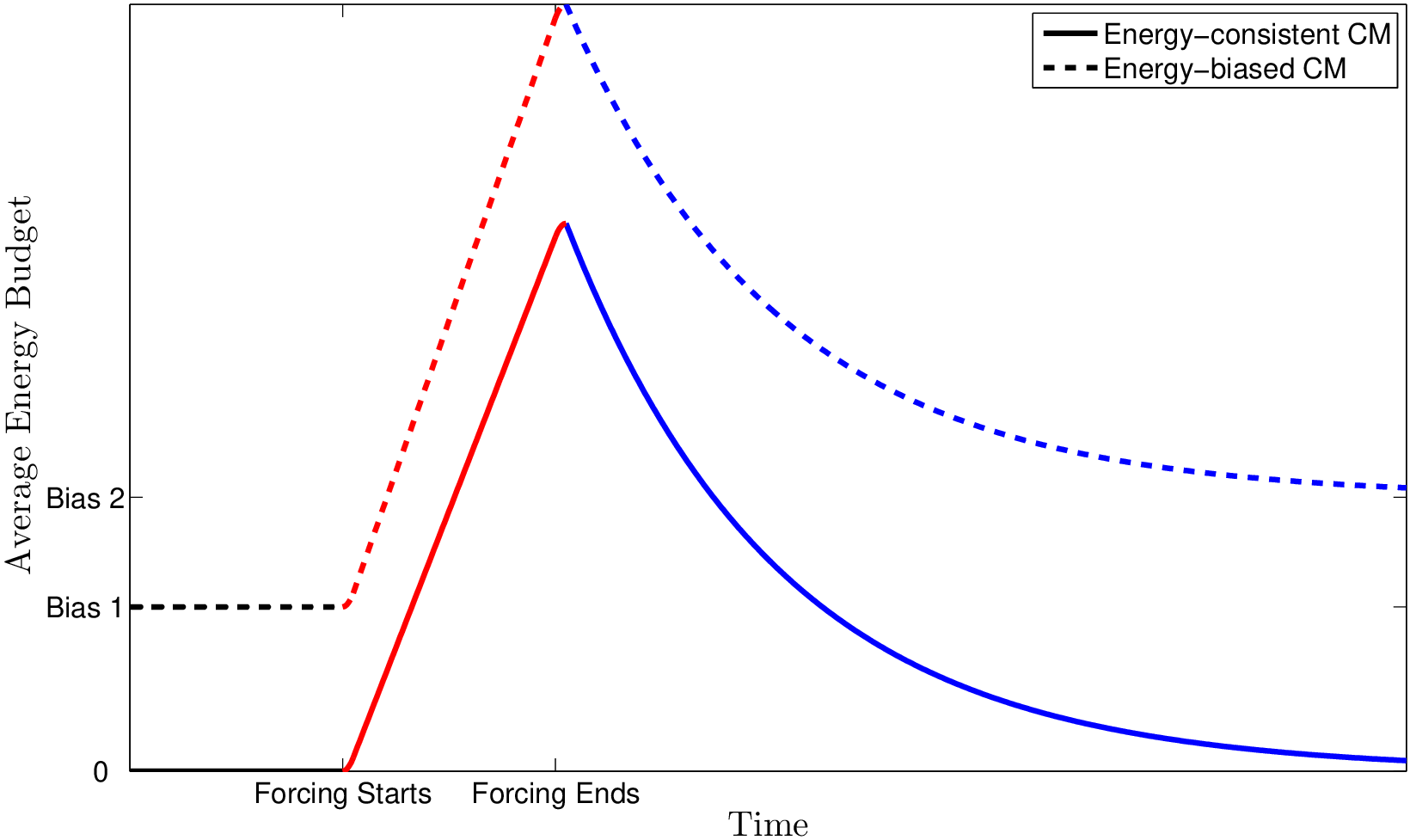}
   \includegraphics[width=0.5\textwidth,angle=270]{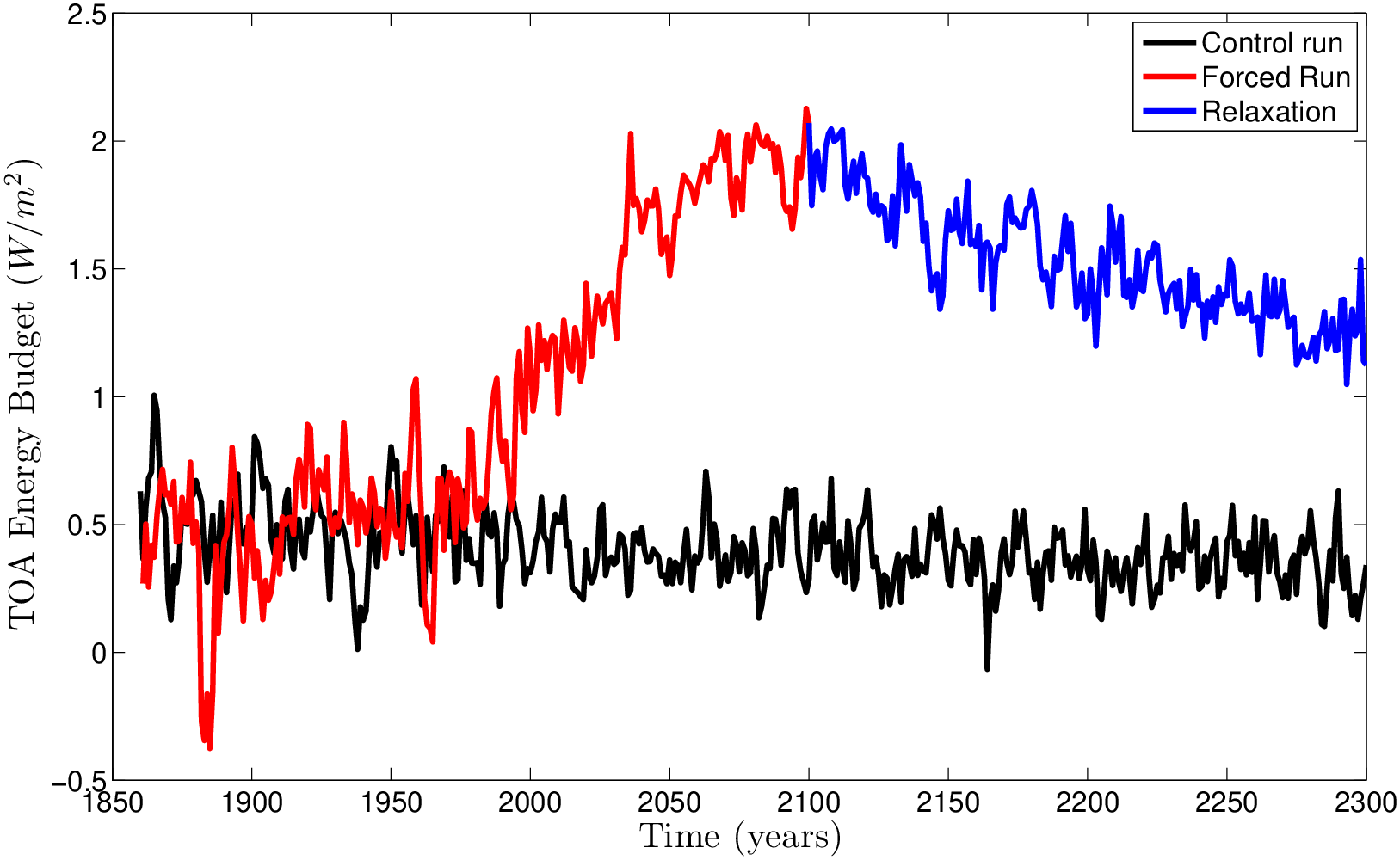}
      \end{center}
    \caption{First Panel: Idealized time dependence of the TOA global energy budget as given by a CM which has a self-consistent representation of energy exchange processes (solid line) versus an energy biased CM. Black, red, and blue lines refer to the initially unperturbed runs, the forced runs, and the runs performed with newly established fixed parameters, respectively. Second Panel: Smoothed (5-year filter) yearly time series of the GFDL2.1 model (CM 9) TOA energy budget. The black line refers to the pre-industrial run. The red line refers to the XX century simulation (started from year 1 of the pre-industrial run) and to the XXI century portion of the SRESA1B simulations (started from the end of the XX century simulation). The blue line refers to the XXII and XXIII century SRESA1B simulation.}
  \label{bias}
\end{figure}

The first panel of Fig. (\ref{bias}) contrasts the idealized behavior of a CM whose energy exchange processes are consistent (solid line) with that of a biased CM (dashed line). In the former case, in the initial steady state conditions (black line) the TOA average energy budget vanishes, whereas in the latter a spurious bias (Bias 1 in the figure) is present. When the parameters of the system are changed with time so that forcing is applied (red line), both models foresee a non vanishing TOA budget (positive in this representation, as in global warming conditions). When the parameters are held fixed to newly established value, the system relaxes toward the new steady state within a a time comparable to a few units of the slowest of the internal time scales (the oceanic one, in usual climate conditions). But whereas the energy-consistent model readjusts towards a zero TOA energy balance, the other model's TOA energy budget will readjust towards a non-vanishing bias, generically different (and so climate-dependent) from the initial one. Note that, whereas in the case of a good model the net total heating resulting from the forcing corresponds to integral of the energy budget, in the case of imperfect model with potentially climate-dependent bias the net heating of the system is in principle ill-defined. Conceptually analogous cartoons actually apply not only for the global energy budget, but to the energy budget of any climatic subdomains, with the crucial difference that the slowest internal time scale will be faster for, e.g., the atmosphere than for the ocean.

Given the complexity of the processes inside the climate system and the fact that large cancelations are involved, the provision of a consistent  or at least reasonable energy budget is in general far from being an obvious task, so that an extensive CMs intercomparison analysis is necessary to understand to what extent the actual CMs will conform to the case depicted in the first panel of Fig. (\ref{bias}).

In the case of renalyses, the bias in the diagnosed global energy budget is of the order of -10 $W m^{-2}$, which is about one order of magnitude larger in absolute value (and wrong in the sign) than the reasonably expected imbalance due to the anthropogenic and natural forcings in the second half of the XX century. Surprisingly, the performance of these datasets for this specific benchmark does not improve even after 1979, which marks the beginning of the massive assimilation of the satellite data \cite{Yang99,ERA40OLR}.

Additionally, in spite of ever increasing and improving sources of observative data from satellites on incoming and outgoing radiation at the top of the atmosphere (TOA), it is really hard to restrict the uncertainties on the observed yearly averaged energy balance of the planet to the order of 1 $W m^{-2}$ \cite{EEB97,EEB09}. Very recently, it has emphasized how important is to correctly track, in models and observations, the Earth's energy for understanding climate change and variability \cite{TF10b}

Some studies have shown that numerical codes commonly adopted for the representation of the dynamics of geophysical fluids are not necessarily energy-conserving, as kinetic energy losses due to friction and diffusion are not exactly fed back in the system as thermal forcing, and that commonly adopted hyperdiffusive schemes seem to be especially problematic in this regard \cite{Becker2003,LorenzVK}.

\subsection{Transports}
When considering global energy balances the CS is reduced to a zero-dimensional system. We then make a step forward and tackle the problem of understanding how models represent the meridional enthalpy transport by the CS as a whole and by the atmosphere and the ocean separately. In order to achieve steady state, a net positive TOA energy budget in a given location must be compensated when long term averages are considered by a net divergence of the enthalpy flux. The TOA planetary budget has to a very good approximation a zonal structure, with low latitude regions featuring a positive budget between the net input of solar radiation (controlled by incoming solar radiation and albedo) and the output of longwave radiation (controlled by surface temperature, water vapour concentration, and clouds). A negative budget is instead observed in high latitude regions \cite{Peixoto}. Therefore, on the average, the fluid envelope of the CS transports large amounts of heat from low to high latitude regions in both hemispheres \cite{Stone78,Tapio}. See \cite{Frierson,Vallis09} for recent theoretical treatments in a rather general setting.

It is then found that in each hemisphere the maximum intensity of the meridional transport is in the long term equal to the absolute value of the total energy imbalance in the low latitude region (or, equivalently, the energy deficit in the high latitude region), whereas its maximum is located at the \textit{transition} latitude where the TOA energy balance vanishes \cite{Peixoto,Tapio}. Therefore, the location and the intensity of the maximum of the transport provide fundamental information on the properties of the CS.

Earlier reconstructions of the meridional transport using interpolations of radiosonde data - with unavoidable uncertainties due to the uneven spatial coverage and homogeneity issues of the datasets - suggested a comparable contribution to the meridional transport by the ocean and by the atmosphere \cite{Peixoto}. Instead, more recent data suggest that, especially in the southern hemisphere, the atmospheric transport definitely plays a dominant role  \cite{Trenberth2001}.

On the other hand, large scale organized atmospheric and oceanic motions are result of the mechanical work (then dissipated in a turbulent cascade) produced by the climate engine thorugh the Lorenz energy cycle, which is fuelled essentially by the presence of temperature differences across the CS \cite{Lorenz}. Recently, a rather complete picture of the Lorenz energy cycle in a simple yet rigorous thermodynamic terms has been provided \cite{Johnson2000} and linked to estimates of the turbulent entropy production of the planet \cite{L09}.

Many properties of the CS play an important role in determining the properties of the meridional atmospheric and oceanic transport, the main being the meridional gradient of the planetary albedo. Moreover, on Earth the mechanisms of transport are rather different in the northern and southern hemisphere. In the northern hemisphere, considerable contributions to the meridional transport are given, separately, by the transient and stationary waves of the atmosphere as well as by the ocean. Instead, in the southern hemisphere, most of the transport is performed by transient atmospheric waves only. Nonetheless, the maximum value and the location of the peak of the transport are very similar in the two hemispheres \cite{Trenberth2001}.

In fact, some basic features of the transport profile seem to be symmetric with respect to the equator as they are constrained by basic fluid dynamical laws and by the geometrical properties of the system. In a classical paper, Stone \cite{Stone78b} showed that in a planet the climatological total meridional transport profile (and especially the position of its peak) is determined, to the first order of approximation, by spherical geometry and latitudinal dependence of TOA incoming radiation only. See also \cite{Ende09} for a recent re-examination of this brilliant idea.

Since we still do not have a comprehensive rigorous theory able to express the peak location and strength of the meridional transport and to accomodate the variety of processes contributing to it, it seems crucial to test the degree of realism and consistency of CMs in representing these basic climatic features.

The direct calculation of the transports from 3D climate fields requires very serious diagnostic tools and high-resolution data (in time), since quadratic quantities are involved. Nevertheless, under the assumption of steady state conditions and of no biases in the energy balances of the whole CS, of the atmosphere and of the ocean, it is possible to rigorously reconstruct the meridional enthalpy transport by atmosphere and ocean from the zonal averages of the heat fluxes at the upper and lower boundaries of the fluid envelope \cite{Peixoto}. Even in the presence of biases in the energy budgets - if these are small enough with respect to the latitudinal variability -  such a reconstruction technique can be safely adopted with minor corrections, as discussed in \cite{Carissimo85,Trenberth2001}, where non steady state conditions like those of the present climate and rather heavily biased energy balances are analyzed. The meridional transport profiles provide a very strong characterization of the CMs, and it seems crucial to intercompare their performances in order to understand how well they agree in the representation of such crucial climatic properties. Moreover, the analysis of the response to $CO_2$ increase of the meridional transport probably provides the most basic way of investigating the adjustment of the dynamics of the fluid components of the CS.

\subsection{Goals and Structure of the Paper}
The overall goal of this paper is to present a complete review of the state-of-the-art of climate modelling in terms of their representation of the CS energetics and of the energetics of ocean and atmosphere seperately. The paper also wishes to stimulate lines of activities aimed at improving the reliability of CMs, which is a task of the uttermost urgency in the context of the preparation of the future IPCC reports.  The paper is organized as follows. In Section \ref{sec2}, starting from the Eulerian energy balance equations, we briefly recapitulate the equations determining the energy budget of the CS and of its subdomains and show how the meridional transport can be inferred. In Section \ref{sec3} we present the data considered in this work and discuss the methods used in their processing. In Section \ref{sec4} we present our results, in Section \ref{sec5} we discuss our findings, and in Section \ref{sec6} we give our conclusions and provide perspectives for future work.

\section{Energy Budget in the Climate System}
\label{sec2}
In this section we review the derivation of the zonally and globally integrated energy balance of the CS and of its subdomains and show how, assuming steady state conditions, it is possible to infer the meridional enthalpy transport by the atmosphere, by the ocean, and by the system as a whole. We also discuss how to derive such transport properties when energy budgets are not perfectly close, due to actual deviations from  steady state or to spurious biases in the data.

The evolution equation in Eulerian form for the total energy density $e$ in the fluid component of the Earth system, given by the sum of the kinetic and moist static potential components, can be written as \cite{Peixoto}:
\begin{equation}
{\frac{\partial \rho e}{\partial t}} = - \nabla\cdot(\rho e \bm{c}) - \nabla \cdot \bm{F_R} - \nabla\cdot\bm{F_S} - \nabla\cdot\bm{F_L} - \nabla\cdot(p\bm{c} + \bm{\tau}\cdot\bm{c})
\label{totalenergy}
\end{equation}
where $\rho$ is the density, $\bm{c}$ is the velocity vector, $\bm{F_R}$, $\bm{F_S}$, and $\bm{F_L}$ are the vectors of the radiative, turbulent sensible and turbulent latent heat fluxes, respectively, $p$ is the pressure and $\bm{\tau}$ is the stress tensor. It is useful to rewrite the previous equation as:
\begin{equation}
{\frac{\partial \rho e}{\partial t}} = - \nabla\cdot\bm{J_h} - \nabla \cdot \bm{F_R} - \nabla\cdot\bm{F_S} - \nabla\cdot\bm{F_L} - \nabla\cdot(\bm{\tau}\cdot\bm{c})
\label{totalenergy2}
\end{equation}
where we have introduced the total enthalpy transport $\bm{J_h}=( \rho e + p ) \bm{c}= \rho h  \bm{c}$ where the standard definition of enthalpy $h=e+p/\rho$ is adopted \cite{Fermi,Kundu}. Note that the more commonly the expression energy transport is adopted \cite{Peixoto}. As discussed in \cite{Ambaum}, Eq. (\ref{totalenergy2}) expresses a local balance corresponding to the first law of thermodynamics for an open system, and enthalpy (rather than energy) transports enter the picture because exchange of mass takes place. Note also that since all terms on the right hand side of Eqs. (\ref{totalenergy}-\ref{totalenergy2}) are written as convergence of vector fields, by using Gauss' theorem it is easy to obtain an expression for the derivative of the total energy content of a subdomain of the CS in terms of the fluxes at the boundary.

Expressing the divergence operator in spherical coordinates, and taking the usual thin shell approximation so that $R+z\approx R$ with $R$ the radius of the planet, we obtain:
\begin{eqnarray}
{\frac{\partial \rho e}{\partial t}} = - {\frac{1}{R \cos{\varphi}}} {\frac{\partial J_{h\lambda}}{\partial \lambda}}  - {\frac{1}{R \cos{\varphi}}}{\frac{\partial J_{h\varphi} \cos{\varphi}}{\partial \varphi}} -  {\frac{\partial J_{h z}}{\partial z}}  +\nonumber \\  -{\frac{\partial}{\partial z}} ( F_{Rz} + F_{Sz}+ F_{Lz} + (\bm{\tau}\!\cdot\!\bm{c})_z),
\label{eq: total energy evolution spherical}
\end{eqnarray}
where in the last term of Eq. \ref{eq: total energy evolution spherical} we have retained only the vertical components of the radiative and turbulent fluxes, and of the contributions due to frictional stresses, since they are largely dominant by various orders of magnitude \cite{Peixoto}.

If we want to characterize the large-scale properties of the CS, we need to analyze the properties of the temporal and spatial averages of the energy budget. It is particularly relevant to exploit Eq. (\ref{eq: total energy evolution spherical}) to study the interaction between the various climatic sub-systems. Let $\Omega$ be a subdomain of the CS. For our purposes, $\Omega$ will refer to the atmosphere ($\Omega$=$A$), the ocean ($\Omega$=$O$), or the land subdomain ($\Omega$=$L$), as well as the complete system ($\Omega$=$T$). Given a generic field $a$=$a(t, \lambda, \varphi, z)$, we denote as \textit{mass weighted vertical integral} of $a$ over $\Omega$ the integral
\begin{equation}
A_\Omega(t, \lambda, \varphi) \equiv \int_{z_b}^{z_t} \rho \, a(t, \lambda, \varphi,z') \,\, dz'
\label{eq: global mean}
\end{equation}
where $z_b$ and $z_t$ are respectively the lower and the upper boundaries that delimit $\Omega$. Note that where the thickness of the subdomain $\Omega$ is vanishing - e.g., $\Omega=O$ and the coordinates $(\lambda, \varphi)$ correspond to continental areas - Eq. (\ref{eq: global mean}) is still well-defined as $A_\Omega(t, \lambda, \varphi)$ results automatically to be zero.
We can then define the \textit{zonal integral}:
\begin{equation}
[A_\Omega](t, \varphi) \equiv  \int _{0}^{2\pi} A_\Omega(t, \lambda', \varphi) R \cos{\varphi} \, d\lambda'
\label{eq: zonal mean}
\end{equation}
%
%
In order to characterize the 1-dimensional structure of the system, we integrate zonally and vertically Eq. \ref{eq: total energy evolution spherical}. For the entire planet ($\Omega=T$) we have
\begin{equation}
\dot{[E_T]} =  - {\frac{1}{R}} {\frac{\partial {T_T} }{\partial \varphi}} + [{F_{R}}]_{TOA}
\label{eq: TOA zonal}
\end{equation}
where $[F_{R}]_{TOA}$ is the zonal mean of $F_R$ evaluated at TOA and
\begin{equation}
T_T(t,\varphi)= \int_{0}^{2\pi} \int_{z_b}^{TOA} {J_{h\varphi}(t,\lambda',\varphi,z')} R \cos{\varphi} \, dz'd\lambda'
\label{eq: T zonal}
\end{equation}
is the total transport of enthalpy realized through the latitudinal wall at latitude $\varphi$, being $z_b$ the bottom of the ocean. Similarly, we can derive the equations for the atmosphere, ocean and land:
\begin{equation}
\dot{[E_A]} =  - {\frac{1}{R}} {\frac{\partial {T_A} }{\partial \varphi}} + [F_{R}]_{TOA} + [ {F_{R}} + {F_{S}} + {F_{L}}]_{surf,A};
\label{eq: atm zonal}
\end{equation}
\begin{equation}
\dot{[E_O]} =  - {\frac{1}{R }} {\frac{\partial {T_O} }{\partial \varphi}}  - [{F_{R}} + {F_{S}} + {F_{L}}]_{surf,O};
\label{eq: ocean zonal}
\end{equation}
\begin{equation}
\dot{[E_L]} =  - [{F_{R}} + {F_{S}} + {F_{L}}]_{surf,L};
\label{eq: land zonal}
\end{equation}
where the subscripts in the last terms on the right hand side of Eqs. (\ref{eq: atm zonal}-\ref{eq: land zonal}) refer to the boundary considered. Moreover:
\begin{equation}
T_A(\varphi)=\int_{0}^{2\pi}  \int_{z_s}^{TOA} {J_{h\varphi}(t,\lambda',\varphi,z')}  R \cos{\varphi} \, dz'd\lambda'
\label{eq: T zonal_atm}
\end{equation}
and
\begin{equation}
T_O(\varphi)= \int_{0}^{2\pi}  \int_{z_b}^{z_s} {J_{h\varphi}(t,\lambda',\varphi,z')}  R \cos{\varphi} \, dz'd\lambda'
\label{eq: T zonal_sea}
\end{equation}
are the atmospheric and oceanic total transport through the latitudinal wall at latitude $\varphi$, respectively. Note that, since land is not a fluid subdomain of the system, the large scale enthalpy transport through this medium is entirely negligible, so that no transport term is present in Eq. (\ref{eq: land zonal}). Note also that we have neglected the transfer of mechanical energy from the atmosphere to the ocean since it is smaller by some orders of magnitude than the other energy exchange terms \cite{Peixoto}. Such a term is instead of great importance when considering the driving mechanisms of the large scale ocean circulation \cite{Kuhl07}. Note that the fact that this term is so small compared to the thermal energy fluxes is related to the very low efficiency of the CS as a whole \cite{L09}.

If we further integrate Eq. (\ref{eq: TOA zonal}) and Eqs. (\ref{eq: atm zonal}-\ref{eq: land zonal}) over latitude from $-\pi/2$ to $\pi/2$, we obtain the global energy budget and the energy budgets for the atmosphere, the ocean, and the land as follows:
\begin{equation}
\dot{\{E_T\}} =  \{F_{R}\}_{TOA},
\label{eq: TOA glob}
\end{equation}
\begin{equation}
\dot{\{E_A\}} =  \{F_{R}\}_{TOA} + \{F_{R}\}_{surf,A} + \{F_{S}\}_{surf,A} + \{F_{L}\}_{surf,A},
\label{eq: atm glob}
\end{equation}
\begin{equation}
\dot{\{E_O\}} =   - \{F_{R}\}_{surf,O} + \{F_{S}\}_{surf,O} + \{F_{L}\}_{surf,O},
\label{eq: ocean glob}
\end{equation}
\begin{equation}
\dot{\{E_L\}} =  - \{F_{R}\}_{surf,L} + \{F_{S}\}_{surf,L} + \{F_{L}\}_{surf,L},
\label{eq: land glob}
\end{equation}
where we have defined:
\begin{equation}
{\{X\}}(t) =  \int_{-\pi/2}^{\pi/2}[X](t,\varphi)R \, d\varphi.
\label{averagingtot}
\end{equation}
for a generic variable $[X](t,\varphi)$.

\subsection{Time averages}
If the system is forced with the usual seasonal cycle and the internal and external parameters of the system are not time-dependent, we expect that after a suitable transient, steady state conditions are realized. As different domains of the climatic system have different internal time scales, the realization of steady state conditions occurs in a large variety of time horizons when the system is initialized from arbitrary initial conditions or when parameters are suddenly altered. In general, when time-dependent forcing in the form of parametric variations are present, a climatic subsystem is in quasi-steady conditions when its internal time scales are much faster than the characteristic time scale of the parametric change. In this case, the subsystem is driven by the parametric modulation of the forcing. The atmosphere and land subdomains have a much shorter time scale than the ocean. The atmosphere and the land can be considered to be balanced over time scales of less than one year, while the ocean features time scales ranging from some years to several hundreds of years. The current climate radiative forcing (and, arguably, those projected under the SRESA1B scenario) are slow enough that the atmosphere can be expected to be in quasi-steady-state conditions, whereas the ocean lags behind \cite{IPCC}. Following \cite{Saltzmann}, we can define the atmosphere and land subdomain as \textit{thin} climatic subdomains, while the ocean is a \textit{thick} subdomain. The presence of interacting subsystem with very different characteristic time scales is one of the key elements of complexity of the CS.

Under steady state conditions, the long-term average of the derivative of any climatic fields needs to vanish, because otherwise trends would be present. When long term averages are applied to Eqs. (\ref{eq: TOA glob}-\ref{eq: land glob}), we obtain the general formula:
\begin{equation}
\overline{\dot{\{E_\Omega\}}} =  \overline{\{B_\Omega\}}=0,
\label{eq: TOA glob_ave}
\end{equation}
where the overbar indicates the time averaging operation, and $B_\Omega$ is the algebraic sum of the fluxes entering the domain thorough the upper and lower boundaries of the domain $\Omega$.
Instead, the time-average of the zonally and vertically integrated energy equation for a generic domain $\Omega$ can be written as:
\begin{equation}
\overline{\dot{[E_\Omega]}(\varphi)} =  - {\frac{1}{R}} {\frac{\partial \overline{T_\Omega} }{\partial \varphi}}+ \overline{[B_{\Omega}]}.
\label{eq: D zonal}
\end{equation}
Under steady state conditions, the left hand side of Eq. (\ref{eq: D zonal}) vanishes, so that the average transport $\overline{T_\Omega}$ can be computed as
\begin{equation}
\overline{T_\Omega}(\varphi)=-\int_{\varphi}^{\pi/2} \! \overline{[B_\Omega]}(\varphi') R \, d\varphi'.
\label{eq: trasporto}
\end{equation}
Since basically no substantial large scale enthalpy transport can take place through a solid medium, in stationary conditions the surface energy fluxes over land are expected to be locally vanishing when long term averages are considered. Instead, non-vanishing sustained local energy imbalances are possible at the atmospheric and oceanic interface and at TOA, since they are compensated by the divergence of the enthalpy transport realized by the motions of the system.

Geometry imposes that we should have zero meridional transport at the poles. In agreement with that, Eq. \ref{eq: trasporto} imposes $\overline{T_\Omega}$=0 at $\varphi=\pi/2$ (North Pole). Moreover, in steady state conditions the energy budget $\overline{\{B_\Omega\}}$ is expected to vanish, so that, consistently, $\overline{T_\Omega}$ is zero also at $\varphi=-\pi/2$ (South Pole).

As suggested in \cite{Carissimo85,Trenberth2001}, the procedure of inferring meridional transports from zonal energy balances can be considered robust enough also in the case of non-steady state conditions and inhomogeneous information, as when gathering from various sources observational estimates on the energy balance in present climate conditions. The condition for the applicability of the inference is that latitudinal variability of $\overline{[B_\Omega](\varphi)}$ should be much larger than $\overline{\{B_\Omega\}}/(2\pi R)$, which is easily met in terrestrial conditions for the outputs of climate simulations (see below). Nevertheless, in order to have vanishing transport also at $\varphi=-\pi/2$ we need to correct Eq. \ref{eq: trasporto}. The simplest way to accomplish that \cite{Carissimo85} is to redefine $\overline{T_\Omega}$ as follows:
\begin{equation}
\overline{T_\Omega}(\varphi)=-\int_{\varphi}^{\pi/2} \! \overline{[B_\Omega]}^{corr}(\varphi') R \, d\varphi',
\label{eq: transporto corretto}
\end{equation}
where
\begin{equation}
\overline{[B_\Omega]}^{corr} (\varphi)=\overline{[B_\Omega]} (\varphi) - \overline{\{B_\Omega\}}/{2\pi R}.
\label{eq: transporto corretto2}
\end{equation}
This amounts to assuming a constant flux correction throughout the surface. Different approaches, based, e.g., on
weighting differently various latitudinal belts on the basis of the a priori knowledge of the data uncertainties, provide rather similar outputs  when biases are overall small as in our case. Therefore, we will stick to the definition given
in Eq. (\ref{eq: transporto corretto}) throughout this paper.

\section{Data and Methods}
\label{sec3}
The PCMDI/CMIP3 (\texttt{http://www-pcmdi.llnl.gov/}) is a CMs intercomparison project run by the Lawrence Livermore National Laboratories (USA) freely providing the access to a vast dataset of outputs of standardized simulations performed by state-of-the-art CMs under various past, present and future climate scenarios. The PCMDI/CMIP3 includes data from over 20 CMs and has been a key contribution to the IPCC4AR, since it has provided a large community of scientists the possibility of analyzing extensively and systematically the performances of many CMs on the same test-beds, and has greatly stimulated the development of metrics to be used for intercomparison and validation studies. Presently, the new initiative PCMDI/CMIP5 (\texttt{http://cmip-pcmdi.llnl.gov/cmip5/}) is being set up along similar lines to support the preparation of the fifth IPCC Assessment Report. Data referring to this new generation of climate simulations should become available in the second half of 2011.

In this paper, we are interested in analyzing the behavior of the CMs in two standard IPCC4AR scenarios:
\begin{itemize}
  \item pre-industrial scenario, where the CS is integrated with fixed parameters and in particular with fixed $CO_2$ concentration set at 280 ppm until steady state conditions with no apparent trend in the climatic variables are observed;
  \item SRESA1B scenario, characterized by a constant increase of $CO_2$ from 2000 to 2100, when 720 ppm concentration is obtained, and kept subsequently constant.
\end{itemize}
It would have been interesting also to compare two different steady states (e.g., preindustrial vs. $2\times CO_2$ conditions) in order to characterize the generalized climate sensitivities of the system, but unfortunately the $2\times CO_2$ control run data provided in the PCMDI/CMIP3 datasets refer to slab ocean integrations, and therefore are of limited use.

In spite of the overall purpose of PCMDI/CMIP3 of providing complete and standardized datasets, we have found some deficiencies in data availability, which lead to difficulties in performing a complete and systematic intercomparison analysis. This calls for further improvements for initiatives following the magnificent efforts by PCMDI/CMIP3. A typical problem has been the lack of one or more surface fields for a CM in a certain scenario, which has restricted the analysis to TOA properties only. In some cases, also TOA fields are lacking, so that the corresponding CM is simply excluded from the analysis for that scenario. Moreover, some CMs have corrupted data for certain fields or incorrect metadata accompanying the datasets. Therefore, some data have bene directly obtained from the modeling centers.

Other inconsistencies are related to the temporal window of the data provided for different CMs for a given scenario. Regarding the PI scenario, the various CMs present time series of data of different lengths. In our case, we have chosen 100 years as the standard length of the time series to be analyzed. However, some CMs present less than 100 years of data. We have decided to perform the analysis also on these CMs, operating on the available data. Regarding the SRESA1B scenario, not all the CMs present data for the XXII \textit{and} for the XXIII century: some CMs' data stop on 2200, others even on 2100 (before stabilization of $CO_2$ has begun). Again, we have operated on all the available data after 2100. Finally, we have taken the radical choice of excluding the models with flux adjustment since their energetics is intrinsically biased - they are in contact with an infinite heat reservoir. An exception has been made for the INM-CM3.0 model since the flux adjustments are present only in a very limited oceanic region, so that no substantial impacts are expected at global level and for zoanlly integrated quantities. An overview of the CMs considered in this study and of the data availability of the scenario runs is given in Table \ref{tab:label}. Each CM is labelled by an identifying number: this convention is used throughout the text and in the figures.

In the optimal scenario, we have considered for both climate scenarios the fields of short and longwave radiation fluxes at TOA and at the surface, plus the fields of latent and sensible heat fluxes at the surface. Radiative fluxes as well as latent and sensible heat fluxes are expressed in $Wm^{-2}$. Moreover, we have taken into account the Land-sea mask, which quantifies the portion of surface covered by sea and by land. Depending on the considered CM, the values of the land-sea mask can be either boolean (0 = sea and 1 = land), or present continuous values between 0 and 1, thus allowing for mixed coverage. Fields relative to the energy fluxes are 2+1 dimensional, being time series of bidimensional fields, while the land-sea mask is a fixed bidimensional field. Since we are interested in yearly averages, we have reduced the computational and storage burden of our facilities by  downloading only pre-processed monthly mean fields, and then used them to compute the annual means. Data have been downloaded in \texttt{netcdf} format, have then been imported in MATLAB $7.0 ^\circledR$ environment with the freely available CSIRO netCDF/OPeNDAP software (\texttt{http://www.marine.csiro.au/sw/matlab-netcdf.html}), and have eventually been processed with in-house scripts.

Note that, in agreement with the actual estimates for the present climate \cite{Peixoto}, the surface average of the transfer of mechanical energy from the atmosphere in the ocean is in all models and in both scenarios positive and of the order of $10^{-2}$ $Wm^{-2}$ at most. This terms is several orders of magnitude smaller than the absolute value of the other terms contributing to the atmospheric and oceanic energy budgets. The tiny bias introduced by neglecting the transfer of mechanical energy does not, as we shall see below, add much to the analysis of the energy budgets, so that we can safely neglect this contribution.

%

\subsection{Statistical Estimators}
In the pre-industrial control runs, since the slowest external forcing to the system is the seasonal cycle (no variability of the incoming solar radiation is considered), steady-state conditions are expected to hold. Therefore, the time series of the yearly zonally integrated  $\langle [{B_\Omega}]\rangle_j$ and globally integrated energy balance $\langle \{{B_\Omega}\}\rangle_j$ (where $j$ refers to the $j^{th}$ of the N years) are expected to have vanishing expectation values for each domain $\Omega$ . Thus, for a given $\Omega$, the yearly meridional transport one-dimensional fields $\langle T_\Omega(\varphi)\rangle_j$  computed by applying Eq. (\ref{eq: transporto corretto}) are expected to be statistically equivalent.

For the XXII and XXIII century SRESA1B scenario outputs, the slowest forcing is also the seasonal cycle, so that generating the time series of yearly averaged values is a sensible operations. Nevertheless, as the system is readjusting to a new steady state, we may expect to see trends in the yearly data related to slow subsystems, i.e., the whole CS and the ocean subdomain. Actually, we find an approximate exponential behavior with decaying time which can be estimated as being of the order of 500-700 years, which nicely matches the characteristic time scale of the large scale ocean circulation \cite{Kuhl07}. For the time scales of our interest (order of 100 years), it can be treated safely as a linear trend.

%
The study of the global energy budgets relies on the analysis of the statistical properties of the mean values and the interannual variabilities of the yearly time series $\langle{\{B_\Omega\}}\rangle_{j}$. The estimate of the yearly average of the energy budget is computed  as the mean over the N years, and the estimate of the interannual variability is evaluated as the usual standard deviation of the time series. In the case the
time series are not stationary, a linear trend has been removed before computing the standard deviation.

Note that, since we are interested in the  statistical properties of the climatology, we need to compute the range of uncertainty of the estimates.  In each case, the statistical properties of the mean and of the interannual average  have been obtained by applying the standard block-bootstrap resampling technique \cite{Wilks,LNS06}, where the resampling block has been chosen to be as long as the decorrelation time of the yearly time series. The 95\% confidence interval has then been computed using a set of 1000 synthetic time series. Again, in the case of not stationary time series, a linear trend has been removed before applying the block bootstrap resampling.

As expected, the lagged correlation of the yearly time series of the atmospheric and land energy budgets result to be statistically non significant as soon as the lag is larger than one year. Since the relaxation time of these subdomains is fast, no significant trend has been found also in the SRESA1B scenario data. Instead, the climatic and oceanic budgets time series feature a decorrelation time of the order of 5 years. Note that this is a simple way to see that the decadal climatic signals are mainly due to the ocean.

In all cases, we find that the confidence intervals found with the block-bootstrap method agree very well with the asymptotic (for $N\gg 1$) results. The asymptotic theory says that in the case of decorrelated time series, the standard deviation of best unbiased estimate of the yearly average is approximately given by the interannual variability times the factor $2/\sqrt{N}$, while, in general, when a decorrelation time $n$ is present, the factor is approximately $2\sqrt{n}/\sqrt{N}$, which is still $\ll 1$ if $N\gg n$ (as in our cases). The standard deviation of the best unbiased estimate of the interannual variability  is a factor of $\sqrt{2}$ smaller than that of the corresponding mean.

The investigation of the meridional transport in the pre-industrial and SRESA1B scenarios is based on the analysis of the time series of the one-dimensional fields $\langle T_\Omega (\varphi) \rangle_j$. The best unbiased estimator of the yearly average of the meridional transport is computed as the average of the N transport profiles obtained for each year, and the best unbiased  estimate of the interannual variability has been evaluated as the usual standard deviation computed for each latitude. Again, trends have been removed when computing the standard deviation. Since when analyzing the meridional transport we are not testing a zero hypothesis, as in the case of the global energy budgets, we are less interested in obtaining detailed confidence intervals of our best estimates. Moreover, since we are dealing with 1D fields with obvious spatial correlation, the applicability of the above mentioned block bootstrap resampling technique is more problematic. Therefore, we will present conservative estimates of the uncertainties by using the standard deviation of the time series as proxy of the uncertainty of the best estimate of the mean. As we shall see below, the interannual variability results to be so small that no piece of information will be lost with this procedure.

%

\section{Diagnostics of the Climate Models}
\label{sec4}

\subsection{Global Energy Budget}

Results are presented in terms of intensive quantities, so that the CS (Fig. \ref{fig:state_TOA_balance}) and atmosphere (Fig. \ref{fig:state_atm_balance}) energy budgets are divided by the total surface of the planet, whereas the ocean (Fig. \ref{fig:state_sea_balance}) and land (Fig. \ref{fig:state_land_balance}) budgets are divided by the total ocean and land surfaces, respectively. We wish to underline that the approach of studying surface energy fluxes is thermodynamically equivalent to but computationally much more efficient than analyzing the variations of the total energy of the geophysical fluids. The latter approach is quite unfeasible when an extensive CMs intercomparison analysis is performed, because it would require the analysis of three-dimensional time-dependent fields for each CM.

We present for each domain the scatter plots of the 95\% confidence interval of the mean value and of the standard deviation of the yearly averaged intensive energy balance in preindustrial  conditions (first panel). We present also a scatter plot of the 95\% confidence interval of the variations of the mean value and the standard deviation from preindustrial to climate change conditions (second panel). The values includued in the first panel have been obtained with the block-bootstrap technique using $100$-year long time series, selecting as a general rule the last $100$ years of the data available in the datasets. In the case of SRESA1B simulations, we have considered the 2101-2200 data, and, when available, the  data referring to the XXIII century. The 95\% confidence intervals have been centered on the best unbiased statistical estimators and their half-width computed as twice the corresponding standard deviation computed as described above.

An immediate outlook on the problem of simulating physically sensible statistical properties of the energy budget is given in the second panel of Fig. (\ref{bias}), where we present the yearly time series (after application of a smoothing filter of 5 years for purely graphical reasons) of the TOA globally averaged energy budget for the GFDL2.1 model (CM 9). The preindustrial control run is depicted with the black line. The run performed under varying $CO_2$ concentration is depicted with the red line. Such run encompasses the XX century simulation (started in 1861 from year 41 of the pre-industrial run) and the XXI century portion of the SRESA1B simulations (started from the end of the XX century simulation), is depicted in red line. Finally, the blue line refers to the XXII and XXIII century SRESA1B run, where the $CO_2$ concentration is held fixed at 720 ppm. The energy balance of the control unperturbed run is biased towards a positive value of about $0.5$ $Wm^{-2}$. After the energy budget increases as a result of the genuine heating due to the enhanced greenhouse effect due to increasing $CO_2$ concentration, the energy budget decreases towards smaller values as the system relaxes towards the newly established steady state. The relaxation process occurs over a time scale of few centuries (not well resolved by the data) and can be safely attributed to the adjusting process of ocean state. Moreover, it is not clear whether the final energy budget bias will be analogous to the one the CM features in pre-industrial conditions. Therefore, as discussed above, it is in principle not trivial at all to estimate the actual net total heating of the planet due to the transition from 280 ppm to 720 ppm $CO_2$ concentration.

\subsubsection{Pre-industrial scenario}

In preindustrial conditions, most CMs feature a positive TOA yearly averaged energy budget, and in not even one CM the confidence interval intersects zero (see Fig. \ref{fig:state_TOA_balance}, first panel). Therefore, we can say that all CMs have a spurious behavior, which is apparently not compatible with steady state conditions. This is not the result of an adjustment in process, since trends are wholly negligible, but is instead a real bias. Threfore none of the CMs seems to conform to the idealized behavior depicted in Fig. (\ref{bias}). The imbalance is typically below $2$ $Wm^{-2}$ in absolute value, apart from 3 outliers (CMs 7, 10, and 15) featuring positive imbalances ranging between $2.5$ and $5$ $Wm^{-2}$. Models 4, 12, 13, 14 and 16 give the best performances, as they feature biases smaller than about $0.2$ $Wm^{-2}$ in absolute value. In principle, an average positive inbalance of $1$ $Wm^{-2}$ at TOA, if no net flux across the land surface and the ocean bottom occurs, is consistent with a drift of the energy content of the fluid envelope of the planet of about $1.6\cdot10^{22}$ $J/year$.

Very large discrepancies exist among CMs in the interannual variability of energy budget, since the computed standard deviations of the CMs' yearly energy budgets range between $0.1$ and $0.6$ $Wm^{-2}$, with typical values of about $0.3$ $Wm^{-2}$. The presence of such a wide span of results requires some attention, as it suggests the existence of inter-model differences in the intensity and/or time scales of the negative feedbacks regulating the establishment of an (approximate) energy balance over multiyear time scales. We find no obvious relation between the intensity of the interannual variability and the intensity of the bias of the energy budget.

In order to disentangle the possible causes for global energy imbalances in each CM we need to analyze the energy budgets of the atmosphere, of the ocean and of land separately. Common sense would suggest under the hypothesis that the CM is energetically consistent, that TOA spurious energy imbalances are basically due to biases in the surface energy budget of an ocean still not close enough from steady state. Under this hypothesis, we should have virtually vanishing energy budgets of the atmosphere and of the land components, as they are thin components of the CS and they are quasi-steady state.

What we instead find is that the biases in the energy budget of the atmosphere - Fig. (\ref{fig:state_atm_balance}) - and land - \ref{fig:state_land_balance} - are of the same order of magnitude of the bias of the energy budget of the ocean - Fig. (\ref{fig:state_sea_balance}). Therefore, it seems physically unreasonable to interpret the imbalances of the CS as a whole and of its subdomains as being only (or mainly) due to still persistent transient conditions. Note that an imbalance of $1$ $Wm^{-2}$ for the atmosphere corresponds to a staggering drift of about $3$ $K/year$ of its average temperature. Since this is obviously not observed in the data, it is clear that some fundamental issues in the dynamics of the CMs need to be addressed. Therefore, it is clear that the observed biases in the global energy budget of the CMs cannot be explained as a byproduct of the fact that the ocean component has not reached a true steady state.

An explanation for the energy imbalance of the atmosphere can be tracked to the lack of closure of the Lorenz energy cycle \cite{Lorenz55,Lorenz}. In fact, CMs in geneeral do not have a detailed parameterization scheme for the reintroduction as thermal forcing of the energy dissipated by viscous dissipation, interaction with the boundary layer, cloud processes, and diffusion. For a specific study in this direction please refer to \cite{Becker2003}, while an example of prototypical value can be found in \cite{LorenzVK}. If in the considered medium the kinetic energy is just lost by dissipatioon and not re-injected in the system as heat (as also in the simple case of Boussinesq approximation), we expect a compensating spurious positive energy budget, which then does not translate into a temperature drift of the atmosphere, but actually compensates a \textit{ghost} energy loss and keeps the system at steady state.

Land modules typically feature positive energy budgets, whereas in steady state conditions the land component of CMs should see, when long term averages are considered, vanishing energy balance not only in the global mean, but in each grid point, since no compensating horizontal transport is possible in a solid medium. Therefore, land modules must feature spurious water budget in the land climate modules \cite{DanubioGCM}, due, e.g., to relaxation schemes in the bottom boundary conditions. Otherwise, they must present biases in the treatment of phase transitions, like those related to the common procedure of calving ice and snow accumulating beyond given thresholds, as done, e.g., in \cite{LR2002}. In CM 22 The spurious energy sink associated to ice calving has been estimated to amount to about $0.3$ $Wm^{-2}$ when globally averaged \cite{TF10}.

We note that the energy balance of the atmosphere (and in principle also of the ocean where, actually, some contribution from a slow temperature drift cannot be ruled out) of the majority of CMs is positive, thus suggesting that typically there is an underestimation of the thermal energy produced at the end of the turbulent cascade dissipating the kinetic energy. Moreover, most land modules also feature a positive budget, suggesting that energy is lost because of inconsistencies in the treatment of phase transitions and water and heat fluxes.

In all CMs the standard deviation of the annual energy budget time series is much smaller for the land and atmosphere components than for the oceanic component, whose interannual variability is similar to that of the TOA budget. This confirms that the ocean acts as an integrator and plays a major role in controlling the variability of the global energy budget over multiannual time scales.

The performance of the various CMs wildly differs from case to case, and, typically, for a given model, the quality of the representation of the energy budget of the various subdomains is not consistent, thus pointing to the modular structure of the CMs. In any case, it should be noted that the biases we find for the atmosphere and ocean subdomains are at least one order of magnitude larger than the mechanical energy transfer term we have neglected, which implies that our approximation is consistent with the scope of our analysis.

We hereby highlight some specifically interesting features. We first note that CMs 4, and 16 feature a energy balance of the atmosphere and of the ocean accurately close to 0, with the land modules of introducing only a minor bias in the global energy budget. As opposed to that, CM 7 features a well-balanced land component, but both the atmosphere and the ocean feature strong positive biases, indicating that the energetics of the fluid media requires attention. On the other hand, the excellent performance on the global budget of CM 14 comes from non-negligible compensating errors in the sea, atmosphere and land modules. The poor performance of CM 15 in terms of global energy budget seems to derive from inconsistencies in the representation of the atmospheric and of the land energy balance. Energy inconsistencies in the atmosphere seem to be very relevant also for CM 1 and CM 10.

\subsubsection{SRESA1B scenario}

\begin{figure}[htbp] 
   \begin{center}

   \includegraphics[width=0.5\textwidth,angle=270]{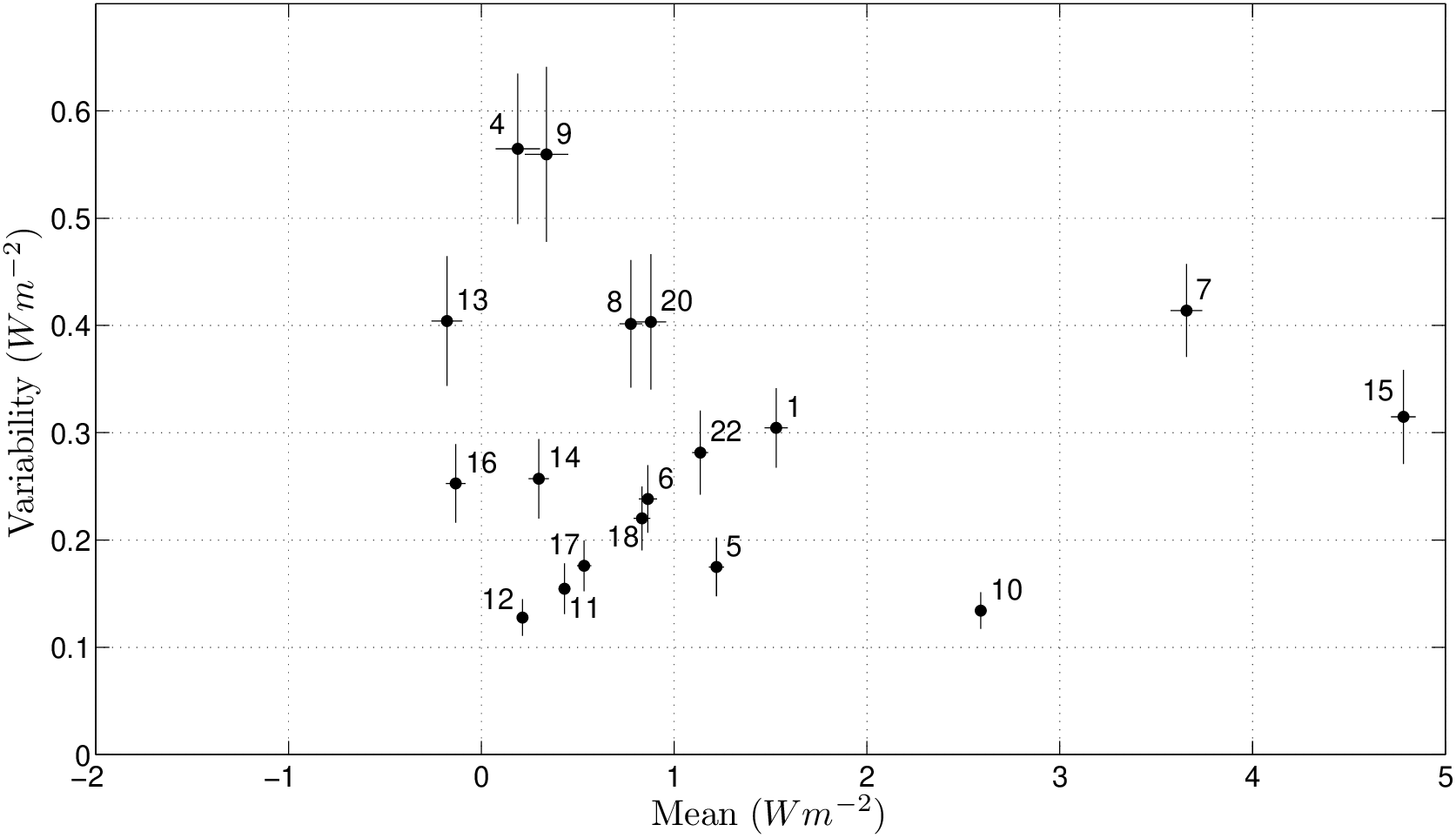}
   \includegraphics[width=0.5\textwidth,angle=270]{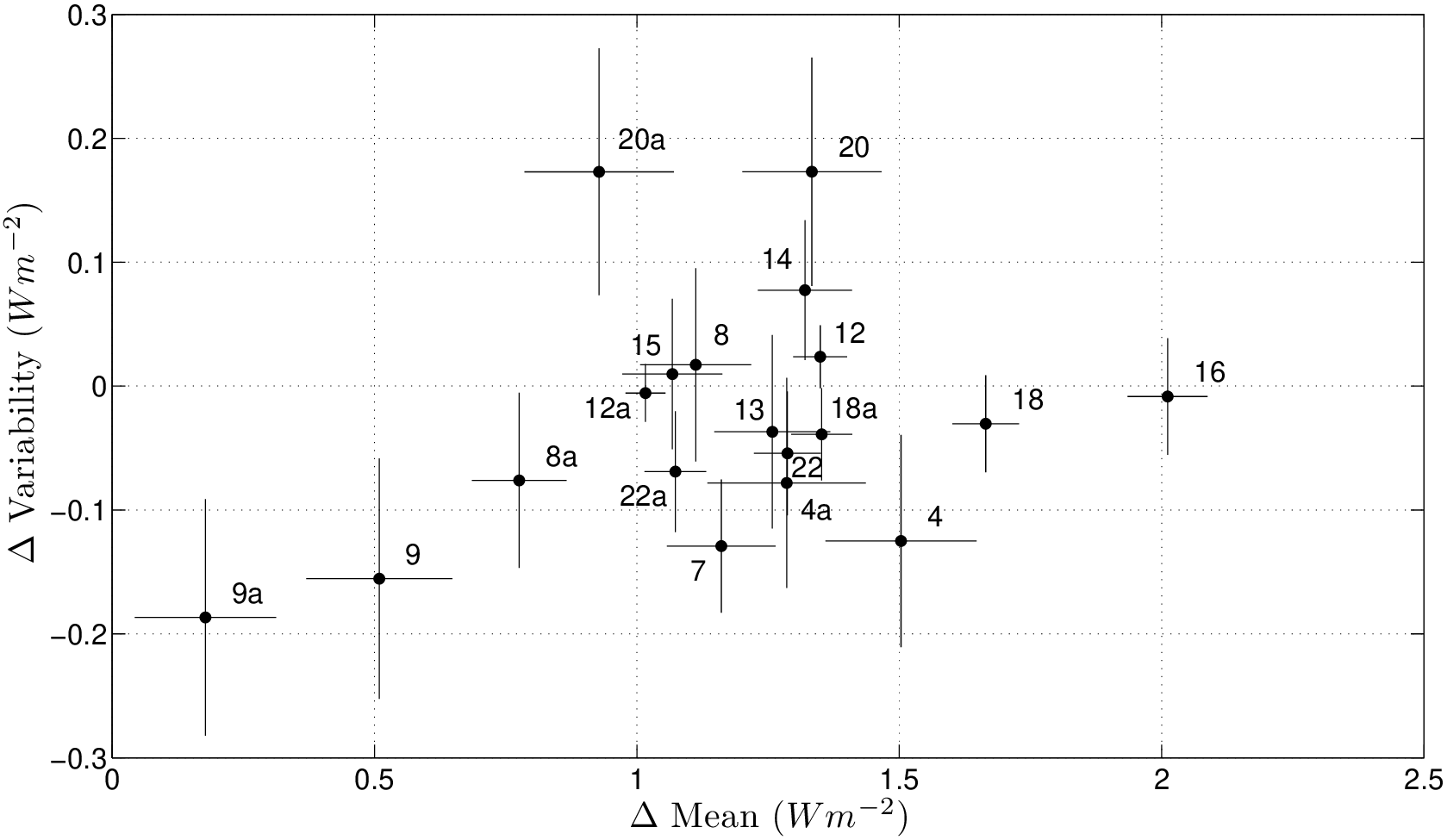}
      \end{center}
    \caption{First Panel: Mean and standard deviation of globally averaged climate energy balance for all the models over 100 years long time periods, in the preindustrial scenario. Second Panel: Difference between the SRESA1B and the pre-industrial scenario.}
  \label{fig:state_TOA_balance}
\end{figure}

\begin{figure}[htbp] 
   \begin{center}
   \includegraphics[width=0.5\textwidth,angle=270]{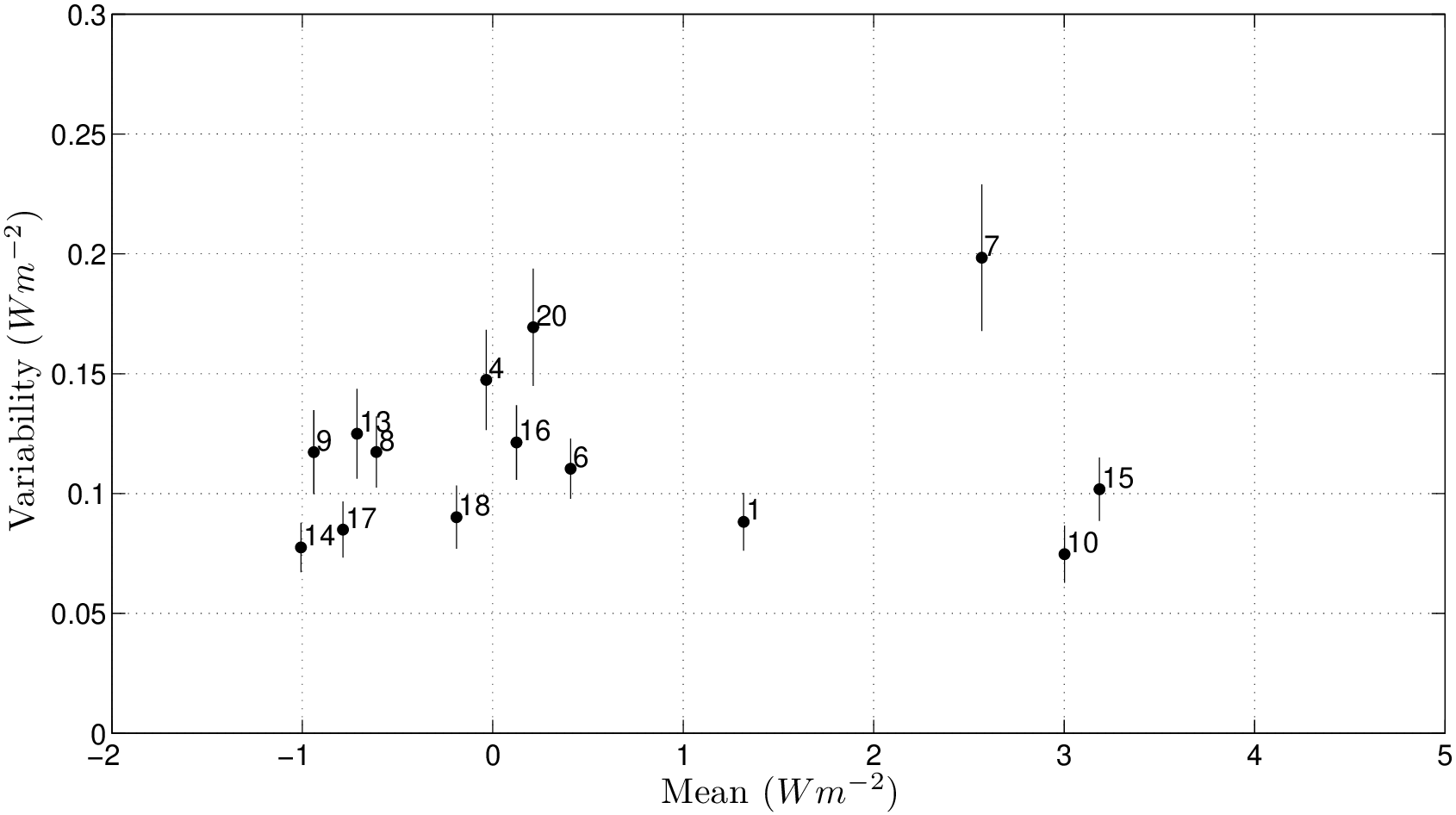}
   \includegraphics[width=0.5\textwidth,angle=270]{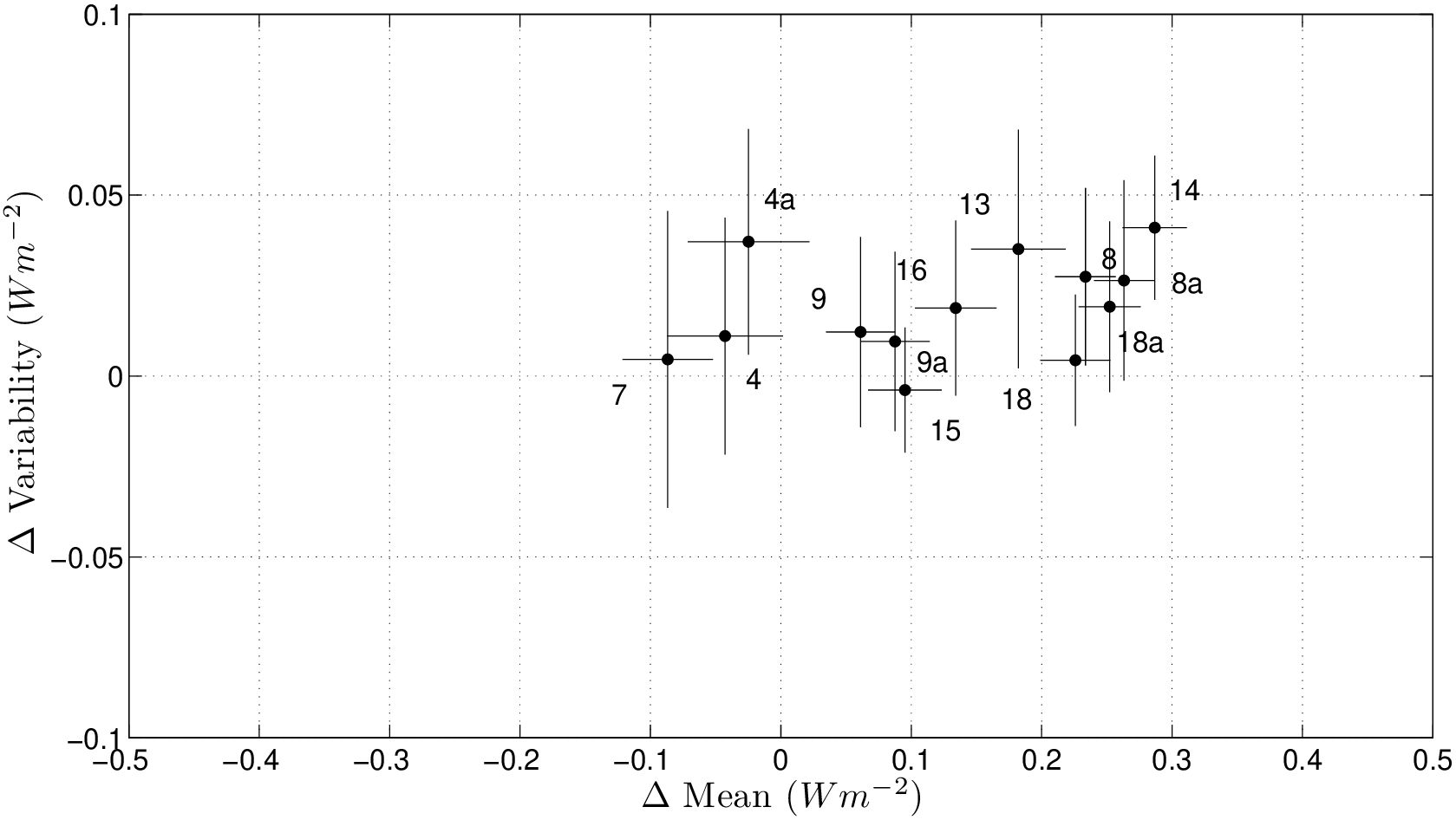}

      \end{center}
   \caption{First Panel: Mean and standard deviation of globally averaged atmosphere energy balance for all the models over 100 years long time periods, in the preindustrial scenario. Second Panel: Difference between the SRESA1B and the pre-industrial scenario. The label \textit{a} refers to data relative to the XXIII century in the SRESA1B scenario.}
   \label{fig:state_atm_balance}
\end{figure}

\begin{figure}[htbp] 
   \begin{center}

   \includegraphics[width=0.5\textwidth,angle=270]{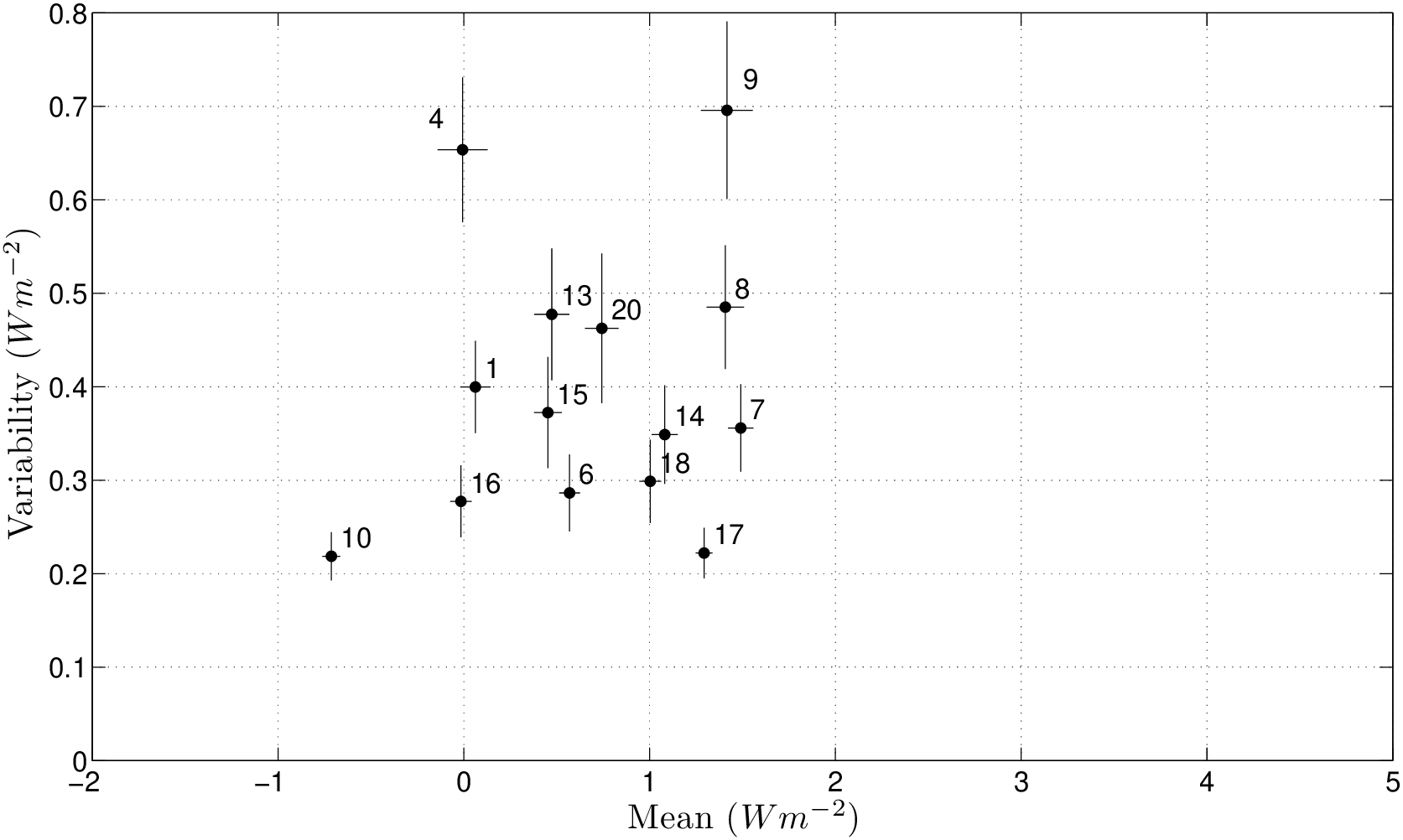}
   \includegraphics[width=0.5\textwidth,angle=270]{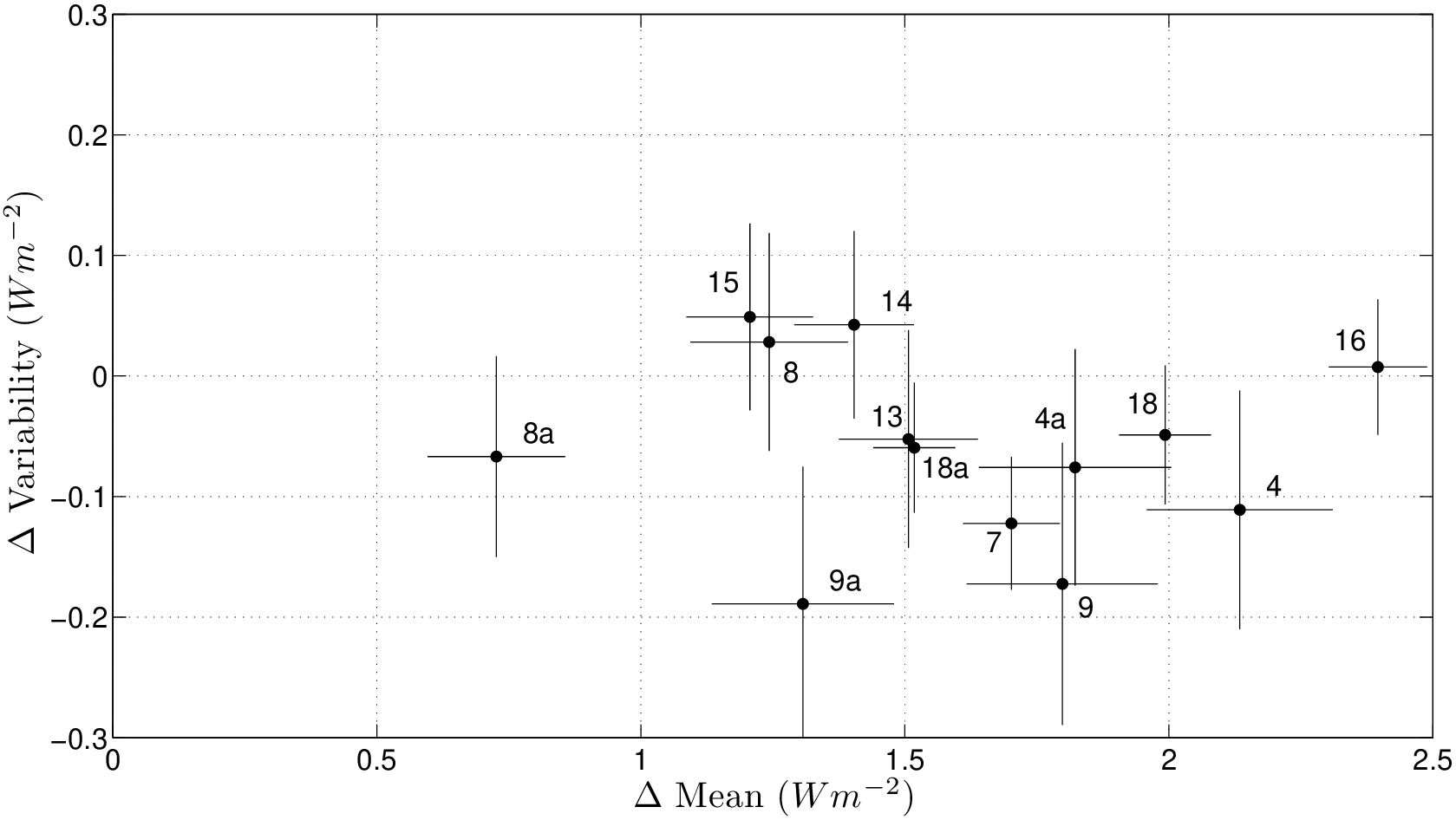}

      \end{center}
   \caption{First Panel: Mean and standard deviation of globally averaged ocean energy balance for all the models over 100 years long time periods, in the preindustrial scenario. Second Panel: Difference between the SRESA1B and the pre-industrial scenario. The label \textit{a} refers to data relative to the XXIII century in the SRESA1B scenario.}
   \label{fig:state_sea_balance}
\end{figure}

\begin{figure}[htbp] 
   \begin{center}

   \includegraphics[width=0.5\textwidth,angle=270]{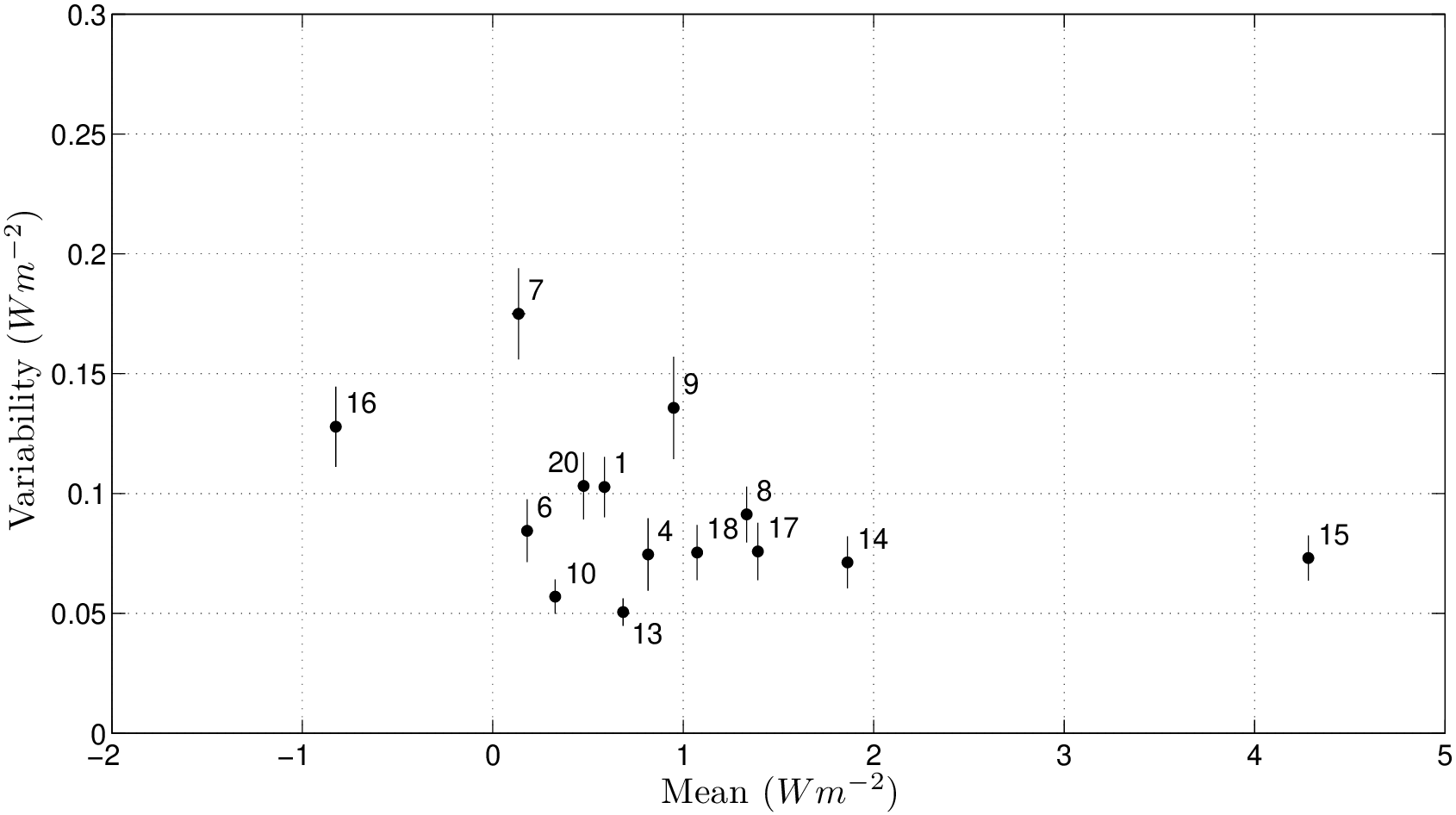}
   \includegraphics[width=0.5\textwidth,angle=270]{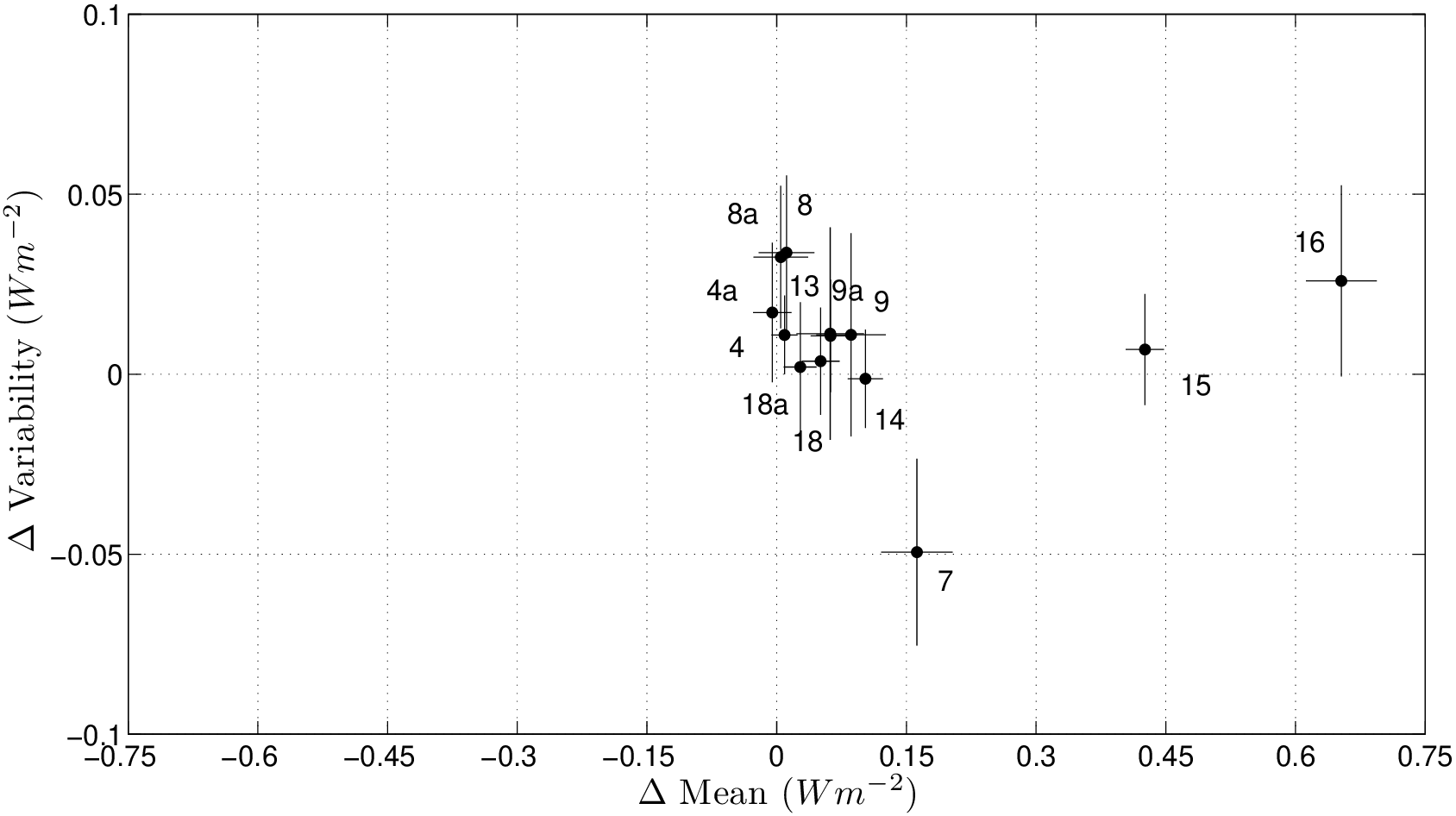}

      \end{center}
   \caption{First Panel: Mean and standard deviation of globally averaged land energy balance for all the models over 100 years long time periods, in the preindustrial scenario. Second Panel: Difference between the SRESA1B and the pre-industrial scenario. The label \textit{a} refers to data relative to the XXIII century in the SRESA1B scenario.}
   \label{fig:state_land_balance}
\end{figure}

While for each CM it is crucial to assess the energy bias in the pre-industrial, steady state conditions, globally and in each subdomain, it is also important to understand how the energy budget changes as a result of a prolonged radiative forcing. The change in the energy budget between the pre-industrial and the SRESA1B scenario are presented in the second panels of Figs. \ref{fig:state_TOA_balance}-\ref{fig:state_land_balance}. Values are shown for data referring to 2101-2200 (just numerical index) and, where available, also to 2201-2300 (same numerical index with lower index 'a').

In the SRESA1B simulations most CMs feature an increase in the total energy budget, ranging from about $0.2$ $Wm^{-2}$ to $2.0$ $Wm^{-2}$ with a
typical value of about $1.0$ $Wm^{-2}$. This can be interpreted as the fact that the system is still gaining energy from outside as steady-state
conditions have not yet been attained. The relaxation process in action is made more evident by the fact that the change in the energy bias for
the XXIII century is 20-30\% smaller than for the XXII century for all CMs where longer integrations are available (4, 8, 9, 12, 18, 20, and 22).
Note also that the changes shown in the second panel of Fig. \ref{fig:state_TOA_balance}b, which should be taken as robust signature of
climate change, are of the same order of magnitude or even larger than the biases found in steady state conditions (see first panel of Fig. \ref{fig:state_TOA_balance}a).

Disentangling this dependence of the bias on the climate state among the various subdomains provides additional insights (second panels in Fig. \ref{fig:state_atm_balance}-\ref{fig:state_land_balance}). We find that the changes in the energy bias of the atmosphere are relatively small, ranging for all CMs between -0.1 $Wm^{-2}$ and 0.3 $Wm^{-2}$. This is consistent with the fact that the atmosphere is in a quasi-steady state and supports the fact that the biases observed in pre-industrial conditions are due to the internal dynamical processes and are relatively state-independent.

Similarly, for almost all models the land energy balance changes by a little amount (less than 0.2 $Wm^{-2}$ except CMs 15 and 16) between SRESA1B and pre-industrial conditions, again in agreement with the fact that we are dealing with a thin, fast system. Also in this case, the biases seem to depend weakly on the climate state. This provides further support to the fact that the discrepancies found in the pre-industrial case are, as for the atmosphere, related to structure of the land modules.

Instead, the ocean acts as a heat \textit{reservoir} for all CMs: the energy balance of the ocean is consistently increased for all the CMs between $0.5$ and $2.5$ $Wm^{-2}$ in the SRESA1B scenario with respect to the pre-industrial one, and it is the key player in determining the change in the global budget shown in Fig. \ref{fig:state_TOA_balance}. Note that values reported in Fig. \ref{fig:state_sea_balance}b are larger than those given for the TOA in Fig. \ref{fig:state_TOA_balance}b because the ocean covers about 70\% of the Earth surface: the extensive values of the change in the energy uptake are, instead, rather similar. The ocean is still relaxing to its steady state and is genuinely taking up heat. The relaxation to the steady state is confirmed by the fact that, as for the TOA budget, the values referring to the XXIII century are smaller than those referring to the earlier century by about 20-30\%.

The importance of discussing the energy budgets for each subdomain separately can be emphasized by the following argument. By checking only the first panel  of Fig. \ref{fig:state_TOA_balance}, we would conclude that there is a relatively large group of CMs (e.g., 4, 9, 11, 12, 13, 14, 16, and 17)  featuring comparable performances in representing correctly the steady state properties of the CS. The analysis of the second panel of Fig. \ref{fig:state_TOA_balance} does not allow us to discriminate among them, as we know that a non-zero response has to be expected. Instead, we  observe that CMs 4 and 16 consistently provide among the smallest biases in energy budget of all subdomains in pre-industrial conditions \textit{and} feature very small changes in the bias for the atmosphere and land components also in SRESA1B conditions, as theoretically expected since these are thin subsystems. Therefore, we conclude that these CM are the best in terms of representing consistently the energy budget of the CS. Instaed, it is quite interesting to note that in the case of CM 9 a global energy balance close to 0 results from an almost perfect cancellation o fthe atmospheric and oceanic biases. Unfortunately, we do not have enough data to assess the quality of CMs 11, 12 and 14 in all subdomains.

\subsection{Meridional Enthalpy Transport}

\begin{figure}[htbp] 
   \begin{center}

   \includegraphics[width=0.5\textwidth,angle=270]{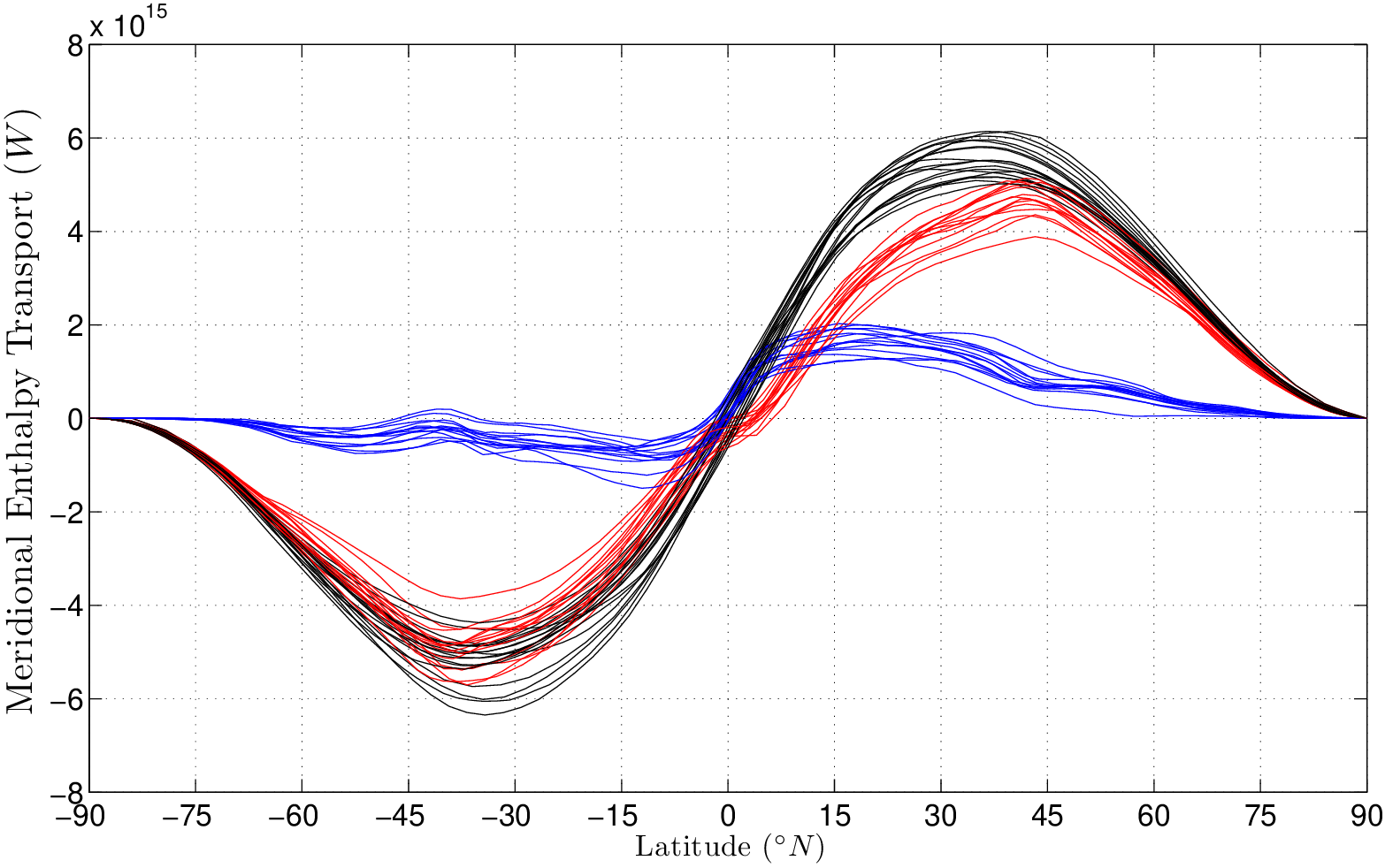}
   \includegraphics[width=0.5\textwidth,angle=270]{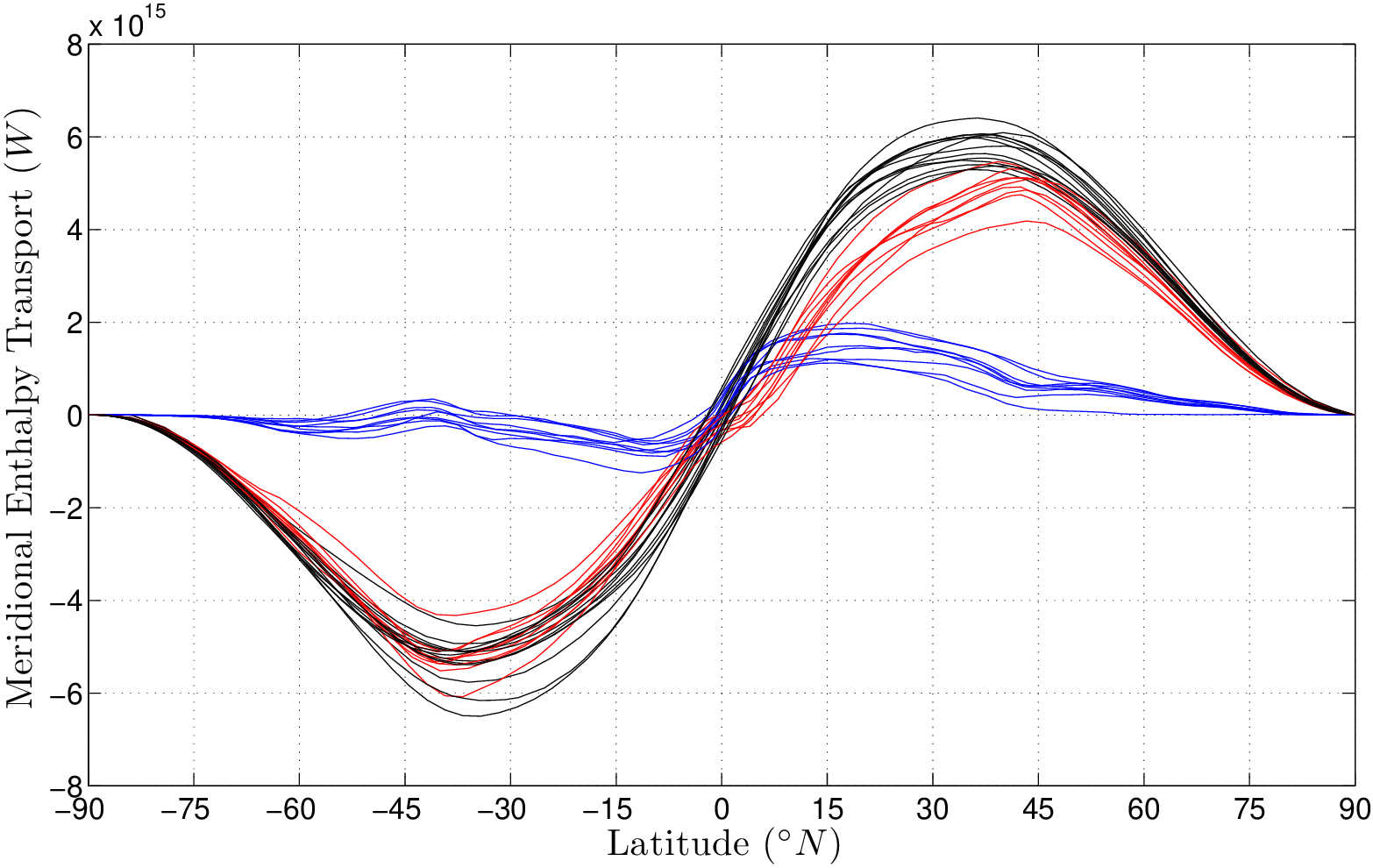}
      \end{center}
    \caption{Mean total meridional enthalpy transport by the climate system (black lines), the atmosphere (red lines), and the ocean (blue lines) in the preindustrial scenario (first panel), and in SRESA1B scenario for the XXII century (second panel) for all models where suitable data are available.}
  \label{fig:spag1}
\end{figure}

Composite plots of the profiles $\hat{\mu}(\langle{T}_\Omega(\varphi)\rangle_{j})$ ($\Omega=T$, $\Omega=A$, $\Omega=O$) obtained for the CMs analysed in this study are shown in Fig. \ref{fig:spag1}a for pre-industrial and in Fig. \ref{fig:spag1}b for SRESA1B conditions. Individual CMs results are not singled out (see below). All CMs feature a very low interannual variability on the meridional transport profiles, with absolute values of the local coefficients of variation (ratio of the standard deviation and of the mean) of the order of 1\% except near the region where meridional transport changes sign. Therefore, the variability is not reported in the figures. In each scenario, CMs agree on the most important features of the meridional  transport, at least qualitatively, and no radical differences emerge between the two scenarios. This shows that the impact of the (robust) $CO_2$ increase is to modulate the global circulation and not to disrupt it. We clearly see that the fluid envelope transport enthalpy from the low- to the high-latitude regions, with the effect of decreasing the meridional temperature gradient and of producing entropy \cite{Peixoto,Ozawa}.

In both scenarios, for most CMs the atmospheric energy transport is, in agreement with today's climate \cite{Peixoto,Trenberth2001,DellAquilaRianalisiSouth}, stronger in the southern hemisphere than in the northern hemisphere. In both hemispheres the peak of the atmospheric transport is located at the mid-latitudes, where we have the maximum of the baroclinic activity. Most CMs feature a slight southward atmospheric transport across the equator, probably due to the fact that the Hadley cell is not perfectly symmetric, since the Intertropical Convergence Zone is located north of the equator \cite{ITCZ}. The atmospheric transport profile is very smooth, thus showing a strong dynamical link between the Hadley circulation and the mid-latitude atmospheric variability \cite{SeamlessTransport}.

The ocean features a  strongly asymmetric meridional transport, which mirrors the asymmetry in the oceanic circulation between the southern and the northern hemisphere \cite{Kuhl07}. In both hemispheres most CM feature two local maxima of transport, corresponding to the gyres. The larger one peaks in the low latitudes, while the secondary one peaks in mid-to-high latitudes. Overall, transport is stronger in northern hemisphere, with vanishing or weak northward cross-equatorial transport. The structure of the (weaker) meridional transport in the southern hemisphere is somewhat more complex, as some CMs feature latitudinal bands of oceanic equatorward counter-transport. As discussed in \cite{TF10}, the representation of ocean transport in the southern hemisphere is problematic aspect of most PCMDI/CMIP3 CMs.

Finally, the total northward transport is almost hemispherically specular, featuring a very smooth shape, not exactly vanishing at the equator for most CMs. Since the peaks of the atmospheric and of the oceanic transport are shifted by more than $20^\circ$, and the atmospheric transport is typically stronger by a factor of $\sim 3-5$, the peak of the total transport is slightly shifted equatorward with respect to the atmospheric one. The most relevant qualitative feature of the meridional transport profiles presented in Figs. \ref{fig:spag1}a-b match remarkably well with those inferred from observations of the present energy budget at TOA and at surface \cite{Trenberth2001}, thus implying that they constitute robust structural properties of the CS which are relatively well captured by the CMs. In particular, it is quite confirting that, in agreement with Stone's results \cite{Stone78b}, for all CMs  the total transport profile is remarkably similar in the two hemispheres in spite of wide difference in the contributions provided by trasnient and stationary waves in the atmosphere and by the oceanic transport.

From a quantitative point of view, the disagreements among CMs are large. In the pre-industrial simulations, the northern hemisphere total and atmospheric transport features discrepancies among CMs up to 10\%, and up to 20\% in the southern hemisphere. Agreement is worse for the oceanic transport, discrepancies being up to 40-50\% in the northern hemisphere. In the southern hemisphere, where the intensity of the oceanic transport is lower, we have discrepancies up to 100\%, and also the qualitative behaviour is less consistently represented \cite{TF10}. In the SRESA1B simulations the situation is rather similar, with the notable difference that, even if the number of models included in the analysis is smaller, the models discrepancies on the ocean meridional transport are larger, probably as a result of the fact that transient time response of the ocean to $CO_2$ is rather uncertain.

Efficient metrics for the evaluation of CMs require the definition of simple but comprehensive indicators able to characterize the main features of the process under investigation. As an example, in a previous work we scaled down the Hayashi spectral density of mid-latitude waves to some integrated variables describing the overall spectral weight of qualitatively different waves \cite{IntercomparisonNorthWinterModels}. In this case, we define a procedure to extract from the 1D fields of meridional transports the essential information. Therefore, from each yearly averaged transport profile $\langle{T}_\Omega(\varphi)\rangle_{j}$ we extract four variables, namely the value of the peak of the northward transport in the northern hemisphere $\langle{T}_\Omega^{NH}\rangle_{j}$, its latitude $\langle{\varphi_\Omega^{NH}}\rangle_{j}$, the peak of the southward transport in the southern hemisphere $\langle{T}_\Omega^{SH}\rangle_{j}$, and its latitude $\langle{\varphi_\Omega^{SH}}\rangle_{j}$. The latitudinal band between $\langle{\varphi_\Omega^{SH}}\rangle_{j}$ and $\langle{\varphi_\Omega^{NH}}\rangle_{j}$ defines the region within $\Omega$ from which a net export of enthalpy is realized (except the case of southern hemisphere oceanic equatorward counter-transport), whereas $\langle{T}_\Omega^{NH}\rangle_{j}$ ($\langle{T}_\Omega^{SH}\rangle_{j}$) describes the total imbalance in the domain $\Omega$ north of (south of) $\langle{\varphi_\Omega^{NH}}\rangle_{j}$ ($\langle{\varphi_\Omega^{SH}}\rangle_{j}$).

The various CMs feature rather different horizontal grids, in terms of resolution and structure (e.g., regular lat-lon vs. Gaussian grids), so that the comparison of local indicators like those introduced above requires the remapping of all profiles $\langle{T}_\Omega(\varphi)\rangle_{j}$ into the same latitudinal grid. Therefore, we have selected the regular $2.5^\circ$ spacing from $-90^\circ N$ to $90^\circ N$ as a common framework , since it approximately corresponds to the average resolution of the CMs, and performed the remapping for all CMs using a linear interpolator. Results are robust with respect to the interpolating algorithm.

\subsubsection{Pre-industrial scenario}

\begin{figure}[htbp] 
   \begin{center}
   \includegraphics[width=0.5\textwidth,angle=270]{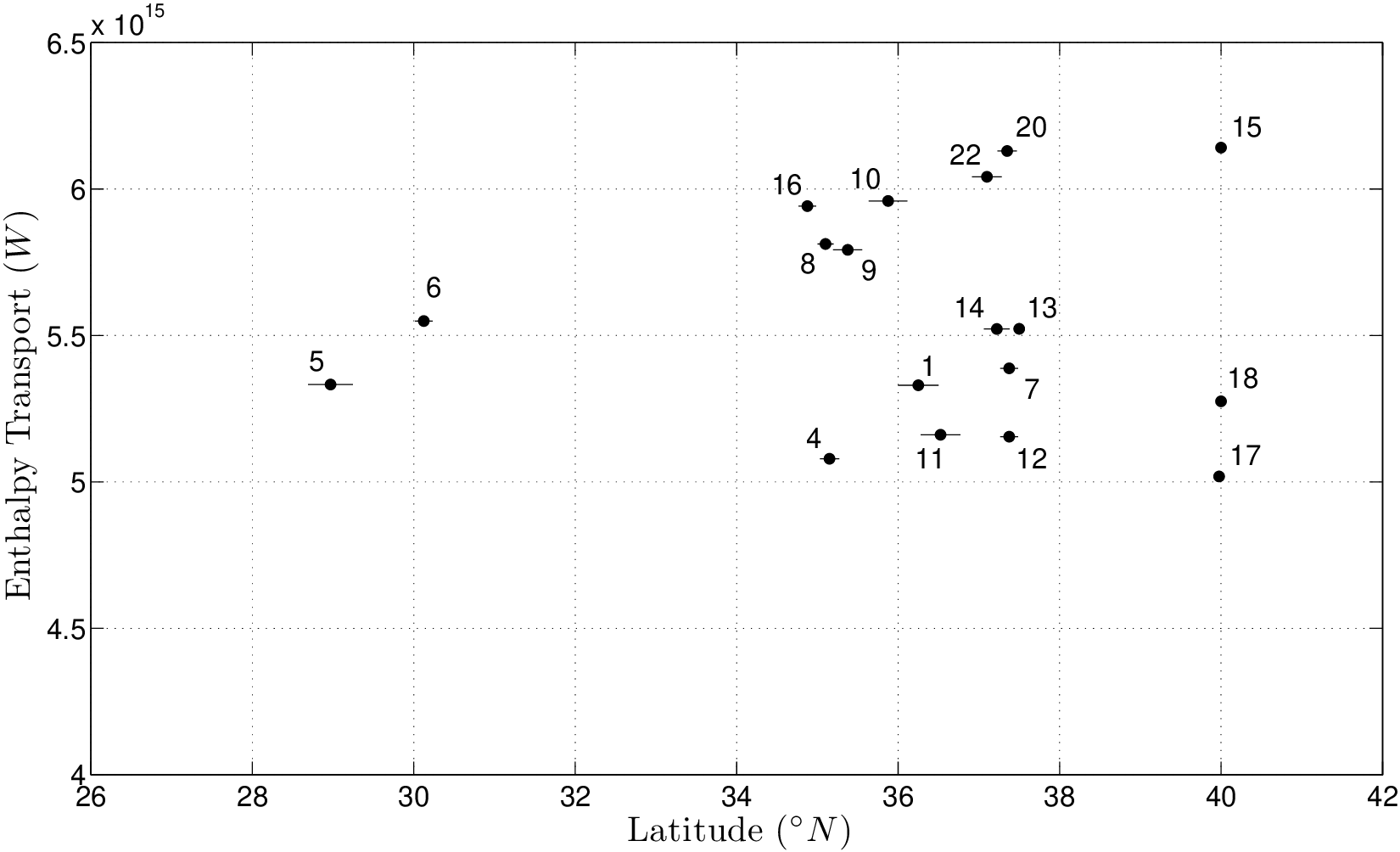}
   \includegraphics[width=0.5\textwidth,angle=270]{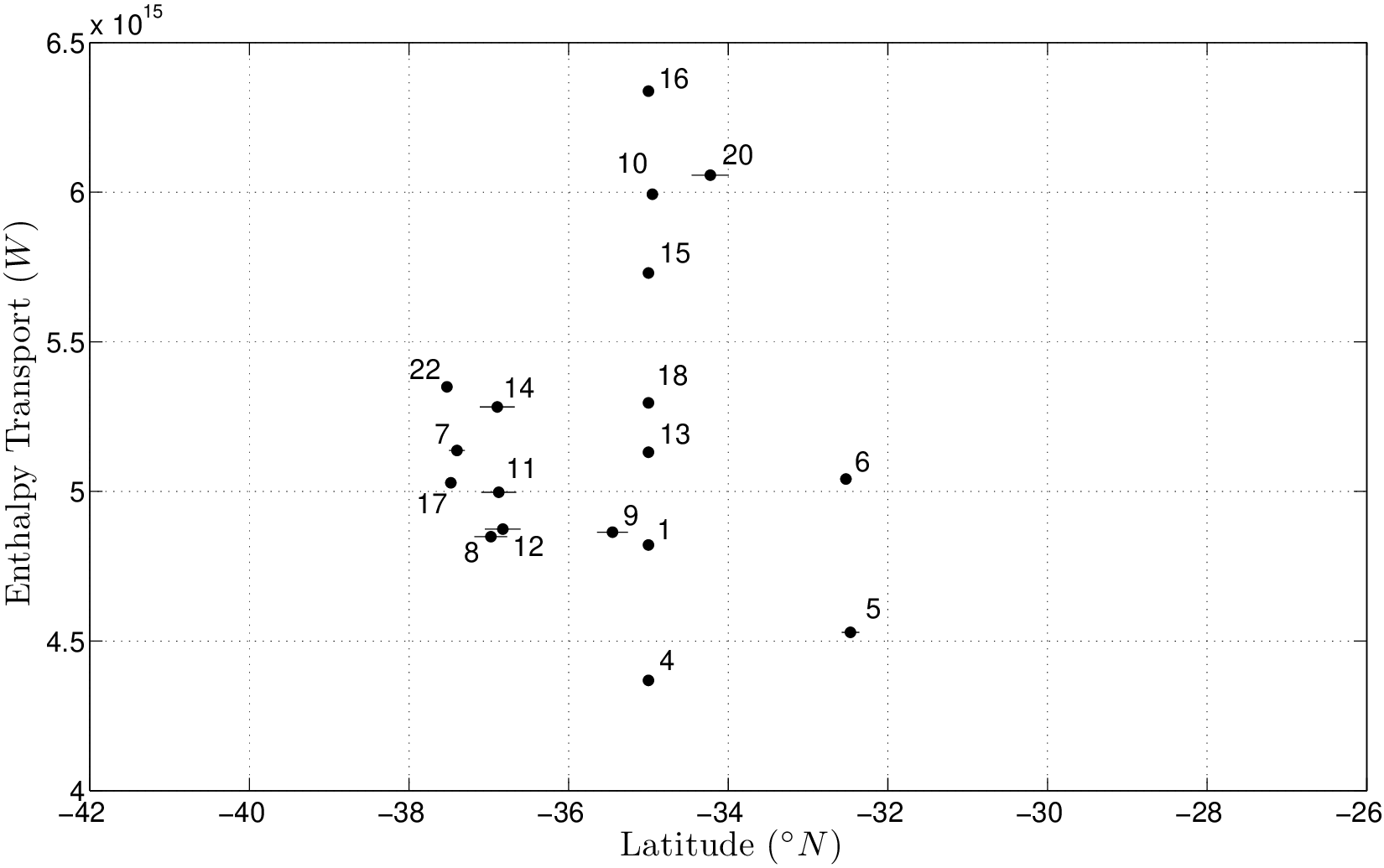}
      \end{center}
   \caption{Value and position of the peak of the total poleward meridional enthalpy transport in the preindustrial scenario for the northern hemisphere (first panel) and for the southern hemisphere (second panel).}
   \label{fig:transtoa}
\end{figure}

We begin our analysis by focusing on the peaks of the total meridional transport in pre-industrial conditions for the northern and southern hemisphere, depicted in Figs. \ref{fig:transtoa}a-b, respectively. We first observe that in both hemispheres most CMs feature transports peaking at $5$ to $6$ Petawatts ($PW$, $1$ $PW=10^{15}$ $W$), so with an effective range of about 20\%, and with maxima situated between $34^\circ$ and $38^\circ$. On the average, the total transport in the sourthern hemisphere is typically slightly smaller by about $0.3$ $PW$, in agreement with what found in \cite{TF10} for present climate conditions on a similar set of CMs. The model-wise disagreements on the localization of the transport maxima are more serious in the northern hemisphere, while disagreements on the peak value of the transport are somewhat wider in the southern hemisphere.

For each CM, the interannual variability of the position and the strength of the transport maximum is very small, so that typically each year the meridional transport peaks at the same grid point, and the transport maximum, as also discussed in the context of Fig. \ref{fig:spag1}a, has weak year-to-year variations.This is hardly unexpected, since we are looking into one of the most robust features of the CS, which should be affected only by large climate fluctuations. We may guess that when the CS is close to tipping points, the enhanced, large scale, organized fluctuations might heavily impact also these indicators. An interesting result is that CMs tend to behave consistently in the two hemispheres. Typically, relatively strong (weak) transports are observed in both hemispheres, as in the case of CMs 10, 15, 16, 20, 22 (CMs 4, 5, 11, 12, 17). Similarly, relative poleward (equatorward) anomalies in the position of the transport maxima are often consistent in the two hemispheres, as in the case of CM 17 (CMs 5 and 6).

\begin{figure}[htbp] 
   \begin{center}
   \includegraphics[width=0.5\textwidth,angle=270]{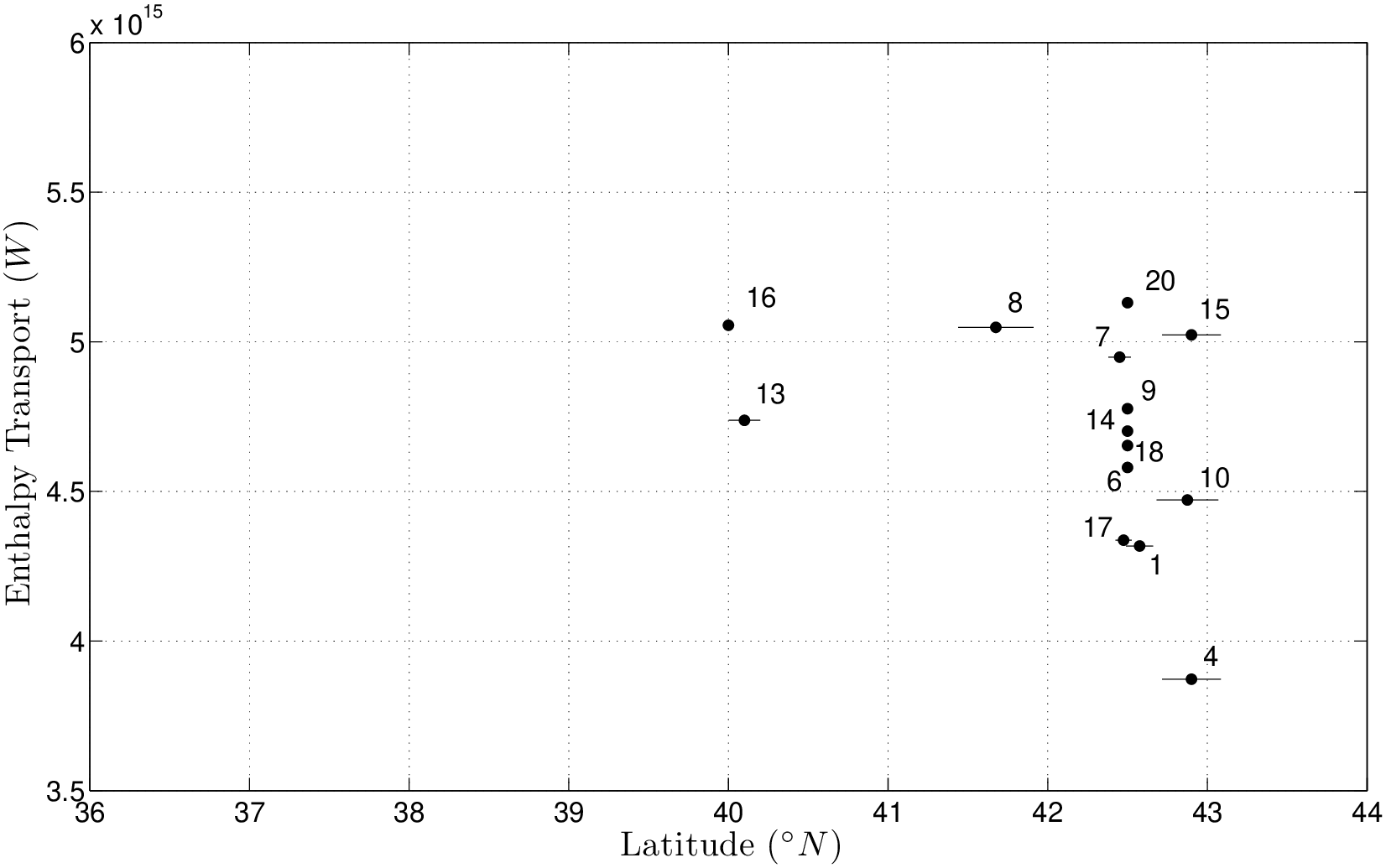}
   \includegraphics[width=0.5\textwidth,angle=270]{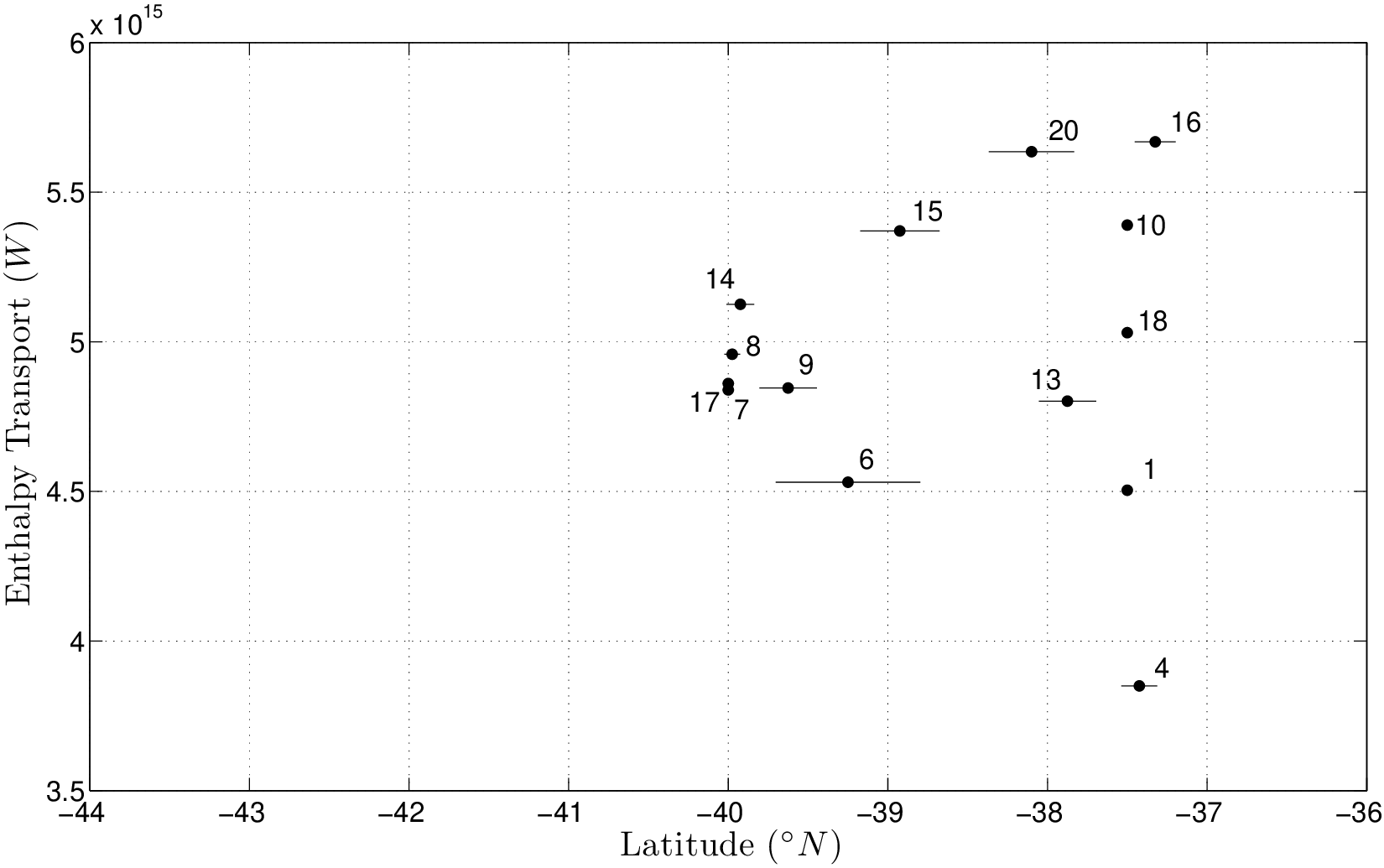}
      \end{center}
   \caption{Value and position of the peak of the poleward meridional enthalpy atmospheric transport in the preindustrial scenario for the northern hemisphere (first panel) and for the southern hemisphere (second panel).}
   \label{fig:transatm}
\end{figure}

The agreement among CMs in the representation of the atmospheric enthalpy transport in the northern (see Fig. \ref{fig:transatm}a) and southern hemisphere (see Fig. \ref{fig:transatm}b is also problematic. In both hemispheres the agreement among CMs is rather good on the position of the maxima (within one grid point), with a slight more poleward position of the maxima in the northern than in the southern hemisphere. Since we are dealing with latitudes of about $40^\circ$, we deduce that a great contribution to the transport is performed by baroclinic activity, and that the position of the peak coincides with that of the storm track. As in the case of the total transport, the interannual variability of the position and the intensity of the peak of the atmospheric transport is very small for all CMs.

Instead, quantitative differences in the intensity of the transport are quite pronounced in both hemispheres. In the northern hemisphere, most values are clustered within $1$ $PW$ around 4.7 $PW$, which means a spread of about 20\%. Such inconsistencies in the representation of the large scale properties of the atmosphere in the northern hemisphere agree with what discussed in \cite{IntercomparisonNorthWinterModels}. In this paper, albeit referred to the present climate, it is shown that large discrepancies among PCMDI/CMIP3 CMs exist in the representation of the northern hemisphere winter atmospheric variability in the mid-latitudes, which is when and where the meridional energy transport is particularly effective.

In the southern hemisphere, even if the total meridional enthalpy transport is typically lower, for most CMs the values of the maximum of the atmospheric transport tend to be somewhat larger than in the northern hemisphere, by up to $0.5$ $PW$. This is in agreement with what found in \cite{DellAquilaRianalisiNorth,DellAquilaRianalisiSouth} for present climate data, and it is basically due by the fact that the baroclinic activity is stronger in the southern hemisphere \cite{TrenberthSouthStormTrack}. The intensity span among CMs is more pronounced than in the northern hemisphere, reaching a value of $1.5$ $PW$ (about $30\%$), thus suggesting more uncertainty among CMs in the representation of the the meridional temperature gradient in the southern hemisphere. We also observe that some CMs feature extremely weak (CMs 1, and 4) and extremely strong (CMs 15, 16, and 20) atmospheric transports in both hemispheres, which hints at the presence of some peculiar structural properties of the atmospheric circulation.

When looking at the CMs performance in terms of representation of the ocean meridional enthalpy transport, the picture is rather different than in the two previous cases. We find results supporting the idea the the discrepancies among CMs in the representation of the ocean circulation - and especially that of the southern ocean - are much larger than in the case of atmospheric circulation.

In preindustrial conditions, in the northern hemisphere the intensity of the maxima of ocean transport are between $1.3$ $PW$ and $2$ $PW$, so the span is about $40\%$ of the typical value. The positions of the peaks are not correlated to their intensity and are distributed between $15^\circ N$ and $25^\circ N$, so with a span of $4$ grid points. The maximum for CM $10$ is relevantly shifted northward, up to around $30^\circ N$. For each CM, interannual variability is small for both intensity and position of the maximum, thus supporting the idea that the peak of the oceanic transport in the northern hemisphere provides a robust characterization of the state of the oceanic system.

As already seen in Fig. \ref{fig:spag1}a, in the southern hemisphere the definition of the peak of the oceanic transport is more problematic due to the presence of two well defined local maxima. We can see that for most CMs the peak of the oceanic transport is located in a relatively narrow band between $12.5^\circ S$ and $10^\circ S$ (1 grid point span), while for CM 6 the peak is shifted poleward between $15^\circ S$ and $20^\circ S$. This CM typically features a very weak transport, peaking typically around $10^\circ S$, but in few specific years the maximum of the transport is realized around $55^\circ S$. Therefore, the mean peak position has an intermediate value. Finally, the bimodality in the location of the maximum is very pronounced in CMs 1, 4, and 10, for which the (rather weak) meridional transport peaks in high latitudes quite often. A more detailed analysis of the distribution of the maxima accounting for the bimodality of the \textit{pdf} of the locations would surely be of interest. Restricting ourselves to the CMs with a clearly located maximum, we can see that the span of the intensities is very large. We observe values ranging from $0.4$ $PW$ to $1.6$ $PW$, so that the span is about $100\%$ of the typical value. Instead, for each CM, the interannual variability on the value and position of the peak is rather small. The existence of such discrepancies among the observed climatologies is in agreement with findings of previous studies, which showed that the PCMDI/CMIP3 models feature marked difficulties in the representation of large scale energy processes in the southern ocean.

\subsubsection{SRESA1B scenario}

Rather interesting differences emerge when comparing the simulations performed with pre-industrial scenarios and those performed with increased $CO_2$ concentration. The first non-trivial information is that for each CM the position of the peak of the total meridional transport does not change significantly, in both the southern and the northern hemispheres. The shift is at most of one grid point and, model-wise, no systematic (poleward or equatorward) variation is found (not shown). The behavior of each CM is in agreement with Stone's results mentioned above \cite{Stone78b} on the stability of the position of the peak of the transport independently, e.g., of the specific atmospheric composition. Therefore, the latitudinal boundary separating the areas of the planet differing on the sign of the difference between the incoming and outgoing  radiation is rather stable with respect to changes in the $CO_2$ concentration.

\begin{figure}[htbp] 
   \begin{center}
   \includegraphics[width=0.8\textwidth,angle=0]{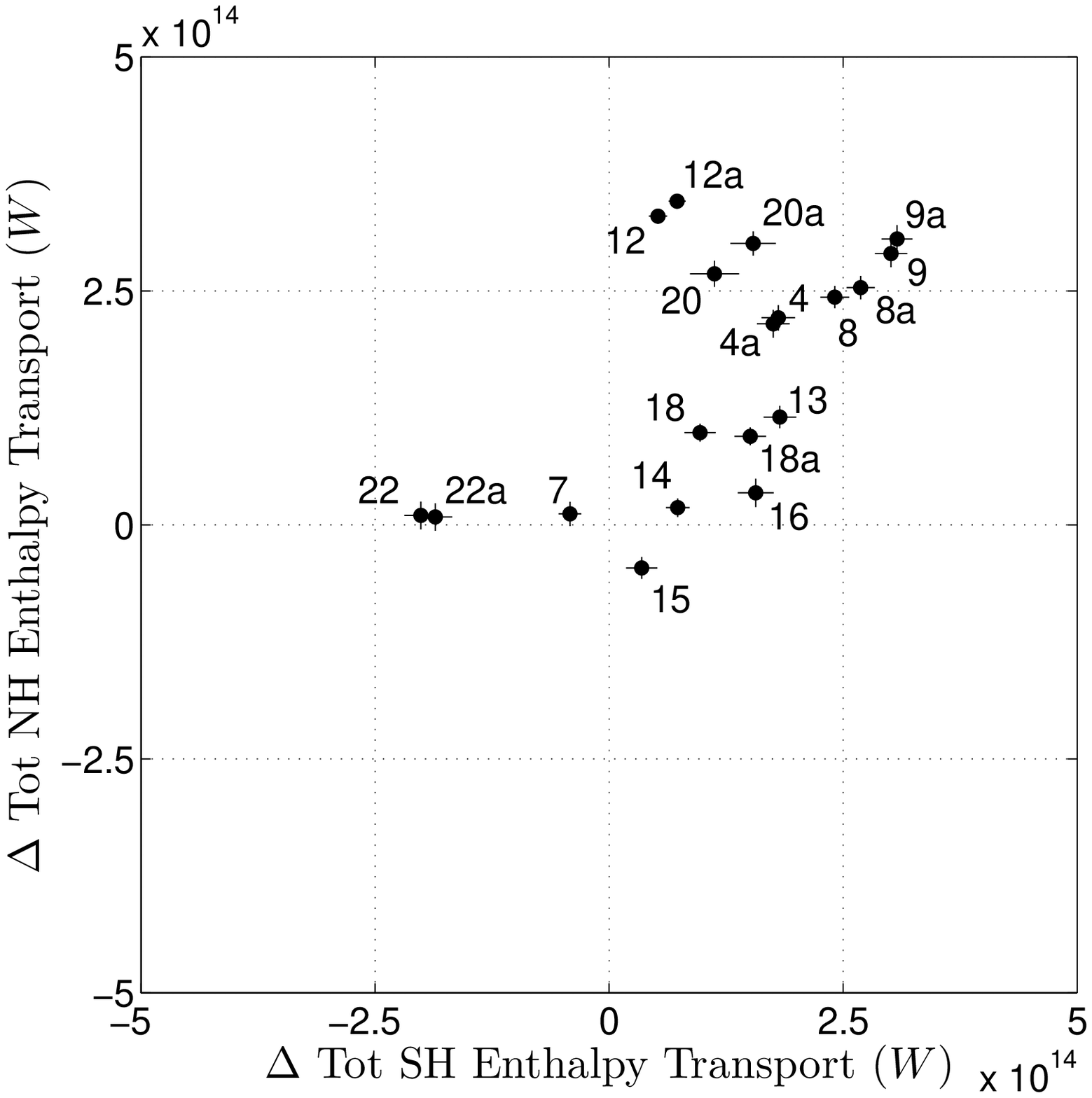}
   \includegraphics[width=0.8\textwidth,angle=0]{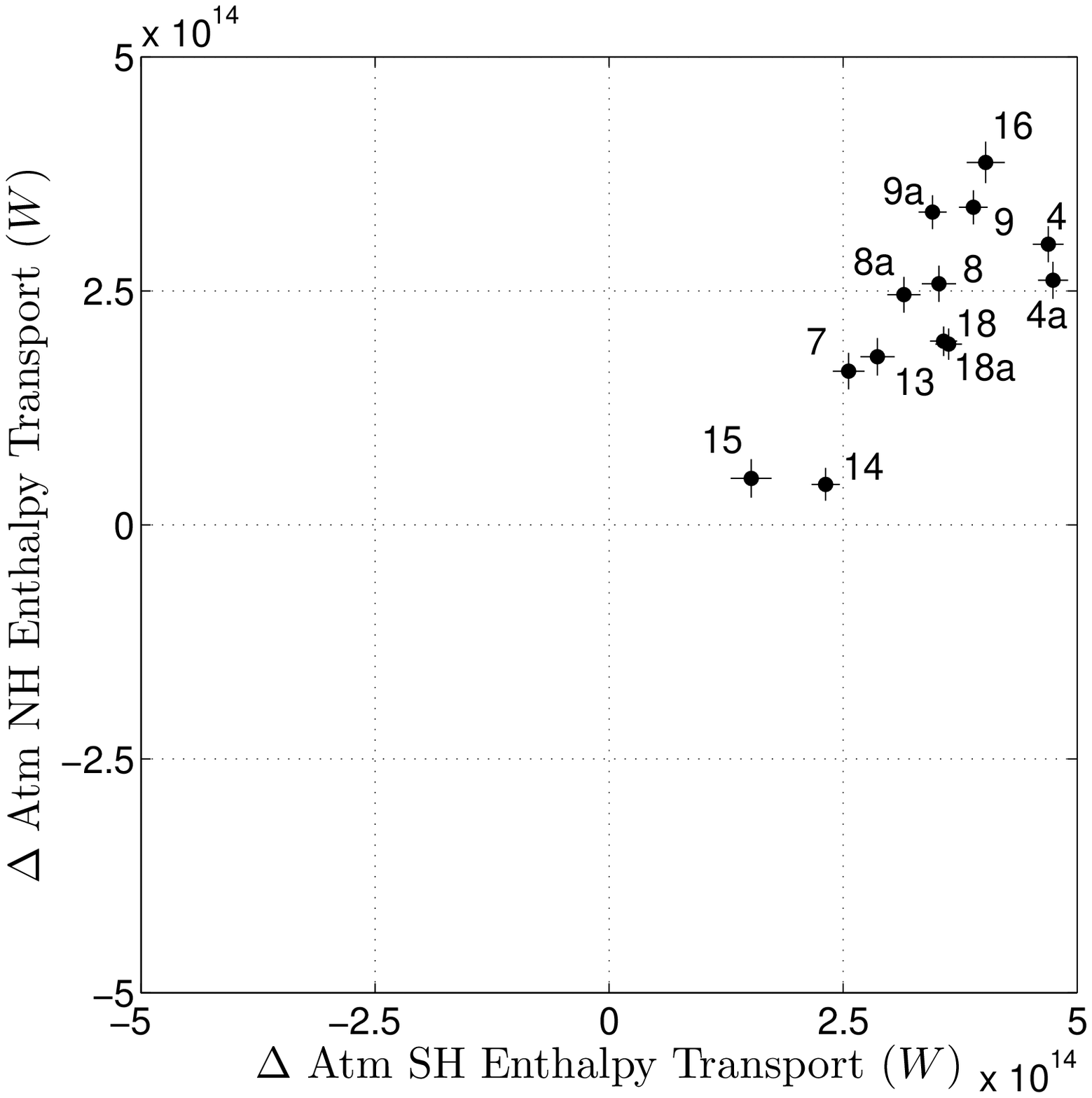}
   \includegraphics[width=0.8\textwidth,angle=0]{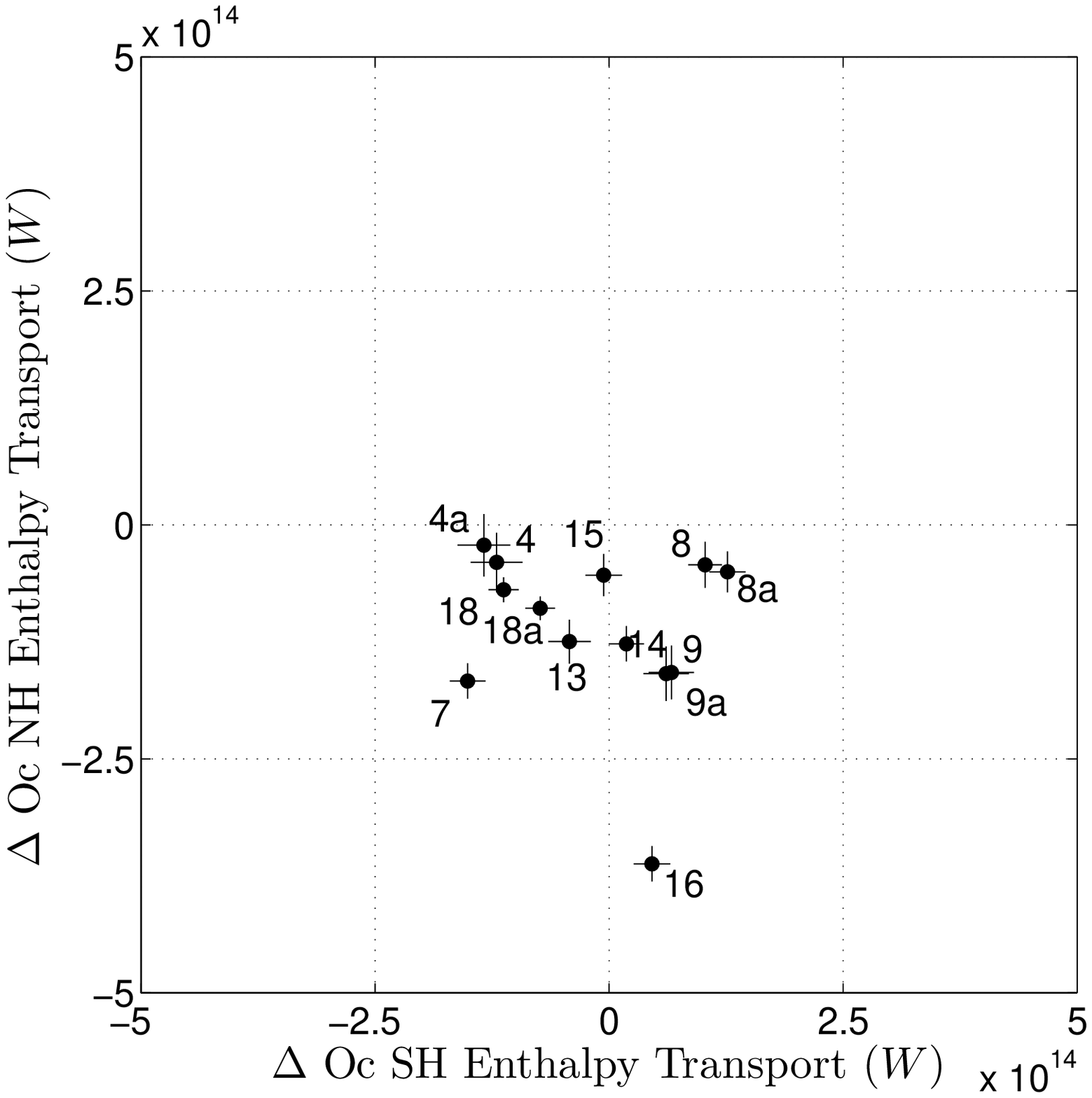}
      \end{center}
   \caption{Difference in the peak value of the poleward meridional enthalpy transport for both hemispheres between the SRESA1B and the pre-industrial simulations. Data are provided for the total transport (first panel), the atmospheric transport (second panel), and the ocean transport (third panel).}
   \label{fig:deltatranspo}
\end{figure}

Changes in the intensity of the meridional transport are more interesting. In Fig. \ref{fig:deltatranspo}a we plot the change in the value of the peak of the meridional transport in the southern hemisphere vs. the change observed in the northern hemisphere. Note that in all cases the 95\% confidence interval is very narrow. We have that most CMs feature positive variations in both hemispheres up to about 10\% of the pre-industrial value (typical changes range around 5\%), which implies that in a warmer climate energy is redistributed more effectively. This contributes to the mechanism of polar amplification \cite{IPCC}, as its additional heat is fed into the high-latitude regions of the globe. Given the smaller thermal inertia of the system, the effect in terms of surface temperature is expected to be stronger in the northern hemisphere. In CM 22 significant deviations from this behavior are observed because, while the change in the meridional transport in the northern hemisphere is almost negligible. This CM foresees a relevant reduction (of the order of 6\%) in the poleward meridional transport in the southern hemisphere. Therefore, in this CM the climate as a whole becomes less efficient in warming up the Antarctic region, and an increase in the relevant meridional temperature gradient is expected unless compensating changes in the meridional albedo gradient occur. Instead, CMs 7, 14 and 15 foresee a negligible change in the meridional transport in both hemisphere, thus featuring a very stable climate.

\begin{figure}[htbp] 
   \begin{center}
   \includegraphics[width=0.5\textwidth,angle=270]{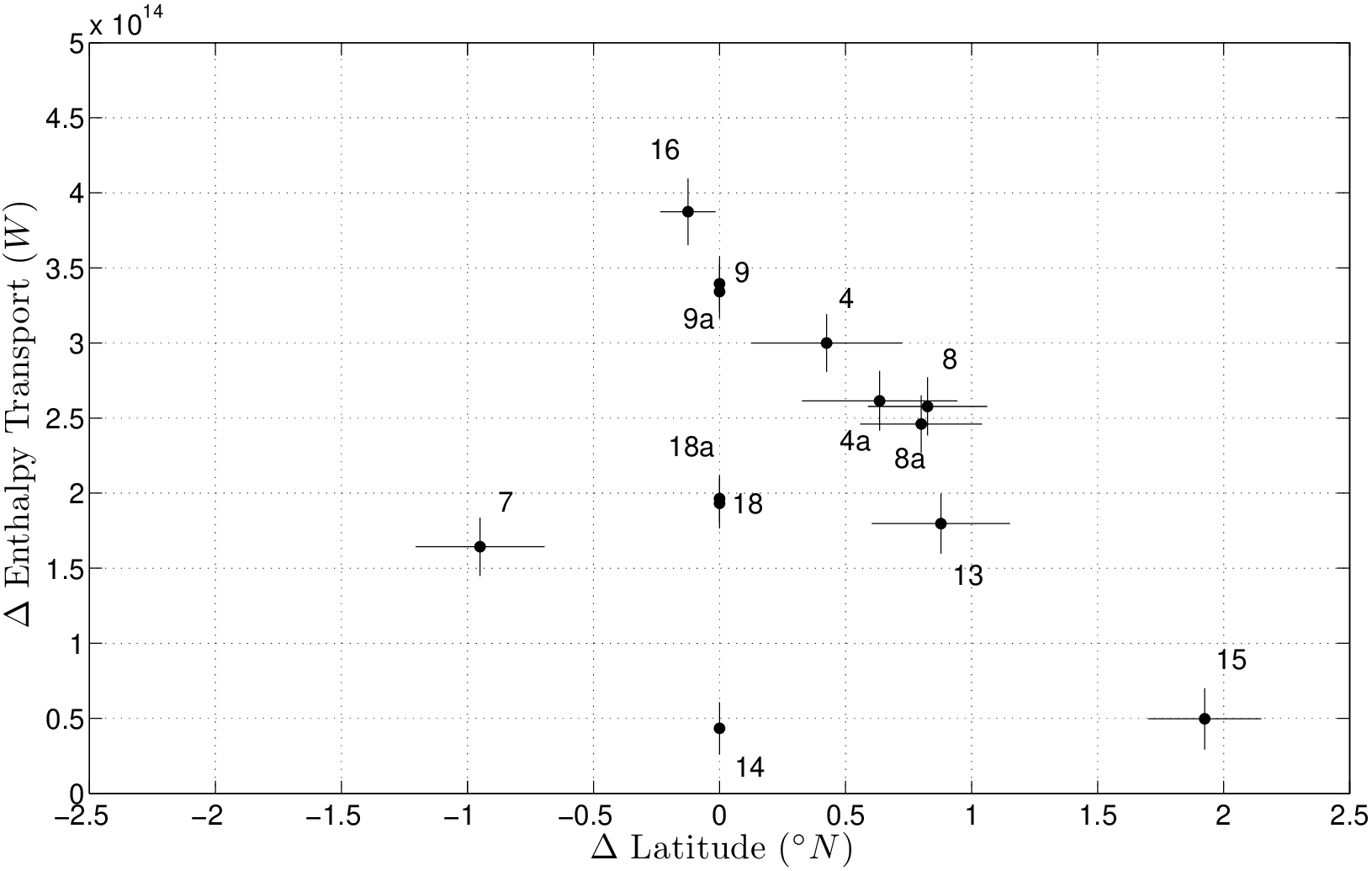}
   \includegraphics[width=0.5\textwidth,angle=270]{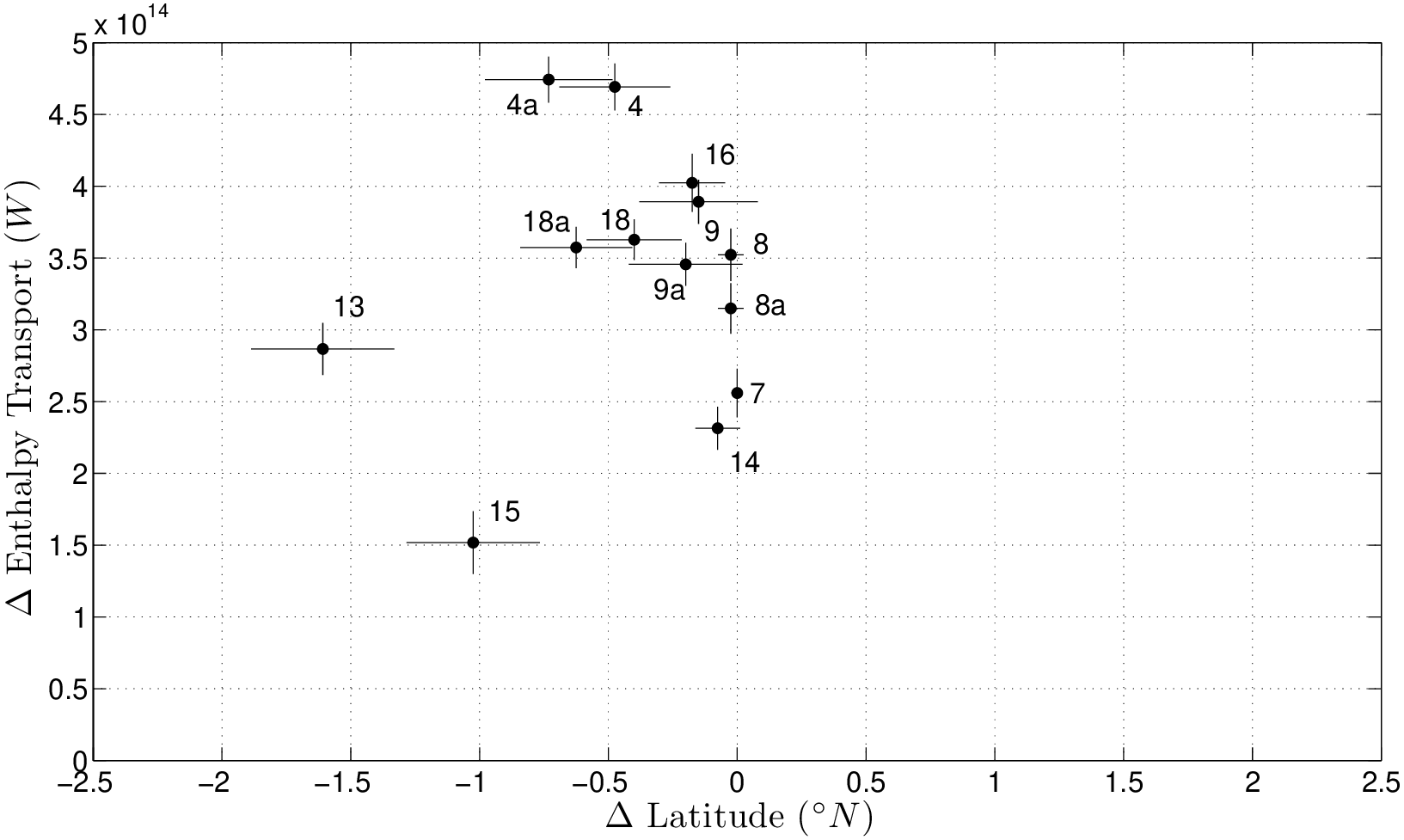}
      \end{center}
    \caption{Difference in the value and position of the peak of the atmospheric poleward meridional enthalpy transport between the SRESA1B  and the pre-industrial simulations for the northern hemisphere (first panel) and southern hemisphere (second panel).}
      \label{fig:deltaatmsresa}
\end{figure}

When looking at changes in the meridional atmospheric transport, we again find that in both hemispheres
(see Fig. \ref{fig:deltaatmsresa}a-b) changes in the position of the peak are small for
all CMs but with few exception systematically poleward. As the atmospheric transport peaks are related to baroclinic activity and storm track,
this results seems to be (at least partially) in agreement with what found in \cite{Yin05}, who observed in XXI century
climate simulations  a poleward shift of the core of the mid-latitude atmospheric variability, with a stronger signal in
the southern hemisphere. Nonetheless, recent analyses have attributed the shift of the peak in the southern hemisphere due to spurious trends in the clouds cover \cite{TF10}.
The impact of $CO_2$ increase is instead relevant when the values of the peak of the atmospheric transport is considered.
In Fig. \ref{fig:deltatranspo}b we show that all CMs feature consistent increases (up to about 12\%) of the atmospheric
transport in both the northern and the southern (where changes are typically larger) hemisphere, with the exception of
CMs 14 and 15, which foresee only a negligible increase of the transport for the northern hemisphere. Our results agree with what shown
in \cite{Held06}, where aggregated results for the CMs' ensemble mean are presented.

%

The response of the ocean to $CO_2$ increase is also of great interest. We observe that typical changes of the order of 0.1-0.2 PW (Fig. \ref{fig:deltatranspo}c). These constitute large relative changes in the ocean transport, ranging from 10\% to over 50\%. Half of the CMs foresee a decrease in the transport in the southern hemisphere, while a decrease in the transport is foreseen for all CMs in the northern hemisphere. The observed (partial) compensation between the changes in the oceanic and atmospheric transports is in overall agreement with the aggregated results presented in \cite{Held06}. There is no obvious link between the observed changes and the unperturbed values. Instead, in both hemispheres most CMs feature (typically small) equatorward shifts in the position of the maxima, typically of the order of half of a grid point. The largest equatorward shifts are realized for CM 18 in the northern hemisphere and CM 4 in the southern hemisphere, which featured among the most poleward position of the maxima in the pre-industrial conditions.

\subsection{Atmosphere-Ocean Transport compensation}
The Bjerknes compensation mechanism suggests a compensating role of the atmosphere and of the ocean in transporting heat poleward \cite{Bjerk}. The mechanism has been the subject of detailed studies performed also on PCMDI/CMIP3 CMs \cite{Bjerk06,Bjerk07}. In the previous subsection we have seen that for higher $CO_2$ concentration for basically all CMs the increase in the atmospheric transport is moderated by the decrease in the oceanic transport in both hemispheres. We test here whether a hemispheric compensating mechanism is in place also on smaller time scales  by computing for both hemispheres the linear correlation between the yearly time series of the maxima of the meridional transport for the atmosphere and for the ocean. Results are shown in Tab. \ref{tab:scattercorr}. Note that the 95\% confidence interval of the null hypothesis corresponds to that of an uncorrelated time series since the atmospheric transport has no memory.  We find that no CM, in no scenario, in neither hemisphere features a statistically significant positive correlation between the two time series. Instead, we find that the majority of the CMs feature a statistically significant negative correlation between the time series in both hemispheres and in both scenarios. CMs 4, 6, 7, 10, 16, and 20 seem to have an especially strong and consistent compensation mechanism. We see that in both hemispheres the negative correlation is often reinforced in the SRESA1B scenario, suggesting that the mechanism may become stronger in a warmer climate. This may be motivated by the stronger coupling due to intensified (thanks to Clausius-Clapeyron relation) latent heat fluxes under global warming conditions. The fact that the Bjerknes compensation mechanism is active in both hemispheres and is evident also when yearly time series are considered further suggests that it acts a local or regional scale. In fact, if the compensation were related to large scale ocean-atmosphere coupled processes, the behavior would be totally different in the two hemispheres, given the strong asymmetry of the oceanic circulation.


\section{Discussion}
\label{sec5}

In this work we have tackled the problem of assessing the consistency of a large set of state-of-the-art CMs in their representation of the energy balance of the whole CS as well as of its main subdomains (atmosphere, ocean, land). Moreover, we have discussed the statistical properties of their meridional enthalpy transports due to the fluid components.

Our analysis has consider the preindustrial and SRESA1B integrations performed following the standards provided by the PCMDI/CMIP3 project, which has greatly contributed to the redaction of the IPCC4AR \cite{IPCC}. The preindustrial scenario has played the role of a control run with fixed $CO_2$ concentration and has been instrumental for defining a baseline for dealing with transient-free climate behavior and for setting the objects of our investigation in the well-defined context of the non equilibrium steady state systems \cite{Gallavotti06}. The other scenario has been selected in order to understand how the energetics of the CS changes as a result of changes in the $CO_2$ concentration .

We find that in the pre-industrial simulations no CM features a global energy balance rigorously consistent with the condition of stationary state, so that a driftless state is realized even if no global balance is realized. The performances of the various state-of-the art CMs are wildly different, both in terms of global budgets, and when budgets of the main subdomains are considered. When global balances are considered, CMs biases are positive in all cases, with values spanning between $0.2 W m^{-2}$ and $2W m^{-2}$, with a few CMs featuring imbalances larger than $3W m^{-2}$.
Typically, this behavior is interpreted as resulting almost entirely from biases in the surface energy budget of an \textit{energy-absorbing} ocean, which has not attained yet its steady state, with an ensuing almost unnoticeable spurious but harmless drift in the climatogical temperature thanks to the ocean's large heat capacity. Instead, the energy imbalances in the land and atmospheric subdomain are of the same order of magnitude (about $1$ $Wm^{-2}$) of those of the ocean, which rules out the usual interpretation. Instead, inconsistencies are mostly related to basic issues in the land and fluid modules of the CMs. Note that for some of the CMs  \textit{good} global balances result from compensating errors in the subdomains. Instead, CMs 4 and 16 perform consistently well in all subdomains, which suggest that their representation of global energy processes is more physically consistent.

Most CMs feature spurious positive energy budgets for the land subdomain, as a result of unphysical biases in water fluxes \cite{DanubioGCM} and in the tretament of phase transitions. As an example, long-term accumulation or loss of 15 $mm/year$ of liquid water (which amounts to about 1.5\% of the intensity hydrological cycle) or 12 $mm/year$ of ice, e.g., due to spurious water infiltration or ice calving, correspond to an energy bias of about 1 $W m^{-2}$ in absolute value.

CMs feature biases in the atmospheric energy budget, which are typically positive and with of the order of $1$ $Wm^{-2}$. It this were related to a genuine energy input to the atmospheric subsystem, it would correspond to a staggering average temperature trend of $3$ $K/year$. As such a trend is definitely not observed, the spurious energy budget must compensate an unphysical energy sink (if the budget if positive) or source (if the budget is negative). The inconsistency in the energy budget of the atmosphere is mainly related to the fact that kinetic energy dissipated by various processes, including viscosity and diffusion, cloud parameterisation, and interaction with the boundary layer, is not exactly re-injected in the system as thermal energy. This has been analyzed in detail in \cite{Becker2003}, whereas an example of prototypical value has been provided in \cite{LorenzVK}. As an example, commonly used hyperdiffusion schemes feature mathematical properties that make them inadequate in terms of representing correctly the energy budget and introduce an artifical energy sink.

Note that these inconsistencies may have relevant impacts in the global properties of the atmospheric and and of the oceanic circulation, as they alter the input-output terms of the Lorenz energy cycle. Recently, in an intermediate complexity CM it has been shown that, when the energetic inconsistencies of the fluid and of the land modules are fixed, the representation of the global energy balance in greatly improved, with an obtained reduction of the the spurious imbalance by about one order of magnitude \cite{Snowball,Climatechange}.

We can link the presence of a bias in the global energy balance of the system $\Delta E $ to a bias $\Delta T_E$ in the radiative temperature $T_E$ of the planet. Using a linear expansion to the black body relation, we find that $\Delta T_E/\Delta E  = - 1/(4\sigma T_E^3)$ ($\approx -0.27 K/(W m^{-2})$ with usual Earth values). Moreover, if we consider the simple but effective parameterisation for the outgoing longwave radiation expressed as the linear relationship $A+BT_S$ with respect to the surface temperature $T_S$, we find that an energy bias $\Delta E$ is related to a surface temperature bias $\Delta T_S/\Delta E  = -1/B$, which, using the - typical - Nakamura \textit{et al.} values \cite{Naka}, results to be $\approx -0.6 K/(W m^{-2})$. Therefore, we have that a positive (negative) imbalance  - which actually \textit{cures} an energy leakage (source) in steady state conditions - is related to a spurious cooling (warming) of the system, both in terms of its thermodynamic temperature and of its surface temperature. As basically models tend to have positive energy budgets in pre-industrial as well as in increased $CO_2$ conditions, we expect them to feature a cold bias. We believe this is an alternative way of rationalizing the long-standing problem of the cold bias of CMs attributed, instead, by Johnson \cite{Johnson1997,Johnson2000} to spurious entropy generation due to numerical noise under the condition of vanishing energy balances.

Another surprising result worth mentioning is that the interannual variability of the energy budgets is model-wise very different, with values spanning almost and order of magnitude, both in the global case and when subdomains are considered. This calls for a detailed re-examination of the differences between the strength of the feedbacks controlling the energy budgets in the various CMs.

When looking at energy biases obtained with the SRESA1B scenario runs, we find that the atmosphere and the land biases change by a rather small amount (typically up to $0.2$ $W m^{-2}$) with respect to the pre-industrial runs. Since we are dealing with thin subsystems with small heat capacity, this suggests that the energy bias in not heavily climate-dependent but actually related to deficiencies in structure of the module. Instead, the ocean energy budget increases for all CMs, in agreement with the fact that in transient global warming conditions the water masses genuinely take up heat since the system is not yet at steady state. It must be observed that the spread of the heat update is very large, going from about 0.2 to over 2 $Wm^{-2}$. Such large discrepancies contribute to explaining the relatively large disagreement in the simulated transient climate change at surface by the various CMs.

The analysis of the inferred meridional enthalpy transports has instead provided us with a picture of how CMs represent the process of re-equilibration by which the fluid component of the CS decrease the temperature difference between the low and high latitude regions in both hemispheres. The qualitative agreement amongst CMs is rather good, and the obtained meridional transport profiles resemble recent estimates obtained with observative data \cite{Trenberth2001} and CMs diagnostics studies \cite{TF10}, thus confirming a fair representation of robust structural properties of the CS.

It is important to note that the indirect calculation of the transport is a linear operation, since it involves the integration of the vertical fluxes of energy at the boundary of the fluid envelopes and at the TOA. Therefore, we need to consider only 2D  diagnostic fields with the temporal resolution of the climatology we wish to compute (yearly data, in this case). As opposed to that, the direct calculation of the transports requires 3D diagnostics fields with high temporal resolution, since quadratic quantities are involved. Therefore, our approach requires storage and processing costs smaller by orders of magnitude than what needed with a direct calculation. While the indirect calculation is rigorously valid only if steady state conditions are precisely obeyed and the integrated energy balances are vanishing, the bias found in the CMs are small enough that only negligible errors are introduced.

Since we are dealing with 1D fields, the necessity to deal with simpler metrics has driven us to consider more manageable indicators. We have then analyzed the statistical properties of the intensity and position of the hemispheric peaks of the atmospheric, of the oceanic and of the total transport.

In pre-industrial conditions, we have found that the CMs do not agree precisely on the estimate of the total transport driven by the CS from the low to the high latitudes regions, as discrepancies of the order of 15-20\% around a typical value of about 5.5 PW (5.2 PW) are found in the northern (southern) hemispheres. Recent studies have suggested that the PCMDI/CMIP3 CMs might slightly underestimate the total transport in the southern hemisphere because too little radiation is absorbed in the tropical regions, due to to a cloud bias \cite{TF10}. When considering the atmospheric transport, we find that in the northern hemisphere the CMs feature discrepancies in the peak intensity of 15-20\% around a typical value of about 4.7 PW, whereas in the southern hemisphere the peaks of the transport are typically larger by about 0.5 PW and the discrepancy among CMs is more evident. Model 4 features an atmospheric transport much weaker than all other CMs. The agreement among CMs is much better for the atmospheric transport when looking at the position of the peak, with disagreements of only one grid point. The atmospheric transport peak is situated between $40^\circ$ N and $43^\circ$ N in the northern hemisphere and $37^\circ$ S and $40^\circ$ S in the southern hemisphere. The peak of the total transport is shifted equatorward in the northern (southern) hemisphere by about $5^\circ$ ($3^\circ$) and the CMs spread is somewhat larger. This confirms that, in spite of the serious discrepancy in the mechanisms contributing to the meridional transport in the two hemispheres, the overall properties are very closely symmetric with respect to the equator. This is in excellent agreement with Stone's theoretical results \cite{Stone78b}. It is important to note that for each CM the interannual variability of the position and intensity of the transport maxima is very small, so that we are dealing with a very robust climatic feature.

The inter-model uncertainties on the position of the peak of the total transport are mainly due to the large disagreement among CMs on the position of the peak of the ocean transport and on its intensity. Previous analyses have shown that the ability of PCMDI/CMIP3 CMs in describing the ocean transport in the southern hemisphere is especially problematic \cite{TF10}. All CMs agree on the fact that the oceanic transport is smaller than the atmospheric transport in both hemispheres, and that the oceanic transport in the southern hemisphere is much weaker that that in the northern hemisphere. Nevertheless, large quantitative inconsistencies emerge: disagreement on its intensity is of the order of 50\% in the northern hemisphere (with values ranging from about 1.3 to about 2.1 PW) and of the order of 100\% in the southern hemisphere (with values ranging from about 0.4 to about 1.6 PW). All CMs also agree on that the peak of the oceanic transport is shifted equatorward with respect to the peak of the atmospheric transport in both hemispheres. Instead, large quantitative disagreements exist on the actual positions of the peaks. In the northern hemisphere, for all CMs except CM 10, whose peak is shifted poleward, the ocean transport peaks between $15^\circ$ N and $25^\circ$ N. In the southern hemisphere, for most CMs the peak is typically situated between $10^\circ$ S and $20^\circ$ S, whereas CMs 1, 4, 6, and 10 feature a much more poleward shifted average position. As opposed to the case of CM 10 in the northern hemisphere, this is not related to a permanent feature of the oceanic circulation. In the southern hemisphere for all CMs the ocean transport has two distinct peaks, one located in subtropical and the other one in high latitude regions. In the case of CMs 1, 4, 6, and 10, the high-latitude peak is typically about as strong as the subtropical one, so that the distribution of location of the yearly maxima of transport is basically bimodal.

When looking at the SRESA1B scenario, we find that all models foresee an increase - ranging from 2\% to 12\% - in the atmospheric heat transport in both hemispheres. It is apparent that the enhancement of the latent heat fluxes due to increased mean temperatures dominates the foreseen reduction of the baroclinicity. A slight poleward shift is found for the peak of the atmospheric transport in both hemispheres, in broad agreement with what found in \cite{Yin05}. We have to remark that recent analyses attribute the poleward shift in the southern hemisphere to spurious representation of the changes in the cloud cover in altered climate conditions \cite{TF10}.

Following \cite{Held06}, the increase in the intensity of the atmospheric transport may be interpreted as the result of a strong cancellation between the large increase in the poleward latent heat transport, due to the strong enhancement of the hydrological cycle, and
the reduction of the sensible heat poleward transport, due to the decrease of the baroclinicity of the system. A recent
publication \cite{Gorman} has shown rather convincingly using a simplified yet Earth-like CM that this compensation mechanism
is in place on rather wide range of climate scenarios.

When considering the ocean component's response to $CO_2$ increase, we find that all CMs foresee a substantial decrease in the intensity of the ocean transport in the northern hemisphere (disagreement exists in the southern hemisphere), and, typically, a relatively small equatorward shift in the position of the peak in both hemisphere. These changes partially moderate the observed increase in the atmospheric transport, so that the total climate transport follow closely but not exactly those of the atmospheric transport: the peak is not shifted systematically in either hemisphere (again, in agreement with Stone's argument) and the transport is typically increased for most CMs up to 10\% in both hemispheres.

The contrasting response to $CO_2$ concentration increase in the atmospheric and oceanic transports suggests that the Bjerknes compensating mechanism \cite{Bjerk} is in place on climatological time scales. The yearly time series of the maxima of the atmospheric and oceanic transports have been used also to test - in a very unsophisticated way - whether the Bjerknes mechanism is active also on shorter time scale. The answer is partially positive, as both in the northern and in the southern hemisphere several CMs feature a statistically significant negative correlation between the time series, the rest features statistically non-significant correlations, while no cases of significant positive correlations are detected. The mechanism seems to become stronger in warmer climate conditions, possibly as a result of stronger atmosphere-ocean coupling due to more intense latent heat fluxes.

\section{Conclusions}
\label{sec6}

These findings suggest that, in spite of the great successes of climate modeling, serious efforts seem to be needed in terms of basic science for improving substantially CMs. This is of great relevance when devising strategies for the future direction to be taken in the development of numerical simulations, and is presently of great interest in the context of the preparation of the 5$^{th}$ Assessment Report of the IPCC. While earth system science requires models of increasing degree of complexity able to simulate correctly the chemical and biological cycles of our planet and their changes in a changing climate, we cannot expect to improve our understanding and ability to describe the CS just by adding and coupling indefinitely new modules. Nor the advocated \textit{quantum leap} in climate simulations can be obtained just by improving the computational capabilities \cite{Shukla}. In particular, the definition of strategies for improving the closure of the energy cycle in the fluid components of the CS \cite{Becker2003} seems to be a crucial step, as important as taking care of representing consistently phase transitions and heat fluxes over land. Some CMs already adopt the strategy of re-injecting in the system the kinetic energy lost to dissipation as a uniform heating term. While this helps in reducing the energy imbalance, it creates regional temperature biases, since climate regions featuring almost inviscid, horizontal dynamics are overheated (e.g., the stratosphere), and defines a spurious source of entropy production, as heat is moved down the gradient of the temperature field \cite{Johnson2000,L09}.

It is apparent that further work is needed to understand why CMs feature in some cases serious biases on the integrated global budgets and disagree quantitatively quite substantially - while agreement on qualitative properties is rather satisfactory - on the representation of a fundamental properties such as the meridional enthalpy transport. Whereas in the atmosphere the agreement in the meridional atmospheric transport is reasonable, in the ocean rather different pictures emerge from the various CMs, especially in the southern hemisphere. In terms of response to $CO_2$ increases, whereas we have relatively good agreement in the CMs response for the atmospheric transport, it is apparent that the ocean response requires a detailed re-examination.

The consideration of multi-model ensemble means has often been used as a way to partially circumvent the problem of dealing with CMs' discrepancies (see, e.g., \cite{Held06,TF09}). The rationale for this choice is that it is often found that such means provide enhanced skill with respect to randomly chosen specific CMs \cite{Gleck08}, even if this is far from being always the case \cite{DanubioGCM}. Even if various statistical techniques have been proposed in place of the the usual arithmetic average \cite{Min06,Min07,Tebaldi07} with encouraging empirical results, it must be emphasized that there is no rigorous statistical ground for combining data from different CMs \cite{Validation} (outputs of different CMs are \textit{not} samples of the same probability space, \textit{no} large numbers law applies \cite{Luca09}), so that a detailed understanding of CMs' uncertainties and discrepancies is of the greatest urgency.

In general, advances in numerical schemes as well as in the representation of physical processes are needed for providing a well-defined and self-consistent representation of climate as a non-equilibrium thermodynamical system when long time scales and large spatial scales are considered. Furthermore, additional work is then needed in the definition of suitable metrics for the investigation of the global thermodynamical properties of the CS and in testing the performances of state-of-the-art and future generations of CMs. Along these lines, and stimulated by recent extensive scholarly reviews \cite{Ozawa,Wu09}, the authors are preparing a companion publication dealing with the analysis of the entropy production \cite{Ozawa,L09} of PCMDI/CMIP3 CMs. Recently, \cite{LF10} have shown that it is possible to estimate the total entropy production of the CS just by analyzing the TOA and surface longwave and shortwave radiative fluxes and to separate the contribution due to vertical, small scale processes from those due to horizontal, large scale processes.

We suggest that the PCMDI/CMIP5 initiative, which will specifically take care of providing the tools necessary for auditing the present and future generation of climate and Earth system models, should take in greater care the provision of data referring to simulations by CMs under steady state with various choice of basic parameters, such as the solar constant (e.g., $\pm$ 2\% of the present one), the $CO_2$ (e.g., 275 ppm, 550 ppm, 1000 ppm), and should make it possible for research groups the access to data necessary for the analysis of the energy and entropy budgets of the planet. While these simulation and datasets  might seem unnecessary and not of direct impact for the assessment of the actual and projected climate change, they would allow for a much more detailed understanding of the performance of CMs. Of course, it is crucial for matters of self-consistency that the energy bias of a CM is as small as possible in all steady state scenarios. It is apparent that climate-dependent energy biases may contribute to increasing the uncertainty in climate sensitivity, and we may conjecture that this may play a role in the relatively large discrepancies among state-of-the-art CMs on the definition of this crucial quantity (still spanning between $2$ and $4.5$ $C^\circ$), and, correspondingly, of the change in the globally averaged surface temperature resulting from historical and projected $CO_2$ concentration increase \cite{IPCC}.

We conclude by noting that most of the results presented here could also be applied for studying the climates, via models and observational data, of celestial bodies such as planets and satellites. This is a rather promising and exciting perspective, because, apart from the everlasting interest in the solar planets and satellites, we live in the dawn of the era of extra-solar planets and satellites science. Therefore, what reviewed here as well as the ongoing studies on the entropy production and efficiency of the Earth system may be of help for studying the thermodynamics of the atmosphere of celestial bodies. Thus, it is encouraging to observe that various models belonging to the PLASIM family have already been adapted to study the atmospheres of Titan \cite{Griger04} and Mars \cite{Stenzel07}, and that intercomparison projects on the modeling of the Venusian atmosphere (see \texttt{http://www.issbern.ch/workshops/venusclimate/}) and Martian atmosphere (see \texttt{http://www.atm.ox.ac.uk/user/\\ newmanc/workshop/intercomp.html}) are ongoing.

\thanks{The authors wish to thank M. Ambaum, E. Cartman, D. Battisti, K. Fraedrich, D. Frierson, J. Gregory, F. Lunkeit, S. Pascale, A. Speranza, and R. Tailleux for useful discussions over these topics.}


\begin{thebibliography}{68}
\providecommand{\natexlab}[1]{#1}
\expandafter\ifx\csname urlstyle\endcsname\relax
  \providecommand{\doi}[1]{doi:\discretionary{}{}{}#1}\else
  \providecommand{\doi}{doi:\discretionary{}{}{}\begingroup
  \urlstyle{rm}\Url}\fi

\bibitem{ERA40OLR}
Allan, R., M.~Ringer, J.~Pamment, and A.~Slingo., {Simulation of the Earth's radiation budget by the European Centre for Medium-Range Weather Forecasts 40-year reanalysis (ERA40)}, \textit{J. Geophys. Res.}, \textit{109}, D18,107,
  2004.

\bibitem{Ambaum}
Ambaum, M., \textit{Thermal Physiocs of the Atmosphere}, Wiley, New York, 2010.

\bibitem{Becker2003}
Becker, E., Frictional heating in global climate models, \textit{Mon. Wea.
  Rev.}, \textit{131}, 508--520, 2003.

\bibitem{Bjerk}
Bjerknes, J., Atlantic air-sea interaction, in \textit{Advances in Geophysics},
  edited by H.~E. Landsberg and J.~van Mieghem, pp. 1--82, Academic Press, New
  York, 1964.

\bibitem{Carissimo85}
Carissimo, B., A.~Oort, and T.~Haar, Estimating the meridional energy
  transports in the atmosphere and ocean, \textit{J. Phys. Oceanogr.},
  \textit{15}, 82--91, 1985.

\bibitem{Groot}
de~Groot, J., and A.~Mazur, \textit{Nonequilibrium Thermodynamics},
  North-Holland, Amsterdam, 1962.

\bibitem{DellAquilaRianalisiNorth}
Dell'Aquila, A., V.~Lucarini, P.~Ruti, and S.~Calmanti, Hayashi spectra of the
  northern hemisphere mid-latitude atmospheric variability in the {NCEP-NCAR}
  and {ECMWF} reanalyses, \textit{Clim. Dyn.}, \textit{25}, 639--652, 2005.

\bibitem{DellAquilaRianalisiSouth}
Dell'Aquila, A., P.~Ruti, C.~S., and V.~Lucarini, Southern hemisphere
  midlatitude atmospheric variability of the {NCEP-NCAR} and {ECMWF}
  reanalyses, \textit{J. Geophys. Res.}, \textit{112}, D08,106, 2007.

\bibitem{Ende09}
Enderton, D., and J.~Marshall, {Explorations of the Atmosphere-Ocean-Ice
  Climates on an Aquaplanet and Their Meridional Energy Transports}, \textit{J.
  Clim.}, \textit{22}, 1593--1611, 2009.

\bibitem{Fermi}
Fermi, E., \textit{Thermodynamics}, Dover, New York, 1956.

\bibitem{Frierson}
Frierson, D., I.~Held, and P.~Zurita-Gotor, A gray-radiation aquaplanet moist
  gcm. part ii: Energy transports in altered climates, \textit{J. Atmos. Sci},
  \textit{64}, 1680--1693, 2007.

\bibitem{Gallavotti06}
Gallavotti, G., Nonequilibrium statistical mechanics (stationary): overview, in
  \textit{Encyclopedia of Mathematical Physics}, edited by T.~S.~T.
  J.-P.~Françoise, G.L.~Naber, pp. 530--539, Elsevier, Amsterdam, 2006.

\bibitem{Gleck08}
Gleckler, P., K.~Taylor, and C.~Doutrioaux, Performance metrics for climate
  models, \textit{J. Geophys. Res.}, \textit{113}, D06,104, 2008.

\bibitem{Goody2000}
Goody, R., Sources and sinks of climate entropy, \textit{Q. J. R. Meteorol.
  Soc.}, \textit{126}, 1953--1970, 2000.

\bibitem{Griger04}
Grieger, B., J.~Segschenider, H.~Keller, A.~Rodin, F.~Lunkeit, E.~Kirk, and
  K.~Fraedrich, {Simulatig Titan's tropospheric circulation with the Portable
  University Model of the Atmosphere}, \textit{Adv. Sp. Res.}, \textit{34},
  1650--1654, 2004.

\bibitem{Held}
Held, I., The gap between simulation and understanding in climate modeling,
  \textit{Bull. Am. Meteo. Soc.}, \textit{86}, 1609--1614, 2005.

\bibitem{Held06}
Held, I., and B.~Soden, {Robust Responses of the Hydrological Cycle over to
  Global Warming}, \textit{J. Clim.}, \textit{19}, 5686--5699, 2006.

\bibitem{Johnson1997}
Johnson, D., General coldness of climate models and the second law:
  implications for modeling the earth system, \textit{J. Clim.}, \textit{10},
  2826--2846, 1997.

\bibitem{Johnson2000}
Johnson, D., Entropy, the lorenz energy cycle and climate, in \textit{General
  Circulation Model Development: Past, Present and Future}, edited by
  D.~Randall, pp. 659--720, Academic Press, New York, 2000.

\bibitem{EEB97}
Kiehl, J., and K.~Trenberth, Earth's annual global mean energy budget,
  \textit{Bull. Am. Met. Soc.}, \textit{78}, 197--208, 1997.

\bibitem{Kleidon}
Kleidon, A., and R.~Lorenz, \textit{Non-equilibrium thermodynamics and the
  production of entropy: life, Earth and beyond}, Springer, New York, 2005.

\bibitem{Kuhl07}
Kuhlbrodt, T., A.~Griesel, M.~Montoya, A.~Levermann, M.~Hofmann, and
  S.~Rahmstorf, On the driving processes of the atlantic meridional overturning
  circulation, \textit{Rev. Geophys.}, \textit{45}, RG2001, 2007.

\bibitem{Kundu}
Kundu, P., and I.~Cohen, \textit{Fluid Mechanics}, Academic Press, San Diego, 2008.

\bibitem{tipping}
Lenton, T., H.~Held, E.~Kriegler, J.~Hall, W.~Lucht, S.~Rahmstorf, and
  H.~Schellnhuber, Tipping elements in the earth's climate system,
  \textit{Proc. Nat. Ac. Sci.}, \textit{105}, 1786–1793, 2008.

\bibitem{Lorenz55}
Lorenz, E., Available potential energy and the maintenance of the general
  circulation, \textit{Tellus}, \textit{7}, 157--167, 1955.

\bibitem{Lorenz}
Lorenz, E., \textit{The Nature and Theory of the General Circulation of the
  Atmosphere}, WMO, Geneva, 1967.

\bibitem{DefCliSci}
Lucarini, V., Towards a definition of climate science, \textit{Int. J. Envir.
  Sci.}, \textit{18}, 413--422, 2002.

\bibitem{Validation}
Lucarini, V., Validation of climate models, in \textit{Encyclopedia of Global
  Warming and Climate Change}, edited by G.~Philander, pp. 1053--1057, SAGE,
  Thousand Oak, 2008.

\bibitem{L09}
Lucarini, V., Thermodynamic efficiency and entropy production in the climate
  system, \textit{Phys. Rev. E}, \textit{80}, 021,118, 2009{\natexlab{a}}.

\bibitem{Luca09}
Lucarini, V., Evidence of dispersion relations for the nonlinear response of
  the lorenz 63 system, \textit{J. Stat. Phys.}, \textit{134}, 381–400,
  2009{\natexlab{b}}.

\bibitem{LorenzVK}
Lucarini, V., and K.~Fraedrich, Symmetry-break, mixing, instability, and low
  frequency variability in a minimal lorenz-like system, \textit{Phys. Rev. E},
  \textit{80}, 026,313, 2009.

\bibitem{LF10}
Lucarini, V., and K.~Fraedrich, Bounds on the thermodynamical properties of the fluid envelope of a planet based upon its radiative budget at the top of the atmosphere: theory and results for Earth, Mars, Titan, and Venus, submitted to \textit{J. Atmos. Sci.}; also available at \texttt{arXiv:1002.0157v2 [physics.ao-ph]}, 2010.

\bibitem{LR2002}
Lucarini, V., and G.~Russell, Comparison of mean climate trends in the northern
  hemisphere between national centers for environmental prediction and two
  atmosphere-ocean model forced runs, \textit{J. Geophys. Res.}, \textit{107},
  4269, 2002.

\bibitem{LNS06}
Lucarini, V., T.~Nanni, and A.~Speranza, Statistical properties of the seasonal
  cycle in the mediterranean area, \textit{Il Nuovo Cimento C}, \textit{29},
  21--31, 2006.

\bibitem{IntercomparisonNorthWinterModels}
Lucarini, V., S.~Calmanti, A.~Dell'Aquila, P.~Ruti, and A.~Speranza,
  Intercomparison of the northern hemisphere winter mid-latitude atmospheric
  variability of the {IPCC} models, \textit{Clim. Dyn.}, \textit{28}, 829--848,
  2007.

\bibitem{DanubioGCM}
Lucarini, V., R.~Danihlik, I.~Kriegerova, and A.~Speranza, Hydrological cycle
  in the danube basin in present-day and {XXII} century simulations by
  {IPCCAR4} global climate models, \textit{J. Geophys. Res.}, \textit{113},
  D09,107, 2008.

\bibitem{Snowball}
Lucarini, V., K.~Fraedrich, and F.~Lunkeit, Thermodynamic analysis of snowball
  earth hysteresis experiment: Efficiency, entropy production, and
  irreversibility, \textit{Q. J. R. Met Soc.}, \textit{136}, 2--11,
  2010{\natexlab{a}}.

\bibitem{Climatechange}
Lucarini, V., K.~Fraedrich, and F.~Lunkeit, Thermodynamics of climate change:
  Generalized sensitivities, \textit{Atmos. Chem. Phys. Disc.}, \textit{10},
  3699--3715, 2010{\natexlab{b}}.

\bibitem{Min06}
Min, S.-K., and A.~Hense, {A Bayesian Assessment of Climate Change Using
  Multimodel Ensembles. Part I: Global Mean Surface Temperature}, \textit{J.
  Clim.}, \textit{19}, 3237--3256, 2006.

\bibitem{Min07}
Min, S.-K., and A.~Hense, {Hierarchical evaluation of IPCC AR4 coupled climate
  models}, \textit{Clim. Dyn.}, \textit{29}, 853--868, 2007.

\bibitem{Naka}
Nakamura, M., P.~Stone, and M.~J., Destabilization of the thermohaline
  circulation by atmospheric eddy transports, \textit{J. Clim.}, \textit{12},
  1870--1882, 1994.

\bibitem{Gorman}
O'Gorman, P., and T.~Schneider, {The Hydrological Cycle over a Wide Range of
  Climate Simulated wih an Idealized GCM}, \textit{J. Clim.}, \textit{21},
  3815--3832, 2008.

\bibitem{Ozawa}
Ozawa, H., A.~Ohmura, R.~Lorenz, and T.~Pujol, The second law of thermodynamics
  and the global climate system: a review of the maximum entropy production
  principle, \textit{Rev. Geophys.}, \textit{41}, 1018, 2003.

\bibitem{Peixoto}
Peixoto, J., and A.~Oort, \textit{Physics of Climate}, Springer, New York,
  1992.

\bibitem{ITCZ}
Philander, S., D.~Gu, G.~Lambert, T.~Li, D.~Halpern, N.~Lau, and R.~Pacanowski,
  Why the {ITCZ} is mostly north of the equator, \textit{J. Clim.}, \textit{9},
  2958–2972, 1996.

\bibitem{Prigogine}
Prigogine, I., \textit{Thermodynamics of Irreversible Processes}, Interscience,
  New York, 1961.

\bibitem{Saltzmann}
Saltzmann, B., \textit{Dynamical Paleoclimatology}, Academic Press, New York,
  2002.

\bibitem{Tapio}
Schneider, T., The general circulation of the atmosphere, \textit{Ann. Rev.
  Earth Plan. Sci.}, \textit{34}, 655--688, 2006.

\bibitem{Bjerk06}
Shaffrey, L., and R.~Sutton, Bjerknes compensation and the decadal variability
  of the energy transports in a coupled climate model, \textit{J. Clim.},
  \textit{19}, 1167--1181, 2006.

\bibitem{Shukla}
Shukla, J., R.~Hagedorn, B.~Hoskins, J.~Kinter, J.~Marotzke, M.~Miller,
  T.~Palmer, and S.~J., Revolution in climate prediction is both necessary and
  possible: A declaration at the world modelling summit for climate prediction,
  \textit{Bull. Amer. Meteorol. Soc.}, \textit{90}, 175--178, 2009.

\bibitem{IPCC}
Solomon, S., D.~Qin, M.~Manning, Z.~Chen, M.~Marquis, K.~B. Averyt, M.~Tignor,
  and H.~L. Miller (Eds.), \textit{{Climate Change 2007: The Physical Science
  Basis. Contribution of Working Group I to the Fourth Assessment Report of the
  Intergovernmental Panel on Climate Change}}, 996 pp., Cambridge University
  Press, Cambridge, 2007.

\bibitem{Stenzel07}
Stenzel, O., B.~Grieger, H.~Keller, K.~Fraedrich, E.~Kirk, and L.~F., {Coupling
  Planet Simulator Mars, a general circulation model of the Martian Atmosphere,
  to the ice sheet model SICOPOLIS}, \textit{Plan. Sp. Sci.}, \textit{355},
  2087--2096, 2007.

\bibitem{Stone78}
Stone, P., Baroclinic adjustment, \textit{J. Atmos. Sci.}, \textit{35},
  561--571, 1978{\natexlab{a}}.

\bibitem{Stone78b}
Stone, P., Constraints on dynamical transports of energy on a spherical planet,
  \textit{Dyn. Atmos. Oc.}, \textit{2}, 123--139, 1978{\natexlab{b}}.

\bibitem{Bjerk07}
Swaluw, E. v.~d., S.~Drijfhout, , and W.~Hazeleger, Bjerknes compensation at
  high northern latitudes: The ocean forcing the atmosphere, \textit{J. Clim.},
  \textit{20}, 6023--6032, 2007.

\bibitem{Tebaldi07}
Tebaldi, C., and R.~Knutti, The use of the multi-model ensemble in
  probabilistic climate projections, \textit{Phil. Trans. R. Soc. A},
  \textit{365}, 2053--2076, 2007.

\bibitem{TrenberthSouthStormTrack}
Trenberth, K., Storm tracks in the southern hemisphere, \textit{J. Atmos.
  Sci.}, \textit{48}, 2159–2178, 1991.

\bibitem{Trenberth2001}
Trenberth, K., and J.~Caron, Estimates of meridional atmosphere and ocean heat
  transports, \textit{J. Clim.}, p. 3433–3443, 2001.

\bibitem{TF09}
Trenberth, K., and J.~Fasullo, Global warming due to increasing absorbed solar
  radiation, \textit{Geophys. Res. Lett.}, \textit{36}, L07,706, 2009.

\bibitem{TF10}
Trenberth, K., and J.~Fasullo, Simulation of present-day and
  twenty-first-century energy budgets of the southern oceans, \textit{J.
  Clim.}, \textit{23}, 440--454, 2010{\natexlab{a}}.

\bibitem{TF10b}
Trenberth, K., and J.~Fasullo, Tracking earth's energy, \textit{Science},
  \textit{328}, 316--317, 2010{\natexlab{b}}.

\bibitem{SeamlessTransport}
Trenberth, K., and D.~Stepaniak, Seamless poleward atmospheric energy
  transports and implications for the hadley circulation, \textit{J. Clim.},
  \textit{16}, 3706–3722, 2003.

\bibitem{EEB09}
Trenberth, K., J.~Fasullo, and J.~Kiehl, Earth's global energy budget,
  \textit{Bull. Am. Met. Soc.}, \textit{90}, 311--323, 2009.

\bibitem{Vallis09}
Vallis, G., and R.~Farneti, Meridional energy transport in the coupled
  atmosphere-ocean system: scaling and numerical experiments, \textit{Q. J. R.
  Met. Soc.}, \textit{135}, 1643--1660, 2009.

\bibitem{WildSurface}
Wild, M., Shortwave and longwave surface radiation budgets in {GCMs}: a review
  based on the {IPCC-AR4/CMIP3} models, \textit{Tellus A}, \textit{60},
  932--945, 2008.

\bibitem{WildSolar}
Wild, M., C.~Long, and A.~Ohmura, Evaluation of clear-sky solar fluxes in
  {GCMs} participating in {AMIP} and {IPCC-AR4} from a surface perspective,
  \textit{Journal of Geophysical Research}, \textit{111}, D01,104, 2006.

\bibitem{Wilks}
Wilks, D., Resampling hypothesis tests for autocorrelated field, \textit{J.
  Clim.}, \textit{10}, 65--82, 2005.

\bibitem{Wu09}
Wu, W., and Y.~Liu, Radiation entropy flux and entropy production of the earth
  system, \textit{Rev. Geophys.}, \textit{48}, RG2003, 2009.

\bibitem{Yang99}
Yang, S.-K., Y.-T. Hou, A.-J. Miller, and K.~Campana, Evaluation of the earth
  radiation budget in {NCEP-NCAR} reanalysis with {ERBE}., \textit{J. Clim.},
  \textit{12}, 477--493, 1999.

\bibitem{Yin05}
Yin, J., A consistent poleward shift of the storm tracks in simulations of 21st
  century climate, \textit{Geophys. Res. Lett.}, \textit{32}, L18,701, 2005.

\end{thebibliography}


\pagebreak[4]

\pagebreak[4]

\pagebreak[4]

%
%

\pagebreak[4]

\pagebreak[4]

\pagebreak[4]

\pagebreak[4]

\begin{figure}[htbp] 
   \begin{center}
   \includegraphics[width=0.5\textwidth,angle=270]{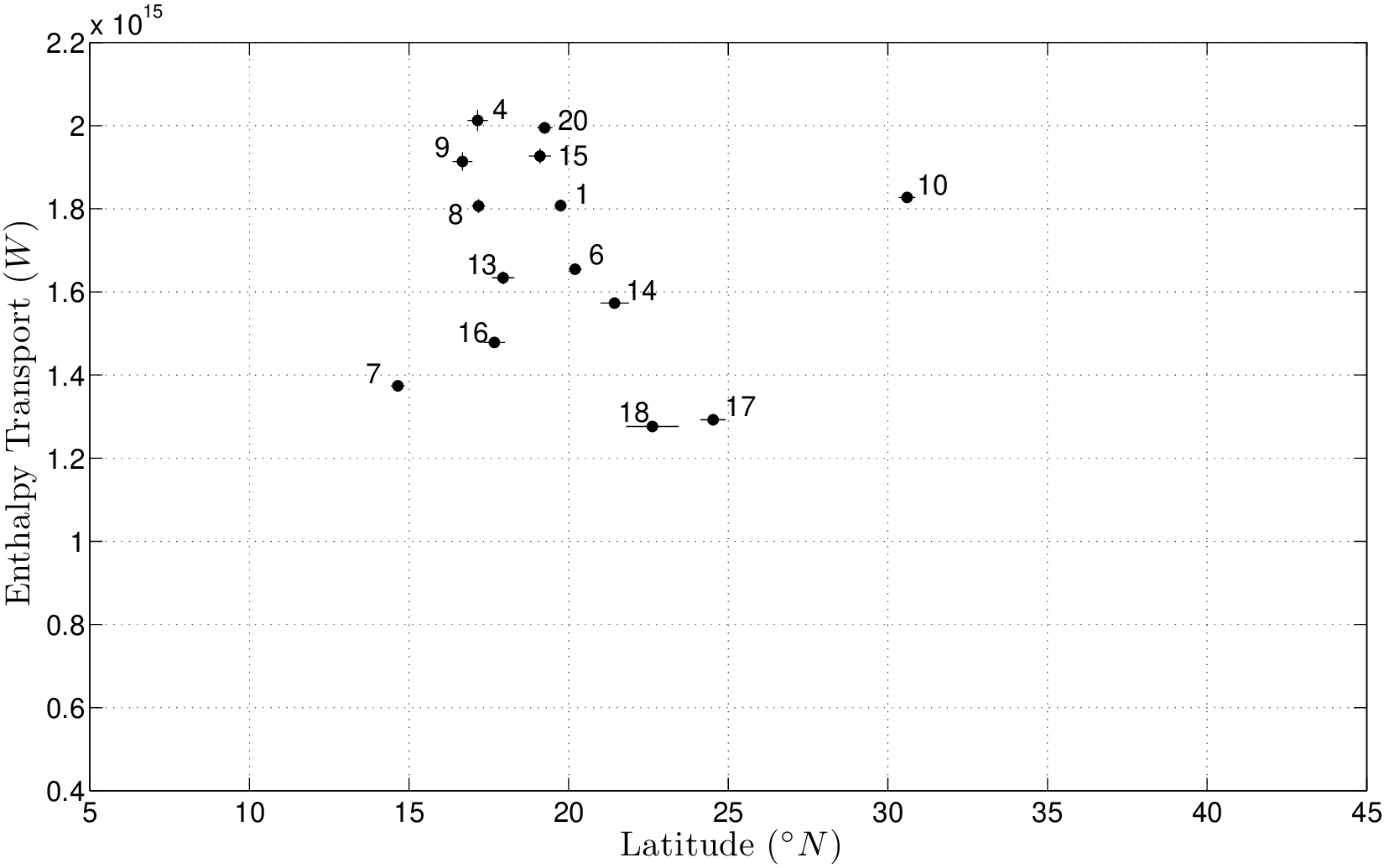}
   \includegraphics[width=0.5\textwidth,angle=270]{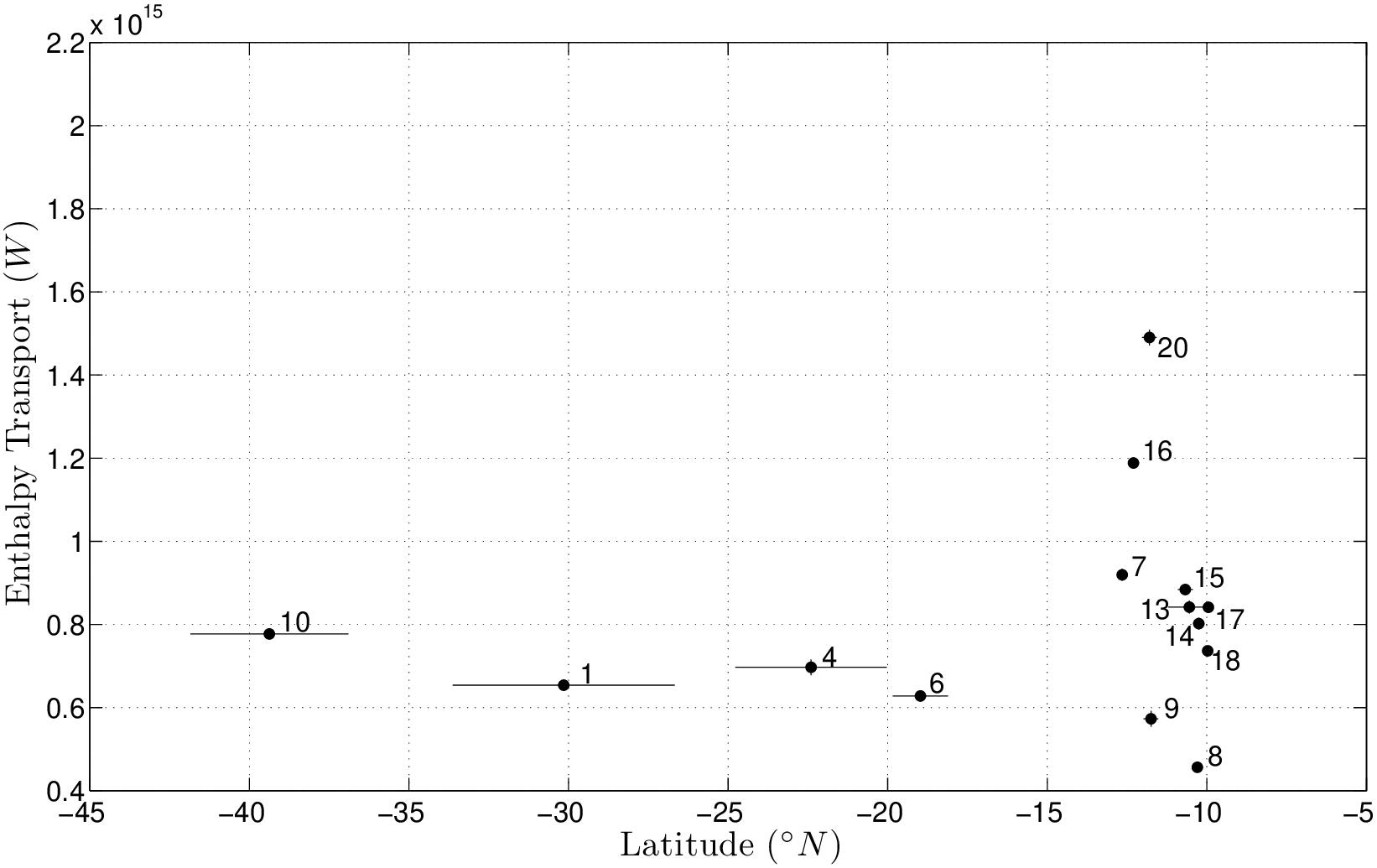}
      \end{center}
   \caption{Value and position of the peak of the poleward meridional enthalpy ocean transport in the preindustrial scenario for the northern hemisphere (first panel) and for the southern hemisphere (second panel).}
   \label{fig:transsea}
\end{figure}

%

\pagebreak[4]

\pagebreak[4]

\pagebreak[4]

\begin{figure}[htbp] 
   \begin{center}
   \includegraphics[width=0.5\textwidth,angle=270]{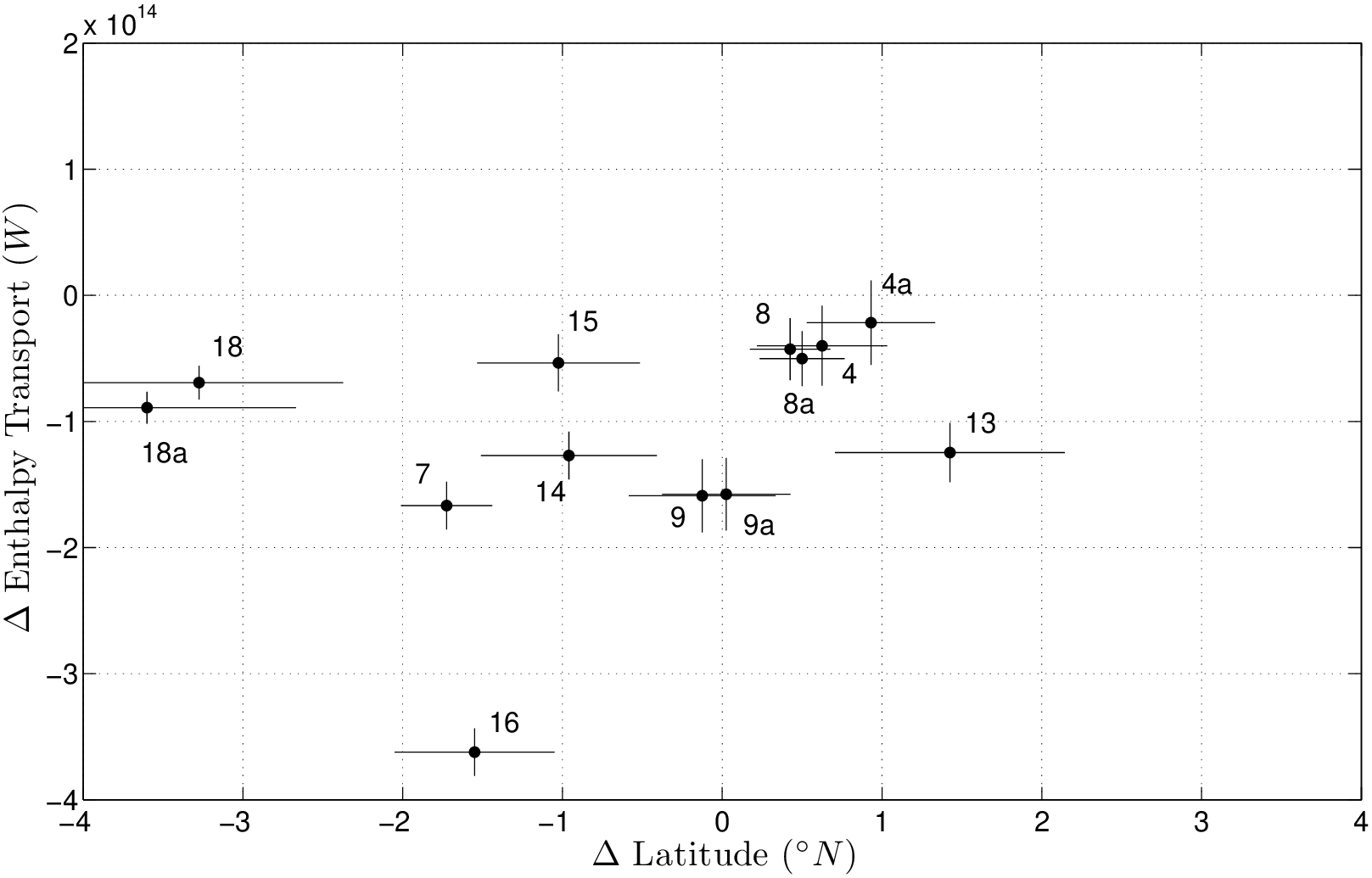}
   \includegraphics[width=0.5\textwidth,angle=270]{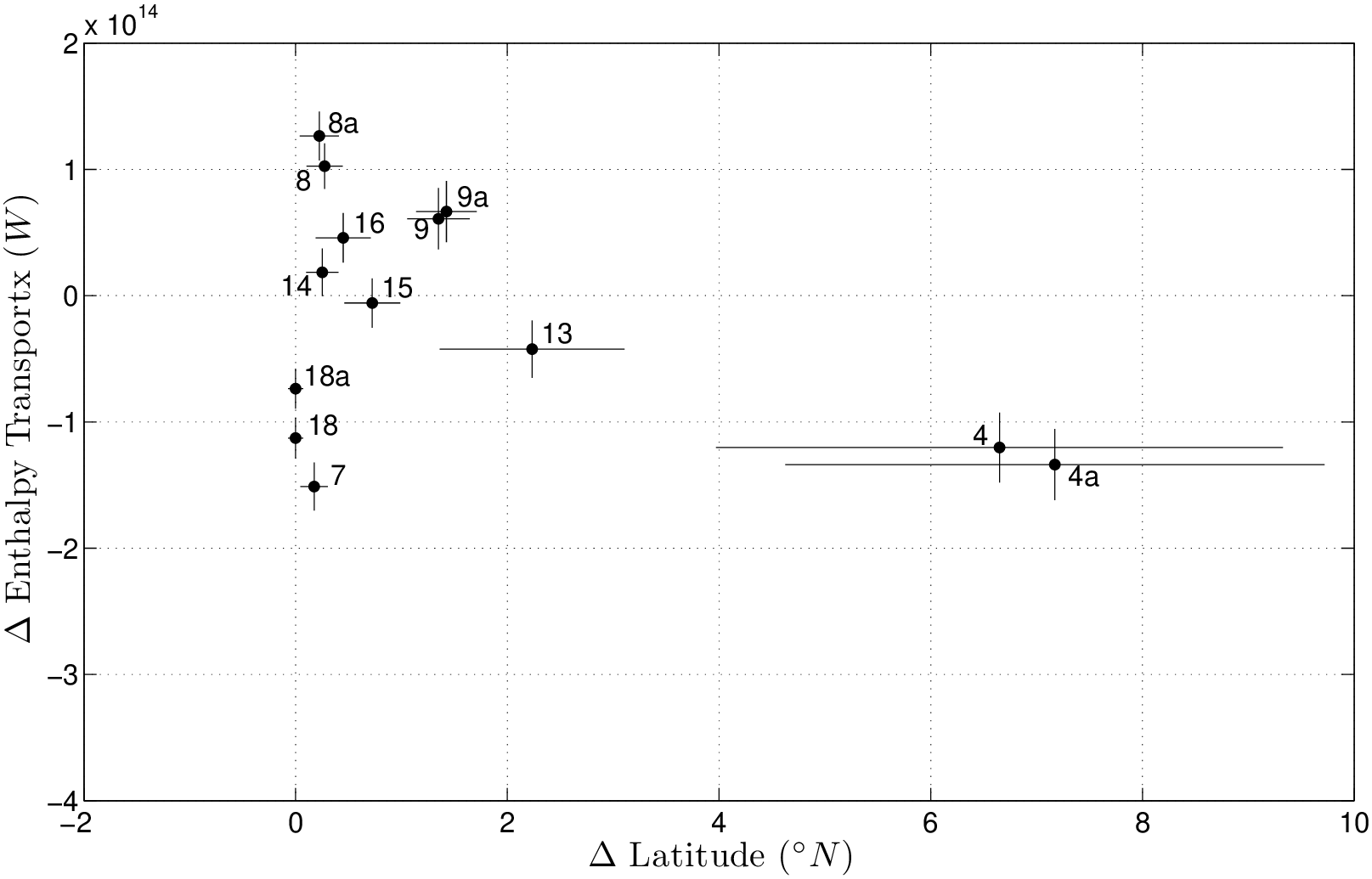}
      \end{center}
    \caption{Difference in the value and position of the peak of the ocean poleward meridional enthalpy transport between the SRESA1B and the pre-industrial simulations for the northern hemisphere (first panel) and for the southern hemisphere (second panel).}
      \label{fig:deltasea}
\end{figure}


%
%


\pagebreak[4]
\newpage

\begin{sidewaystable}[htbp]
  \caption{Models considered in this work, institutions and data availability in each considered time period. Each model is labeled with a number.}
  \centering
  \begin{tabular}{|c|l|l|l|l|l|}
    \hline
     & \textbf{Model} & \textbf{Institution} & \textbf{PI} & \textbf{SRES A1B} & \textbf{SRES A1B} \\
     &                      &                            &                & \textbf{2101-2200}  & \textbf{2201-2300} \\
    \hline
    1 & BCCR-BCM2.0 & Bjerknes Center, \textit{Norway} & yes & no data & no data\\
    \hline
    2 & CGCM3.1 (T47) & CCCma, \textit{Canada} & flux adjustment & flux adjustment & flux adjustment  \\
    3 & CGCM3.1 (T63) &                                    & flux adjustment  & flux adjustment  & flux adjustment  \\
    \hline
    4 & CNRM-CM3 & Met?o France, \textit{France} & yes & yes & 2101-2299\\
    \hline
    5 & CSIRO-Mk3.0 & CSIRO, \textit{Australia} &  80 years, no surf & no data & no data \\
    6 & CSIRO-Mk3.5 &                                     & yes & no data & no data \\
    \hline
    7 & FGOALS-g1.0 & LASG, \textit{China} & yes & yes & no data \\
    \hline
    8 & GFDL-CM2.0 & GFDL, \textit{USA} & yes & yes & yes \\
    9 & GFDL-CM2.1 &                              & yes & yes & yes\\
    \hline
    10 & GISS-AOM & NASA-GISS, \textit{USA} & yes & no data & no data \\
    11 & GISS-EH    &                                       & no surf& no data & no data \\
    12 & GISS-ER    &                                       & no surf & no surf &  2201-2297, no surf \\
    \hline
    13 & HADCM3  & Hadley Center, \textit{UK} & yes &  2108-2199 & no data \\
    14 & HADGEM &                                        &  99 years &  2101-2199 & no data \\
    \hline
    15 & INM-CM3.0 & Inst. of Num. Math, \textit{Russia} & yes & yes & no data \\
    \hline

    16 & IPSL-CM4 & IPSL, \textit{France} & yes & yes &  2201-2231 \\
    \hline
    17 & MIROC3.2 (hires)     & CCSR/NIES/FRCGC, \textit{Japan} & yes & no data & no data \\
    18 & MIROC3.2 (medres) &                                                      & yes & yes & yes \\
    \hline
    19 & ECHO-G & MIUB/METRI/M\&D, \textit{Germany/Korea} &  flux adjustment & flux adjustment & flux adjustment \\
    \hline
    20 & ECHAM5/MPI-OM & Max Planck Inst., \textit{Germany} & yes & no surf & no surf \\
    \hline
    21 & MRI-CGCM2 & Meteorological Research Inst., \textit{Japan} & flux adjustment & flux adjustment  & flux adjustment  \\
    \hline
    22 & NCAR CCSM & NCAR, \textit{USA} & no surf & no surf & no surf \\
    23 & NCAR PCM   &                              & no data &  2101-2199, no surf & no data \\
    \hline
  \end{tabular}
  \label{tab:label}
\end{sidewaystable}

\begin{table}[htbp]
  \caption{Correlation between annual maxima of atmospheric and oceanic poleward transport in each hemisphere. The $95\%$ confidence interval width is $\pm0.20$ in all cases. PI and SRESA1B (2101-2200) scenarios.}
  \centering
  \begin{tabular}{|c|c|c|c|c|}
    \hline
     \textbf{Model} & \textbf{PI-SH} & \textbf{PI-NH} & \textbf{SRESA1B-NH} & \textbf{SRESA1B-SH}\\
    \hline
    1 & $-0.30 $ & $-0.14 $ & - & - \\
    \hline
    2 &  - & - & - & - \\
    3 &  - & - & - & - \\
    \hline
    4 &  $-0.32 $ & $-0.29 $ &  $-0.46 $ & $-0.42 $\\
    \hline
    5 &  - & -  &  - & - \\
    6 &  $-0.36 $ & $-0.26 $ &  - & -  \\
    \hline
    7 &  $-0.48 $ & $-0.20 $ &  $-0.45 $ & $-0.20 $ \\
    \hline
    8 &  $+0.05 $ & $-0.34 $ &  $-0.39 $ & $-0.37 $ \\
    \hline
    9 &  $-0.27 $ & $-0.35 $ &  $-0.21 $ & $-0.12 $ \\
    \hline
    10 &  $-0.62 $ & $-0.36 $ &  - & -  \\
    11 &  - & - &  - & - \\
    12 &  - & - &  - & - \\
    \hline
    13 &  $-0.11 $ & $-0.09 $ &  $-0.26 $ & $-0.21 $ \\
    14 & $-0.08 $ & $+0.01 $ & $-0.27 $ & $-0.24 $\\
    \hline
    15 &  $-0.20 $ & $-0.46 $ &  $-0.05 $ & $-0.28 $  \\
    \hline
    16 &  $-0.31 $ & $-0.33 $ &  $-0.48 $ & $-0.43 $  \\
    \hline
    17 &  $-0.09 $ & $+0.15 $  &  - & -  \\
    18 &  $-0.27 $ & $-0.17 $ &  $-0.11 $ & $-0.08 $ \\
    \hline
    19 &  - & -  &  - & -  \\
    \hline
    20 & $-0.48 $ & $-0.15 $  &  - & -  \\
    \hline
    21 &  - & - &  - & -  \\
    \hline
    22 &  - & -&  - & -   \\
    23 &  - & - &  - & -  \\
    \hline
  \end{tabular}
  \label{tab:scattercorr}
\end{table}

\end{document}